%% file: main.tex
\documentclass[letterpaper, 10pt, final, journal]{IEEEtran}

\IEEEoverridecommandlockouts  

\usepackage{tikz}
\usepackage{algorithmic}
\usepackage[ruled,vlined]{algorithm2e}
\usepackage{url}
\usepackage{times}
\usepackage{amsmath}
\usepackage{amssymb}
\usepackage{color, colortbl}
\usepackage{subfigure}
\usepackage{listings}
\usepackage{multicol}
\usepackage{fancybox}
\usepackage{gensymb}
\usepackage{makecell}
\usepackage{caption}
\usepackage{multirow}
\usepackage{ctable}
\usepackage{comment}
\usepackage{tabularx}
\usepackage{fontawesome}
\usepackage{threeparttablex}
\usepackage{tcolorbox}
\usepackage{threeparttable}
\usetikzlibrary{trees, calendar, fpu}
\usepackage{lipsum,adjustbox}
\usetikzlibrary{shadows}
\graphicspath{{./figures/}}

\usepackage{hyperref}
\usepackage{xcolor}

\definecolor{mygreen}{rgb}{0,0.6,0}
\definecolor{mygray}{rgb}{0.5,0.5,0.5}
\definecolor{mymauve}{rgb}{0.58,0,0.82}

\newcommand\email[1]{\href{mailto:#1}{#1}}
\newcommand\cbox[1]{\begin{tcolorbox}#1\end{tcolorbox}}
\title{Security and Privacy in the Emerging Cyber-Physical World: A Survey}

\author{
\IEEEauthorblockN{
      {Zhiyuan Yu},~\IEEEmembership{Student Member,~IEEE,}
      {Zack Kaplan}, ~\IEEEmembership{Student Member,~IEEE,} \\
      {Qiben Yan},~\IEEEmembership{Senior Member,~IEEE,} 
      {Ning Zhang},~\IEEEmembership{Member,~IEEE}
      } 
      
      \thanks{Z. Yu, Z. Kaplan, N. Zhang is with Washington University in St. Louis, MO, USA. Email: \email{yu.zhiyuan@wustl.edu}, zack.kaplan@wustl.edu, zhang.ning@wustl.edu. \textit{Corresponding author}: Zhiyuan Yu.}
    
      \thanks{Q. Yan is with Michigan State University, MI, USA. Email: qyan@msu.edu.}
}
\begin{document}

\pagenumbering{gobble}
\onecolumn
\begin{center}
{\Large Security and Privacy in the Emerging Cyber-Physical World: A Survey}\\
~\\
{\large Zhiyuan Yu, Zack Kaplan, Qiben Yan, Ning Zhang}\\
~\\
{in {\em IEEE Communications Surveys \& Tutorials}}
\end{center}

\vspace*{\fill}
\noindent
\copyright 2021 IEEE. Personal use of this material is permitted. Permission from IEEE must be obtained for all other uses, in any current or future media, including reprinting/republishing this material for advertising or promotional purposes, creating new collective works, for resale or redistribution to servers or lists, or reuse of any copyrighted component of this work in other
works. DOI:\href{http://doi.org/10.1109/COMST.2021.3081450}{10.1109/COMST.2021.3081450}
\vspace*{\fill}
\clearpage
\pagenumbering{arabic}
\twocolumn

\maketitle
\input{abstract}
\input{intro}
\input{relatedwork}

\input{section2}
\input{section3}

\input{section4}
\input{section5}
\section*{Acknowledgments}
The authors would like to thank the anonymous reviewers for their constructive comments on this paper.
This work was supported in part by US National Science Foundation under grants CNS-1837519, CNS1950171, CNS1949753, CNS-1916926, and CNS-2038995.

\bibliographystyle{IEEEtran}
\bibliography{bibliography.bib}

\input{bio.tex}

\end{document}

%% file: abstract.tex
\begin{abstract}

With the emergence of low-cost smart and connected IoT devices, the area of cyber-physical security is becoming increasingly important. Past research has demonstrated new threat vectors targeting the transition process between the cyber and physical domains, where the attacker exploits the sensing system as an attack surface for signal injection or extraction of private information. Recently, there have been attempts to characterize an abstracted model for signal injection, but they primarily focus on the path of signal processing. 
This paper aims to systematize the existing research on security and privacy problems arising from the interaction of cyber world and physical world, with the context of broad CPS applications. The primary goals of the systematization are to (1) reveal the attack patterns and extract a general attack model of existing work, (2) understand possible new attacks, and (3) motivate development of defenses against the emerging cyber-physical threats.

\begin{IEEEkeywords}
Security, Signal Injection, Side-Channel Information Leakage, Cyber-Physical Systems, Sensors
\end{IEEEkeywords}

\end{abstract}

%% file: intro.tex
\section{Introduction}\label{sec:intro}

Cyber-physical system (CPS) has been an active area of research for a decade, integrating and connecting techniques in embedded systems, communications, and controls to address unique challenges in marrying cyber systems with physical world processes~\cite{cintuglu2016survey}. In recent years, with increasingly lowered cost and improved computational capability, embedded computing systems are pervasively deployed in broad fields, such as aerospace, automotive, chemical production, civil infrastructure, energy, healthcare, manufacturing, materials, and transportation~\cite{7935369}. The recent boom in consumer-facing Internet of Things (IoT) technology has played an important role in the introduction of CPS into the consumer space. It is predicted that there will be a total of 29 billion connected devices by 2022, of which around 18 billion will be related to IoT~\cite{2019_IoTForecast}. Furthermore, the global cyber-physical system market is expected to witness a compound annual growth rate (CAGR) of 8.7\% from 2018 to 2028 and reach \$137,566 million by the end of 2028~\cite{2018-2028}.

As cyber-physical systems play critical roles in our day-to-day lives, attacks on CPS can have particularly severe impacts on human lives and the environment. Recognizing the pressing need to understand the attacks on safety-critical cyber-physical systems, there have been numerous efforts to systemize and categorize the field of attacks and defenses on CPS. For instance, \cite{7935369} and \cite{Lun2016CyberPhysicalSS} quantitatively analyze the broad research trend of CPS security to show the distribution of research topics. \cite{7805372}, \cite{6129371}, and \cite{6622964} categorize CPS research based on different application scenarios such as power grid and industrial control systems (ICS). From the perspective of defenses, \cite{Wu2016} focuses on attack detection and secure estimation techniques, while \cite{6942184} and \cite{Mitchell:2014:SID:2597757.2542049} summarize automatic intrusion detection in CPS. More recently, \cite{giechaskiel2019sok}, \cite{giechaskiel2019framework}, and \cite{yan2020sok} systemize the attacks that leverage physical signals to affect the electronic components (mainly sensors) in cyber-physical systems. However, one of the unique properties of cyber-physical systems is their interaction with the physical world, and it has not been the focus of existing work to analyze the security and privacy problems stemming from such interaction. As the cyber world and physical world are becoming increasingly intertwined, it is important to understand the potential security issues that arise from this evolving threat landscape with new technological developments. 

\begin{figure}[t]
\includegraphics[width=8.4cm]{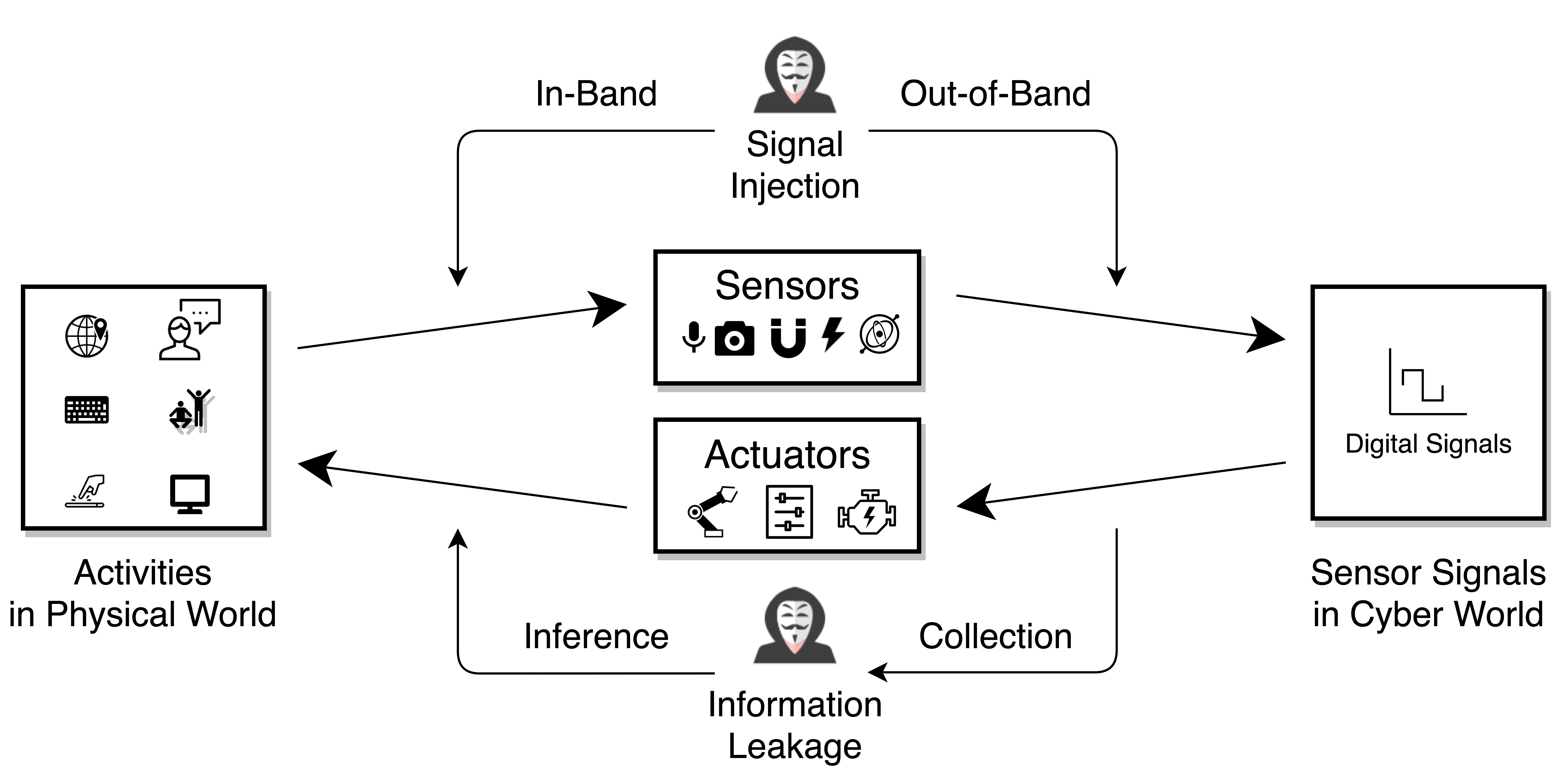}
\caption{The attacker either exploits the transition from the physical world to the cyber domain for \textit{signal injection} attacks, or the transition from the cyber domain to the physical world for \textit{information leakage} attacks.}
\label{fig:fig1intro}
\end{figure}

To fill this gap, we present a comprehensive survey of attacks targeting the interaction between the cyber world and physical world in this paper. More specifically, we explore attacks that target the transition between phenomena in the physical world and signals in the cyber domain, where attackers exploit unintended functionality in the sensing subsystem of CPS to deliver an attack or leverage sensors to pick up unintended emanations from the system. As shown in Fig.~\ref{fig:fig1intro}, we broadly refer these two attacks as \textit{signal injection} and \textit{information leakage}. In a signal injection attack, the attack goal is to manipulate the physical environment, such as acoustic signals, so that the target system obtains a false perception of the physical world, i.e. \textit{exploiting the transition from the physical world to the cyber domain}. In an information leakage attack, the attack goal is to infer sensitive information, such as keystrokes, cryptographic keys, or user activity by sensing and analyzing output from unintended emanations of the target system. Contrary to signal injection attacks, the attacker here intends to infer physical world information by analyzing sensor data, i.e. \textit{exploiting the transition from the cyber domain to the physical world}.

In this paper, we refer to these attacks as \textbf{cyber-physical attacks}, which exhibit the following unique characteristics: (1) the attack surface via sensors is the link between cyber and physical domains; (2) the attackers target the transition process between physical and cyber domains; and (3) different from conventional cyber-domain attacks, these two types of attacks exploit both physical and cyber domains. Therefore, we do not include the following topics in this paper: (1) attacking other cyber components within the CPS, such as the CAN bus \cite{5504804}~\cite{6894181}~\cite{nie2017free}, electronic control unit (ECU) \cite{10.1145/3133956.3134001}~\cite{7518052}, networks \cite{nie2017free}~\cite{hoppe2011security}~\cite{xu2006jamming}, realtime operating systems~\cite{10.1145/2451116.2451145}, and firmware~\cite{cui2013firmware}~\cite{basnight2013firmware}; (2) using sensor properties to fingerprint devices~\cite{Bojinov2014MobileDI}~\cite{8737572}~\cite{8835276} or physical objects \cite{10.1145/3243734.3243735}~\cite{belikovetsky2017detecting}~\cite{VINCENT201577}. 

For \textit{signal injection attacks}, after describing the abstracted model for these attacks, existing researches on signal injection attacks are then categorized and discussed based on the physical vectors being manipulated. Despite remarkable progress in the past few years, there are still many challenges in this area for both attacks and defenses, thus we end the discussion with limitations and future opportunities. For \textit{information leakage attacks}, different from signal injection, they often have more common objectives rather than the exploited physical channels, therefore we group and discuss them based on the attack goal. Lastly, we conclude our review with an in-depth discussion on new opportunities in both attacks and defenses. 

\textbf{Contributions} - The main contributions of this paper are:

\begin{itemize}
  \item We develop a general system model for cyber-physical attacks, and formalize attack requirements and different threat models. 
  
  \item We categorize and discuss existing research in both physical-to-cyber signal injection and cyber-to-physical information leakage attacks as well as the corresponding defenses, followed by discussion on the limitations and opportunities in both areas. 
  
  \item We highlight how latest development in computer network and embedded system will change the evolving landscape of attack and defense in the emerging tightly connected cyber-physical world. 
\end{itemize}

\textbf{Organization} - This paper will be organized as follows: Section \ref{sec:relatedwork} will summarize and categorize related surveys about CPS security. Section \ref{sec:architecture} will give a generalized system model of broad CPS and outline cyber-physical interfaces. Section \ref{sec:siginject} will develop a general model for signal injection attacks and survey the existing literature. Section \ref{sec:infoleakageatks} will develop a general model for information leakage attacks and survey the existing literature. Detection methods and prevention techniques against cyber-physical attacks are presented in Section \ref{sec:defense}. Besides, Section \ref{sec:futureOutlook} will provide research opportunities in both cyber-physical attacks and defenses. Finally, Section \ref{sec:conclusion} will summarize the survey and draw conclusions.

%% file: relatedwork.tex
\section{Related Work}\label{sec:relatedwork}

\subsection{Existing Literature}
Due to the rapid integration of various CPS into society, the security and safety properties of these systems are gathering increasing attention. There have been several studies that attempt to systematize the security challenges and opportunities in cyber-physical systems from different perspectives. As we summarize in Table \ref{tab:relatedwork}, they generally fall into four categories: CPS security overview, domain-specific analysis, algorithm-level analysis, and sensing pipeline analysis. However, our survey systematizes existing research from a fundamentally different angle of cyber-physical interaction, focusing on both attacks from the physical world to the cyber world - \textit{signal injection}, and attacks from the cyber world to the physical world - \textit{information leakage}. This difference in systematization leads to a complementary analysis of the field and offers a distinct perspective for how new defenses can be constructed. In the rest of this section, we will present these existing works and the problems they address.

\input{Table_relatedwork2}

\subsubsection{CPS Security Overview}
The first category of surveys investigate and summarize the entire CPS security field from a high level. Some of these surveys aim to comprehensively present security issues in each layer of CPS architecture~\cite{humayed2017cyber}~\cite{wolf2017safety}~\cite{alguliyev2018cyber}. Although the threats embedded within sensors have been included in these surveys, the discussions are limited to provide breath on in-band injection attacks and examine the impacts of classical attacks such as replay attacks~\cite{alguliyev2018cyber} and signal jamming~\cite{humayed2017cyber}. Besides, as privacy leakage from sensing components is not the focus, therefore the discussion is minimal. Some other surveys investigate research topic distributions and trends by quantitatively analyzing the representative researches and surveyed papers. As an example, Lun et al.~\cite{Lun2016CyberPhysicalSS} use systematic mapping to create a quantitative analysis among a collection of 118 papers concerning CPS security in which publication trends, focus of existing research, and strategies used to validate the existing mechanisms are statistically analyzed. In the same vein, Giraldo et al.~\cite{7935369} investigate existing surveys about CPS security in terms of attacks, defenses, network security, security level implementation, computational strategies, and also quantitatively summarize research trends. 

\textbf{Gap Analysis: }The related work in this area aims to provide breath on various security issues, from network security to application software security, from technological to operational, and from sensors to algorithms. The large scope of the study limits the level of detail that can be presented and analyzed in these surveys. Our survey is different in that there is a strong focus on cyber-physical interaction, which results in an in-depth analysis of the embedded security issues, while other aspects such as traditional network security are not included in the discussion of our survey.

\subsubsection{Domain-Specific Analysis}
Some surveys aim to systematize the attacks and defenses in specific application domains of CPS. For example, \cite{7805372}, \cite{6129371}, \cite{Jawurek2012PrivacyTF}, and \cite{pasqualetti2011cyber} focus on power grids, \cite{6956585} and \cite{Camara2015SecurityAP} focus on medical devices, \cite{6622964} and \cite{Urbina2016SurveyAN} focus on ICS, \cite{Altawy:2016:SPS:3015781.3001836} focuses on drones, etc.

\textbf{Gap Analysis: }As each of them focuses on one specific CPS application domain, the insights and conclusions drawn are often domain-specific. For example, switching attacks targeting circuit breakers and electricity market attacks discussed in the context of power grids~\cite{7805372}~\cite{6129371} are highly scenario-dependent, and are not included in the security discussions of other types of CPS. In this work, instead of analyzing the security of a specific domain, we organize around the physical medium with which an attacker can target the cyber-physical systems, i.e. the cyber-physical attack vectors. For each attack vector, we discuss how existing works are leveraging it to compromise all types of cyber-physical systems. Therefore, some insights into new directions of attacks and defenses can be generalized to cyber-physical systems in other domains with some adaptation.

\subsubsection{Algorithm-Level Analysis}
There are also some surveys that focus on the exploration of different existing algorithms for intrusion detection and anomaly mitigation. Wu et al. \cite{Wu2016} demonstrate recent research concerning topics of \textit{attack detection} and \textit{secure estimation and control} techniques, while \cite{6942184} and \cite{Mitchell:2014:SID:2597757.2542049} look into the self-detection of intrusion in CPS. Giraldo et al. \cite{giraldo2018survey} summarize works where time-series models are created via monitoring the physical world to identify anomalies, control commands, or sensor readings. Algorithmic analysis for CPS primarily focuses on the detection of deviation and mitigation of control for systems under attack. For the algorithmic solution to be effective, there is often an assumption about the physical model, but the overall design philosophy still generally applies. However, an implicit assumption behind these solutions is an accurate or at least error-bounded perception of the real world. 

\textbf{Gap Analysis:} Our survey focuses on the cyber-physical interactions, which is a key foundation for forming an accurate perception of the physical world. As a result, our survey and the algorithmic approach are complementary. Approaches in the algorithmic survey can be effective defenses against some of the attacks we discuss, such as false data injection~\cite{mo2010false}, and the attacks in this survey can in turn inform the algorithmic research of the emerging threat models.

\subsubsection{Sensing Pipeline Analysis}
Lastly, after several demonstrations of the feasibility of the manipulation of physical world signals to attack cyber systems~\cite{Hormuz_GPS}~\cite{Iran_GPS}, some efforts have been made to systematize how maliciously crafted physical environments can affect electronic components such as sensors in cyber-physical systems. Giechaskiel et al. \cite{giechaskiel2019framework} first develop a general model for signal injection attacks, and then propose an algorithm to calculate the security level of real systems such as smartphones. Following this work, they go deeper and formalize a sub-domain of signal injection, i.e. out-of-band signal injection attacks, where the injected signal is out of the intended range of receivers~\cite{giechaskiel2019sok}. Yan et al. \cite{yan2020sok} broaden the scope of out-of-band signal injection and consider the problem along the entire signal processing pipeline, which they call \textit{transduction attacks}, where the victim sensor transduces physical signals to analog output with the possibility of combining effects from multiple processing stages. However, these works focus exclusively on signal injection attacks. 

\textbf{Gap Analysis:} While how the physical world could maliciously impact cyber data is analyzed, attacks leveraging cyber world capabilities to eavesdrop on the physical world are not considered in these works.

\subsection{Gap Analysis and Our Focus}
While the area of CPS security has attracted significant attention, none of the existing work structures analysis based on cyber-physical interfacing. Unlike traditional security issues, such as network and control security, the emerging threats surrounding the transmission of signals from and to the physical world, namely, physical-to-cyber signal injection attacks and cyber-to-physical information leakage attacks, have not been studied systematically yet. The absence of knowledge systemization can significantly hinder the development of novel defenses for these new threats, where traditional defenses have limited effectiveness on. To bridge this gap, we aim to summarize and generalize attacks targeting sensing components from the perspective of cyber-physical interaction. To this effect, we develop a generalized system model of CPS as well as an attack model for each attack, which reveals the capabilities of these attacks when applied to broad CPS including cellphones, autonomous vehicles, drones, UAVs, medical implant systems, voice-controlled systems, and others. More importantly, we provide a new way to understand and motivate effective defenses based on our systemized threat model. By viewing security issues from this brand new perspective with a specific focus on signal attacks, we believe this work will help fill the knowledge gap and motivate new research in this emerging area.

%% file: Table_relatedwork2.tex
\begin{table}[t]
\scriptsize
\centering
\caption{Summary of Related Surveys in CPS Security}
\begin{tabular}{|c|c|c|}
\hline
Scope                                & Specific Domain              & Reference                                                 \\ \hline
  & CPS security summary        & \cite{humayed2017cyber},\cite{wolf2017safety},\cite{alguliyev2018cyber}        \\ \cline{2-3} 
                                           &CPS security surveys          & \cite{7935369}                                                   \\ \cline{2-3} 
                                            \multirow{-3}{*}{CPS security overview}&CPS security research trend           & \cite{Lun2016CyberPhysicalSS}                                    \\ \hline
  & Power grid                   & \cite{7805372},\cite{6129371},\cite{Jawurek2012PrivacyTF},\cite{pasqualetti2011cyber} \\ \cline{2-3} 
                                           & Medical device              & \cite{6956585},\cite{Camara2015SecurityAP}                              \\ \cline{2-3} 
                                           & Industrial control system   & \cite{6622964},\cite{Urbina2016SurveyAN}                                \\ \cline{2-3} 
                                           \multirow{-4}{*}{Domain-specific analysis}& Drone                        & \cite{Altawy:2016:SPS:3015781.3001836}                           \\ \hline
 & Secure estimation \& control & \cite{Wu2016}                                                    \\ \cline{2-3} 
                                           & Intrusion detection          & \cite{6942184},\cite{Mitchell:2014:SID:2597757.2542049}                 \\ \cline{2-3} 
                                           \multirow{-3}{*}{Algorithm-level analysis}& Anomaly identification       & \cite{giraldo2018survey}                                         \\ \hline
\multirow{2}{*}{Sensing pipeline analysis} & Out-of-band injection        & \cite{giechaskiel2019sok},\cite{giechaskiel2019framework}              \\ \cline{2-3} 
                                           \multirow{-2}{*}{Sensing pipeline analysis}& Transduction attack          & \cite{yan2020sok}                                                \\ \hline
Cyber-physical interaction                   & \begin{tabular}[c]{@{}c@{}}Signal injection\\ \& Information leakage\end{tabular} & \textbf{Our survey}                          \\ \hline

\end{tabular}
\label{tab:relatedwork}
\end{table}

%% file: section2.tex
\section{CPS System Model}\label{sec:architecture}

In this section, we first discuss the abstracted CPS architecture and the interaction between components. This is followed by a description of the cyber-physical interface, which consists of the sensing system and actuator. Lastly, we will discuss the unique threats stemming from such cyber-physical interfacing as compared to conventional cyber attacks on networks and software systems.

\begin{figure}[h]
  \includegraphics[width=8.4cm]{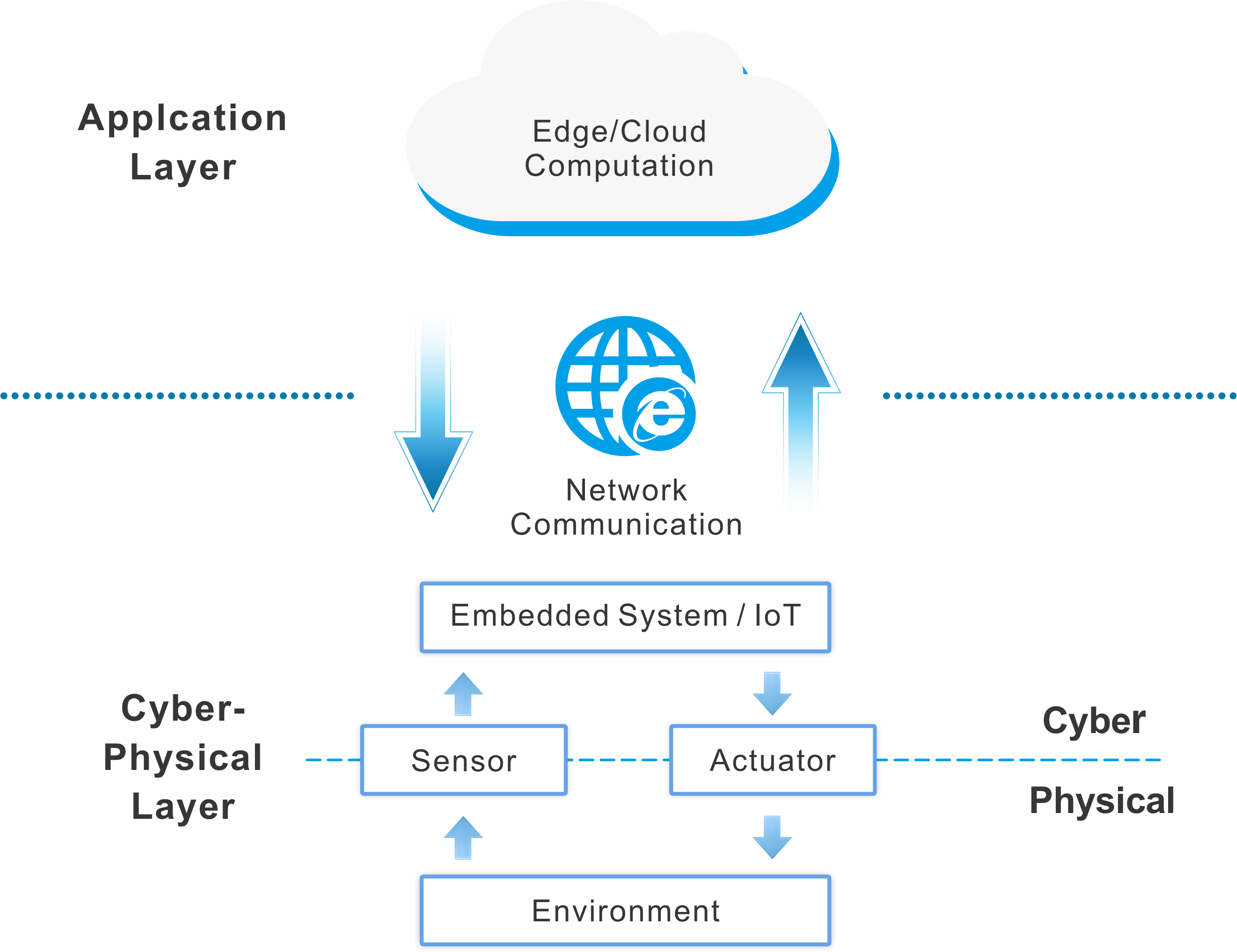}
  \caption{High-level CPS architecture. CPS generally consists of the \textit{application layer} and \textit{cyber-physical layer}. These two layers communicate via networks, and the \textit{cyber-physical layer} can be further divided into \textit{cyber} and \textit{physical} domains.}
  \label{fig:fig2CPSArchitecture}
\end{figure}

\subsection{High Level Architecture}

As shown in Fig. \ref{fig:fig2CPSArchitecture}, the general model for a CPS consists of two layers: the \textit{application layer} and the \textit{cyber-physical layer}. 
The \textit{application layer} facilitates different ways for the \textit{cyber-physical layer} to interact with local or external applications. The key components in the \textit{cyber-physical layer} we consider consist of the physical environment, sensor, actuator, embedded system, and IoT device. Generally, sensors will collect physical signals from the environment and digitize them for the embedded system, which consists of the software or hardware controller that receives and processes digital signals from the sensors. Based on the control algorithm, the embedded controller then outputs the corresponding control signals to the actuators. This feedback loop often has predictable timing behaviors to ensure that the information is processed and acted upon in a safe manner. If the CPS is connected to a network, the embedded system is also responsible for transmitting and receiving network signals to communicate with various services via cloud technologies. 

Since this paper focuses on the security problems that arise from the cyber-physical interactions, we will focus on the discussions relevant to the cyber-physical layer. The following Section \ref{subsec:sensingSys} and \ref{subsec:UniqueThreats} will present how cyber data and physical phenomena interact and the resulting unique threats. 

\subsection{Cyber-Physical World Interfacing}\label{subsec:sensingSys}

Sensors and actuators are the two main components that enable cyber-physical world interfacing. Sensors in CPS generally serve as physical-interfacing nodes that interact with environments, components, or other miscellanea that are not a part of that system~\cite{beavers2017intelligence}. They provide usable outputs in response to specific measurands~\cite{McGrath2013}, by which cyber-physical systems form accurate perceptions of the surrounding environment, leading to more autonomous, intelligent, and reliable systems. Actuators, on the other hand, enable cyber systems to change the physical state of their surroundings. Besides traditional actuators in larger-scale CPS, emerging IoT devices also include actuation components to affect their environments and coordinate among various IoT devices. However, malicious attacks are generally initiated from user input, and the actuator doesn't take input directly from the physical world, therefore the majority of existing attacks still focus on injecting false commands in order to perform cyber-to-physical exploitation. This false command injection can either be realized using a false cyber-domain message, such as an injected network message or control flow hijacking in software, or via exploitation of an unintended receiver in the electronic component to deliver malicious signals. Besides, as the sensor nodes are located at the border between what belongs to CPS and what does not, these sensors are more susceptible to attacks because of their exposure to both cyber and physical domains~\cite{beavers2019intelligence}. Therefore, we will primarily focus our discussion on the sensing system.

Sensors can be categorized into two types based on the way they probe surroundings: \textit{passive sensors}, and \textit{active sensors}. 
\begin{itemize}
    \item \textit{Passive sensors}, such as microphones and MEMS motion sensors, receive pre-existing physical signals and do not emit external stimuli. They are essentially listening devices for physical phenomena, whose purpose is to simply relay the measured signals to upper-layer software.
    \item \textit{Active sensors}, such as radar and LiDAR, probe their surroundings by actively emitting signals to evoke physical responses (i.e., reflected signals), from some measured entity such as surrounding objects. The reflected signal is then measured by the receiver. The sensor contains an on-board controller, which analyzes the relationship between the transmitted and received signals to infer features of the surrounding environment, such as the proximity, relative angle, and shape of nearby objects.
\end{itemize}

Even given high diversity and heterogeneity of sensors, they often consist of the same basic components: sensing mechanism, amplifiers, filters, and analog-to-digital converters.


\begin{figure*}[h]
\centering
\includegraphics[width=16.8cm]{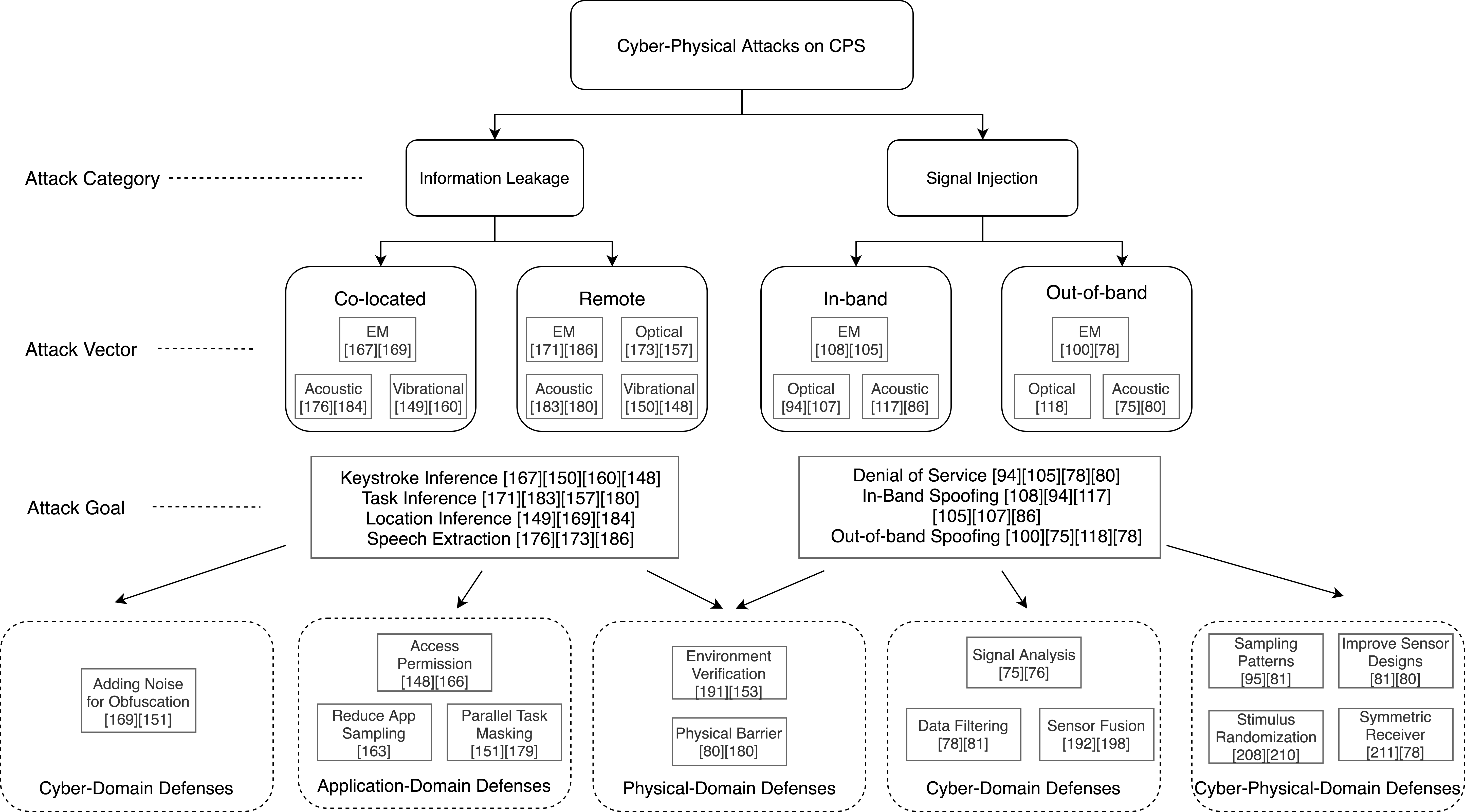}
\caption{Taxonomy of cyber-physical attacks and defenses}
\label{fig:fig10Taxonomy}
\end{figure*}

\subsubsection{{Sensing mechanism}} 
While active sensors are more complex in their operation, their signal reception subsystems are similar to that of passive sensors. The sensing mechanism is the component that responds to changes in a given physical parameter, $s(t)$, and generates an output voltage, $v(t)$, which is directly proportional to the physical signal. This is mathematically represented as $v(t) = \alpha \cdot s(t)$, where $\alpha$ is the proportionality constant, or the \textit{sensitivity} of the sensor. This relationship is referred to as \textit{linearity}, and is a desired quality in a sensor. In reality, sensors exhibit non-linear behavior, which means for certain ranges of the physical signal, the sensor output is not directly proportional to the input. Despite being a non-ideal characteristic of sensors, attackers can actually use non-linearity to their advantage when attempting to inject out-of-band signals. This is further discussed in Section \ref{subsubsec:sysmodelForOOBSigInj}.

\subsubsection{{Amplification and filtering}}
In some sensors, the voltage signal, $v(t)$, is amplified to reach the voltage specification of the microcontroller, and filtered to remove any unwanted frequency components. When injecting out-of-band signals, the filtering stage can pose a challenge for attackers, since the injected signals can be severely attenuated by the filter.

\subsubsection{{Analog-to-digital conversion}}
The analog-to-digital converter (ADC) converts the analog voltage signal to a discrete digital signal by sampling the signal at a certain frequency, which is a parameter known as the ADC's \textit{sample rate}, and converts samples to certain-bit numbers, which is denoted as its \textit{resolution}. More details are presented in Section \ref{subsec:techOOBSigInj}.

\subsection{Unique Threats of Cyber-Physical Interfacing}\label{subsec:UniqueThreats}

\subsubsection{Unique Attack Vector}
One unique aspect of cyber-physical attacks is their transition from one domain to the other domain, in other words, it is possible for a cyber attack to have a physical effect. This unique attribute of these attacks has a significant impact on the risk it presents to our society. While there have been studies pertaining to traditional cyber security issues in the cyber-physical system applications domain, such as power grids~\cite{7805372}~\cite{6129371}~\cite{Jawurek2012PrivacyTF}~\cite{pasqualetti2011cyber}, it is not our focus since they are conventional security problems. More recently, people are increasingly aware of threats brought about by IoT devices, such as leveraging IoT devices to perform DDoS attacks~\cite{kolias2017ddos}~\cite{de2018ddos}. However, this remains a network security problem, widely accessible and vulnerable IoT devices just amplify the attack impact. In this survey, we focus on a new attack vector, cyber-physical interfacing. It is unlike traditional network or software attacks, which are well-known with mature defensive technologies such as firewalls, network intrusion detection systems, and virus scanners. Attacks that target vulnerabilities of the physical construction of computing systems completely evade traditional cyber defenses, due to the fundamental limitation on the amount of information received by the cyber domain after going through layers of communication abstractions (e.g., signal demodulation). Therefore, it is important for the community to understand this changing landscape of threats.

Fig.~\ref{fig:fig10Taxonomy} shows a taxonomy of existing cyber-physical attacks and defenses on CPS we systematize. The attackers can either manipulate cyber data via injecting physical emanations, or eavesdrop on physical information from sensor data. As the approaches taken by the attacker are closely related to the type of signals and adversary goals, we will formalize and demonstrate various attacks using these attack vectors and attack goals in the following Section \ref{sec:siginject} and \ref{sec:infoleakageatks}. Additionally, we summarize the existing defenses that protect CPS from cyber-physical attacks. They are categorized based on the domain where they are implemented because defenses in the same domain share similar principles and limitations, which will be further discussed in Section \ref{sec:defense}.

\subsubsection{Emergence of Consumer IoT}
The concept of IoT emerged from the radio-frequency identification (RFID) community and initially focused on the ability to track the location and status of physical objects~\cite{ashton2009internet}~\cite{ibarra2017tracking}. Today, with the rapid development of wireless sensor networks (WSN)~\cite{raghavendra2006wireless} and cloud computing~\cite{armbrust2010view}, the term IoT has been expanded to refer to networked devices and systems~\cite{brambilla2014simulation}~\cite{greer2019cyber}, and can be used to describe systems as large as smart farms~\cite{ryu2015design}~\cite{jayaraman2016internet}, cities~\cite{mehmood2017internet}, and as small as individual nodes in a sensor swarm~\cite{gunasagaran2015internet}. The concept of IoT overlaps significantly with cyber-physical system~\cite{greer2019cyber}. This is because many key technologies that are essential to CPS also form the foundation of IoT, such as machine-to-machine (M2M) communication~\cite{theoleyre2013internet}~\cite{wan2013machine}, device-to-device (D2D) communication~\cite{bello2014intelligent}, and sensing~\cite{yan2020sok}. For example, industrial internet of things (IIoT) is an important area that intersects significantly with CPS~\cite{xu2018survey}. It is poised to bring revolutionary changes to realize the vision of the next-generation industrial revolution (known as Industry 4.0)~\cite{lu2017industry}. With massive numbers of smart sensors and actuators interconnected and integrated into industrial systems, the M2M communication and the cloud/edge intelligence play a key role in enabling intelligent manufacturing~\cite{coda2019big}~\cite{trinks2018edge}. While IIoT has become an increasingly important area where security protection would be extremely critical~\cite{sadeghi2015security}, the attacks and defenses discussed in our survey could be applied to systems with physical interfaces including IIoT. 

To be more precise in our discussion in this survey, we will use IoT to refer to the consumer electronics that enable smart connections. As consumer IoT devices become ubiquitous, they are significantly changing how our cyber world and physical world interleave in day-to-day life. As a matter of fact, while the cyber-physical interface has always been there, the security aspect of these systems has not received much attention until recently. The large-scale cyber-physical systems such as aircraft and IIoT systems are often less accessible to the general public and security researchers. As a result, much of the existing research focuses on security problems (either injecting signals from the physical world, or stealing physical information from the cyber world) of IoT consumer devices. However, the principles and techniques derived from research on IoT devices are often generally applicable to the larger domain of cyber-physical systems.

Lately, IoT has been in the news, either as the victim of cyber attacks or as the enabler for attackers. Cyber-physical attacks under study in this survey are related to these \textit{IoT attacks}. However, they are different in several respects. Cyber-physical attacks refer to attacks on cyber-physical interfacing, therefore traditional network attacks that compromise IoT devices are not in the scope of cyber-physical attacks. On the other hand, cyber-physical attacks generally apply to any system that interfaces with the physical world, which include consumer IoT devices. 

\subsubsection{Distinguish Attacks on IoT and CPS}
Although attacks can be generalized to target both IoT devices and broad cyber-physical systems, they can differ in deployment architecture, attack impact, and physical environment.

First, the software architecture for consumer IoT and CPS can be significantly different. Consumer IoT devices often adopt the architecture with a lightweight embedded device coupled with a cloud backend to provide smart-living services to consumers. However, large cyber-physical systems often have different types of desktop/server class control stations for operational management by a dedicated team. This results in significant differences in the attack surface between the two types of systems, leading to distinct exploitation and defense techniques both technically and operationally. 

Second, the impact of an attack can be different. Many consumer IoT devices are not safety-critical, so an attack may not directly induce significant harm. However, for larger CPS, many of them are critical systems, such as power grids, which have been the unfortunate targets of multiple recent attacks that paralyze even an entire region~\cite{case2016analysis}. Notably, consumer IoT attacks have also changed the threat model in cyberspace. Due to their pervasive deployment, there is a large number of unprotected IoT nodes on the Internet that can be used to form a botnet. Unlike a cyber-physical attack that compromises a single or small number of CPS within close physical proximity, vulnerabilities in IoT can lead to manifestation of large-scale traditional security problems, such as distributed denial of service (DDoS) attacks on dynamic DNS (DynDNS) service~\cite{antonakakis2017understanding}. As a matter of fact, IoT devices have been the primary force behind the DDoS botnet attacks~\cite{scott2020IoT}, where giant botnets formed of vulnerable IoT devices significantly increase the attack strength and the number of compromised services~\cite{cisco2020cisco}. Besides, attacks on IoT devices can go beyond network security and cause widespread effects. An example is the mobile crowdsensing system, where individual devices contribute to a collective result of the current physical environment~\cite{zhang2015incentives}. In this scenario, a small number of compromised IoT nodes can launch data poisoning attack to influence the result of the output~\cite{miao2018towards}~\cite{tahmasebian2020crowdsourcing}~\cite{miao2018attack}.

Lastly, differences can also lie in the physical environment in general cases. The physical protections for complex CPS can vary significantly. For example, the physical perimeter protection for a self-driving car is different from a gas line crossing rural regions. This diversity raises unique challenges for a generalized cyber-physical attack and defense model. Strategies will have to be devised for individual cases based on the expected threat and risk if compromised. On the contrary, consumer IoTs often have well-defined environments both in the physical world (physical perimeter of a house or worksite) and the cyber world (lightweight frontend and cloud backend). The homogeneity of deployment environment for consumer IoTs makes it easier to generalize approaches for both the attack analysis and protection.

%% file: section3.tex
\section{Physical to Cyber: Signal Injection Attacks}\label{sec:siginject}

Signal injection attacks are attacks in which the attacker injects maliciously crafted signals to the target CPS, inducing unintended behavior of the system. On the one hand, these attacks are designed to disrupt cyber systems via physical signals, therefore they can be regarded as \textit{physical-to-cyber} attacks. On the other hand, the attackers have to proactively manipulate the physical world (often the physical properties of the signal), thus they can be regarded as \textit{active} attacks. An adversary can inject the signals directly from within the intended reception frequency of sensors embedded in the CPS, or outside the intended range. These two types of attacks are coined in \cite{giechaskiel2019sok} as (1) \textit{in-band signal injection}, and (2) \textit{out-of-band signal injection}, respectively.

\begin{figure}
\includegraphics[width=8.4cm]{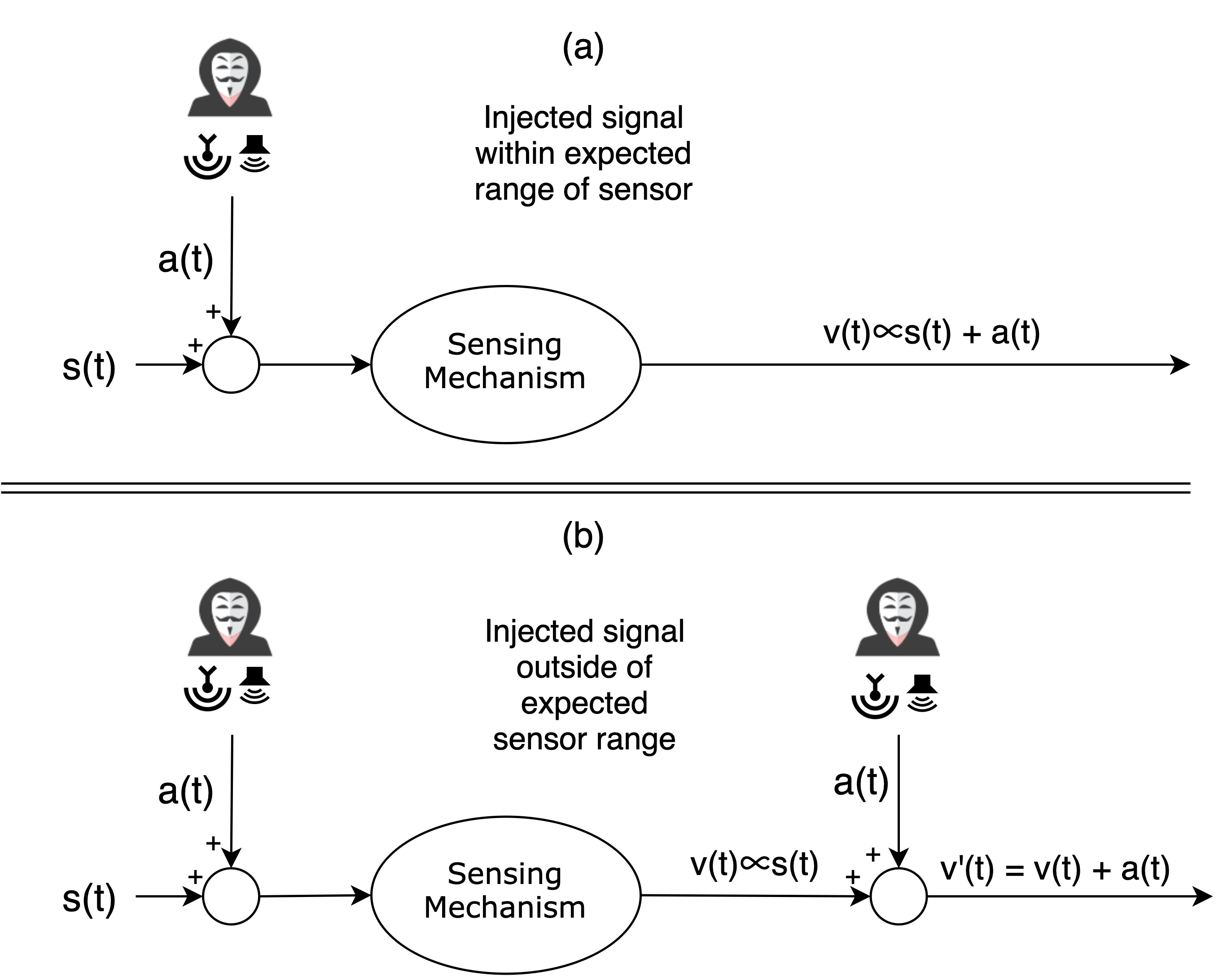}
\caption{(a) \textit{In-band signal injection attacks}. The injected signal manipulates the physical input with the same type of signals. (b) \textit{Out-of-band signal injection attacks}. The injected signal is outside the expected range of the sensor, and can be injected before (targeting microphones, motion sensors, etc.) or after (targeting PCB traces) the sensing mechanism.}
\label{fig:fig4SignalInjectionAttack}
\end{figure}

\begin{enumerate}
    \item \textit{In-band signal injection: }Attacks in which the adversary injects signals that are within the exploited sensor's intended frequency of operation, such as using a bright LED to blind a light sensor. By injecting in-band signals, the adversary can only affect sensor readings by manipulating the physical phenomenon being sensed. 
    \item \textit{Out-of-band signal injection: }The adversary exploits a hardware flaw to inject signals that are out of the intended range into the sensing system or components not intended as receivers, often with the goal of gaining implicit control over the sensor's readings. Contrary to in-band injection attacks, the attacker does not manipulate the property being measured by the sensor.
\end{enumerate}

Fig. \ref{fig:fig4SignalInjectionAttack} demonstrates these two types of attacks. Fig. \ref{fig:fig4SignalInjectionAttack}(a) shows an in-band signal injection attack, where the attacker is directly manipulating the physical property being measured by the sensor, causing the sensing mechanism to output a voltage proportional to the sum of the injected signal, $a(t)$, and the physical signal, $s(t)$. Fig. \ref{fig:fig4SignalInjectionAttack}(b) shows an out-of-band attack, in which the maliciously crafted signal $a(t)$ is injected into $v(t)$ by either injecting into (e.g., cast ultrasound on microphone~\cite{zhang2017dolphinattack}~\cite{roy2018inaudible}~\cite{yan2019feasibility}) or bypassing the sensing mechanism (e.g., inject EMI signal into leads of ECG~\cite{kune2013ghost}, attack voice control systems by injecting EMI into power socket~\cite{latestemi}, etc.)

\subsection{Threat Model} \label{subsec:SigInjThreatModel}

This section will develop a generalized attack model for signal injection attacks. Specifically, we summarize the requirements of the attackers, describe the attack vectors, and present attack impacts based on sensor feedback.

\subsubsection{Attack Requirements}
Attack requirements are the criteria that must be met for the attack to be effective. We propose two general requirements for signal injection attacks:

\begin{itemize}
    \item \textbf{Plausible input}: This requires the maliciously crafted signals to be effectively received by the system. For example, for in-band attacks, this requires the attacking signals are measured by sensors as valid input; while for out-of-band attacks, the injected signals are required to induce effects on various unintended receivers.
    \item \textbf{Meaningful response}: This requires the injected signals to be reflected in the behaviors of the system. Ultimately, the goal of the attacker is to attack the system in a meaningful way, so simply causing changes in the sensor mechanism is not an attack, unless the sensor output in some way dictates the overall behavior of the system. For instance, casting weak light on the camera will be received as \textit{plausible input} by the CPS, but will not affect the system behavior; contrarily, casting a strong light on a camera will blind it and cause DoS to the target.
\end{itemize}

\subsubsection{Attack Vector}
In a signal injection attack, the attack vector is described by the frequency characteristics of the signal (in-band or out-of-band) and the type of signal. Existing work primarily focuses on three categories: \textit{acoustic, electromagnetic}, and \textit{optical}. As a result, the attacker is able to induce either \textit{sensor spoofing} or \textit{denial of service (DoS)}. To put these two attacks into concrete application domains, we will discuss them in the following \textit{attack goals} subsection.

\subsubsection{Attack Goals}\label{subsec:SensorFeed}

To understand the goals of signal injection attacks on CPS, it is important to consider how feedback from individual sensors influences the overall behavior of the system, i.e. \textit{how the sensor feedback affects system behavior and the physical environment}. There are two ways attacks can manifest themselves, either by influencing the physical world or the cyber world, which we summarize as manipulate physical space - \textit{control actuation}, and manipulate cyber space - \textit{control application layer software}. The specific outcome of exercising either type of control over CPS is highly system-dependent, but it is still useful to highlight the most common outcomes of signal injection attacks.

\paragraph{\textbf{Control Actuation}} Many cyber-physical systems use feedback from sensors to determine the control signals that are sent to the actuators. Thus, if an attacker can control the sensor readings, they can potentially trigger or control actuation in the system, which poses a significant threat. The control logic can be implemented locally in the embedded controller, such as altitude control in a drone. More complex control and coordination can also be implemented in a more powerful edge node or cloud node, where the data from sensors are wirelessly transmitted to a remote node where the control computation occurs. Depending on the processing pipeline of the cyber-physical controls, the falsified sensor data could be crafted to trigger unintended behavior exhibited in the final actuation output. For example, an attack can leverage acoustic signals to maliciously manipulate the gyroscope on a drone, which ultimately affects the aviation actuation~\cite{son2015rocking}.

\begin{itemize}
    \item \textbf{Motion control:} For mobile CPS such as radio-controlled cars (RC cars)~\cite{trippel2017walnut} and UAV drones~\cite{son2015rocking}~\cite{davidson2016controlling}~\cite{kerns2014unmanned}, sensor feedback is used to control the system's movement. An attacker can inject signals into these sensors to directly control the system's motion.
    \item \textbf{Physical harm to humans:} In medical devices such as implantable defibrillators and medical infusion pumps~\cite{6956585}~\cite{kune2013ghost}, sensor feedback is used to control actuation that influences physiological processes. Thus, with injected signals, an attacker can directly cause physical harm by interfering with medical-related functionalities.
\end{itemize}

\paragraph{\textbf{Control Application Layer Software}} The second type of adversary impact lies within application layer software in the cyber world rather than influencing physical states. For instance, both \textit{dolphin attack}~\cite{zhang2017dolphinattack} and \textit{surfing attack}~\cite{SurfingAttack} utilize ultrasound to deliver hidden commands, which will further be recognized by voice-controlled systems and executed in applications. The induced actions via injected inaudible commands include calling someone, opening websites in the browser, turning on airplane mode, and navigation. They reveal that with injected signals, the actions of installed applications can be manipulated for malicious goals. The above example demonstrates that for cyber-physical systems in which sensor feedback is used only to influence the behavior of application layer software, the degree to which sensor-based vulnerabilities pose a threat highly depends on the software's purpose and core functionality.

\begin{itemize}
    \item \textbf{Malicious code execution:} For cyber-physical systems that use sensor feedback to control application layer software, an attacker can inject signals into the sensor to trigger the execution of privileged or malicious code. For example, researchers have demonstrated the ability to inject inaudible speech into the microphones of voice-controlled systems~\cite{vaidya2015cocaine}~\cite{yuan2018commandersong}~\cite{abdullah2019practical}, which allows an attacker to bypass permissions and run a variety of applications on the system. 
    \item \textbf{Misleading system operators:} In some human-in-the-loop CPS, sensor feedback is presented via applications to inform a human operator of the system's state. By injecting signals to cause false sensor readings, the attacker can mislead the operator, imposing a false perception of the system's state. Typical examples include GPS spoofing attacks against aircraft \cite{2011}~\cite{schmidt2016survey}, train \cite{volpe2001vulnerability}, and ship \cite{volpe2001vulnerability}~\cite{risk} navigation applications, where the drivers are misled and therefore make false driving operations.
\end{itemize}

\subsection{Techniques for In-Band Signal Injection}

In-band signal injection attacks aim to induce malicious effects within CPS by transmitting signals at the intended range (often frequency) of receivers.
The implementation of an in-band signal injection attack depends on the sensor being attacked. More specifically, the attack methodology differs depending on if the target sensor is \textit{passive} or \textit{active}.   

\subsubsection{Targeting Passive Sensors}
Attacking passive sensors using in-band signals is often less challenging depending on the threat model and attack goal. For example, an attacker can utilize bright light to blind camera~\cite{petit2015remote}~\cite{yan2016can}, or cast an intense infrared ray to saturate an infrared sensor~\cite{park2016ain}.

\subsubsection{Targeting Active Sensors}

Signal injection attacks targeting active sensors are less trivial compared to passive sensors, since different active sensing technologies adopt different methods for transmitting, receiving, and processing the probing signals. Many active sensing systems such as LiDAR or ultrasound-based sensing leverage feedback from echoes/reflections for perception. Since the speed of these transmitted waves is known, the \textit{time difference} between the transmitted and reflected signals reveals the distance between the transmitter and the object from which the received signal is reflected. An attacker can create or replay reflected signals from different distances and angles to fool the receiver side of active sensors. More specifically, the attacker can transmit pulses using a generator to cause fake echoes, thereby misleading object recognition systems to detect objects located at the wrong place~\cite{yan2016can} or even detect objects that do not exist~\cite{davidson2016controlling}.

\subsection{Techniques for Out-of-Band Signal Injection}\label{subsec:techOOBSigInj}

Since out-of-band signals are outside of the normal operating range of the sensor, the attackers first need to guarantee the \textit{plausible input} requirement, i.e. the injected signals must survive the signal conditioning paths to be considered valid. To achieve this, there must be some components in the system that act as an \textit{unintended receiver} whose imperfect design helps to capture out-of-band signals; as well as a \textit{down-converter} that converts the out-of-band signal into in-band, so that it will be considered by the system as plausible input.

The next challenge for the attacker is to meet the \textit{meaningful response} requirement. The approach taken to address this challenge depends on the goal of the adversary. If their goal is to simply disrupt the sensor readings by adding noise, most of the existing works make use of an unintended receiver and a down-converting component to inject a simple constant out-of-band signal~\cite{son2015rocking}~\cite{217565}~\cite{shahrad2018acoustic}~\cite{roy2017backdoor}. 
In this scenario, the attacker does not need strong control over the shape of the down-converted signal, as long as the injected noise is crafted to be ``effective" in terms of sufficient intensity or strength so as to incite a meaningful response from the system. 

On the other hand, if the attack goal is to inject a precise waveform into the sensing system to incite a specific response from the system, the down-converted signal must closely resemble a sensor output that is known to trigger some desired behavior in the system~\cite{zhang2017dolphinattack}~\cite{nashimoto2018sensor}~\cite{kasmi2015iemi}~\cite{10.1145/3319535.3354195}. Thus, the injected out-of-band signal must be shaped such that when it is down-converted, it is transformed into the attacker's desired sensor output. This is generally achieved by using an out-of-band signal as the carrier wave, with the intended sensor output modulated over it. In this case, the down-conversion process is better defined as \textit{demodulation}; a demodulator not only down-converts the injected signal into in-band, but also recovers the modulated intended sensor output signal. In the type of attack where the adversary's goal is to precisely control the sensor output, this general approach is used by all existing literature. Thus, the goal of this section is to consolidate the approaches in this existing literature into a single, generalized mathematical model for sensor spoofing attacks.

The remainder of this section will be organized as follows: Section \ref{subsubsec:sysmodelForOOBSigInj} will propose and describe a general system model for out-of-band signal injection attacks, and Section \ref{subsubsec:DCModOOBSigs} will describe and technically characterize the various components in the sensor's signal conditioning path that can act as down-converters/demodulators.

\subsubsection{System Model for Out-of-Band Signal Injection}\label{subsubsec:sysmodelForOOBSigInj}
The goal of the model is to mathematically characterize the system's response to injected out-of-band signals. It is based on the model proposed in \cite{giechaskiel2019framework}, but adapts its characterization of each component's transfer characteristic. Each component corresponds to a component in a sensor's signal conditioning path, and responds to the attacker's injected signal in a certain way. Overall, the model consists of six main components: (1) \textit{modulated signal} to be transmitted and injected, (2) transmitted signal will suffer \textit{signal attenuation}, (3) transmitted signal will be picked up by \textit{unintended receiver}, and processed by (4) \textit{amplifier}, (5) \textit{low-pass filter}, and (6) \textit{ideal ADC}.

\textbf{Modulated Signal: }
The transmitted signal, $a(t)$, consists of the attacker's intended sensor output signal, $m(t)$, multiplied by an out-of-band carrier wave with frequency \(f_c\). This represents an \textit{amplitude modulated} waveform formulated as:
\begin{equation}\label{eq:eq1ModulatedSignal}
a(t)=(1+m(t))A_{c}sin(2\pi f_ct+\phi).
\end{equation}

The reason why the injected signal should be modulated for a successfully attack will be discussed in detail below.

\textbf{Signal Attenuation: }In transit to the target system, the malicious signal is attenuated by several factors that mitigate the signal, which are integrated into and denoted as \(A_d\). The specific value of \(A_d\) is a function of the transmitted signal affected by multiple factors including the physical properties of the surrounding environment, the distance of the attacker from the target system, etc. Therefore, the attacker has to figure out the configuration which minimizes the signal attenuation.

\textbf{Unintended Receiver: }The signal is then received by the unintended receiver of the system, which is described by its transfer function, \(H_{UR}(s)\). It is the component in the \textit{cyber-physical layer} that acts as a receiver for out-of-band signals. In the existing literature, there are three main types of components that act as unintended receivers.
\begin{itemize}
    \item \textit{Microphones with acoustic signals: } Although microphones are meant to sense audio signals audible to humans, they are also sensitive to higher frequency audio signals in the ultrasonic range due to non-linearity. This allows attackers to inject out-of-band acoustic signals in the ultrasonic band~\cite{zhang2017dolphinattack}~\cite{SurfingAttack}~\cite{roy2018inaudible}~\cite{yan2019feasibility}. 
    \item \textit{MEMS sensors and HDDs with acoustic signals: } MEMS accelerometers and gyroscopes~\cite{trippel2017walnut}~\cite{217565}~\cite{nashimoto2018sensor}, as well as the read/write heads of hard-disk drives (HDDs)~\cite{shahrad2018acoustic}, are shown to be sensitive to acoustic signals at their natural frequencies. These frequencies, also known as \textit{resonant frequencies}, are specific to the sensor. Once an attacker has determined this frequency for a specific sensor, they can play audio signals at this frequency, injecting a sensor output that oscillates at this resonant frequency.
    \item \textit{PCB traces with electromagnetic interference: } The PCB traces connecting sensors to the processor have been shown to act as unintended low gain, low power antennas, making them vulnerable to intentional electromagnetic interference~\cite{kune2013ghost}. These PCB traces also exhibit resonance at certain high frequencies, the specific frequency depending on the length of the trace. Though it is possible to inject in-band electromagnetic interference, signals with non-resonant frequencies will be highly attenuated. Therefore, in order to achieve effective attacks, the injected signals must be out-of-band at resonant frequencies. As such, an attacker can inject electromagnetic signals at one such frequency, resulting in an induced oscillating voltage signal at the output of the sensor.
\end{itemize}

In general, the magnitude of the unintended receiver's frequency response, \(|H_{UR}(j\omega)|\), can be described as a band-pass filter with a low frequency bound, \(f_L\), and a high frequency bound, \(f_H\). Signals outside of this frequency range are significantly attenuated. Thus, the frequency of $a(t)$ must be within this range for it to be received by the system. To this end, the injected signal must be modulated in order to pass the band-pass filter and therefore make the attack successful. For example, in an electromagnetic interference attack, the wires connecting the sensors with the ADC act as unintentional receivers and pick up signals with low gain. The frequencies that can be effectively received are related to the inverse of the wire length, while signals of other frequencies will be tremendously attenuated \cite{809842}. As the resonant frequencies of short wires are in the GHz range, the injected signal $m(t)$ has to be modulated over a high-frequency carrier wave. Then the unintended receiver transforms the attenuated attacker signal, \(\hat{a}(t)\), into an analog voltage signal, $v(t)$.

\textbf{Analog-to-Digital Converter: }The analog-to-digital converter is a core component in the digitization process that converts analog signals to digital. Systems with sensors generally contain one or more ADCs, and some ADCs are even directly integrated into the sensor chip. Although ADCs differ from each other according to their types, each ADC mainly consists of three basic components: (1) the \textit{track-and-hold circuit}, which is also called a \textit{sample-and-hold circuit} that conducts the sampling and holds the sampled signal for digitization; (2) the \textit{analog-to-digital converter} that transfers signals from the analog to digital realm; and (3) the \textit{level-comparison mechanism}~\cite{PelgromAtoDConversion}. These three components show different features within signal conditioning path: the \textit{sample-and-hold circuit} acts as a low-pass filter; the \textit{ideal ADC} simply implements analog-digital conversion; and the \textit{level-comparison mechanism} functions as an amplifier. Based on their different functionalities, we model and demonstrate them separately as independent system counterparts.

\textbf{Amplifier: }As previously mentioned, the level-comparison mechanism within an ADC serves as a non-linear amplifier that induces DC offsets. Besides, amplifiers widely exist in most signal conditioning processes outside ADCs. The amplifier can be characterized by its \textit{gain}, denoted by \(G(x)\). Ideally, the amplifier multiplies its input signal by a constant factor $A$, meaning \(G(v(t))=A\cdot v(t)\). But in reality, amplifiers are non-linear, and can be more accurately characterized as having a second-order response \cite{kune2013ghost}~\cite{sutu1984demodulation}~\cite{7567399}, denoted in Eq. (\ref{eq:eq2Amplifier}) as:

\begin{equation} \label{eq:eq2Amplifier}
G(v(t))=Av(t)+Bv^2(t),
\end{equation}
where \(A\) and \(B\) are the coefficients that describe the degree to which the amplifier exhibits non-linear behavior, and can depend on factors such as the frequency and magnitude of the input signal. In this paper, a constant called the \textit{non-linearity factor} is proposed, which is denoted by \(\alpha=\frac{B}{A}\), and describes the degree to which the amplifier's non-linear characteristic affects its overall response under specific input conditions.

\textbf{Low-Pass Filter: }As mentioned before, the sample-and-hold circuit within ADC can be modeled as a low-pass filter, which also exists outside the ADC and constitutes various sensors such as microphone and camera. It can be characterized by a transfer function \(H_{LP}(s)\). The most important parameter of the low-pass filter is its \textit{cutoff frequency}, denoted by \(f_{cut}\). Frequencies above \(f_{cut}\) will be highly attenuated by the filter.

\textbf{Ideal ADC: } The ideal ADC digitizes the induced analog signal by sampling it at a certain frequency, which is often done in a sample-and-hold circuit, and then converts each sample to an $N$-bit number. The value of $N$ is known as the \textit{resolution} of the ADC. The ADC is thus characterized by its sampling frequency, denoted by \(F_s\), and its resolution $N$.

\subsubsection{Demodulate Out-of-Band Signals}\label{subsubsec:DCModOOBSigs}

Consider an attacker whose goal is to induce a specific in-band voltage signal, $m(t)$ at the sensor's output. As previously mentioned, this desired in-band signal must be modulated over an out-of-band carrier, so that the unintended receiver can respond to the injected signal, which is then demodulated by some component in the system to recover $m(t)$. The specific form of the attacker's modulated injected waveform depends on the component in the circuit that is used as the \textit{demodulator}. In general, the two components that can act as demodulators are the \textit{amplifier} and the \textit{analog-to-digital converter}.

\textbf{ADCs as Demodulators:}
In \cite{giechaskiel2019framework}, Giechaskiel et al. conduct extensive experiments to demonstrate that all major ADCs can demodulate amplitude-modulated (AM) signals. Assume that the injected signal successfully passes the unintended receiver which is modeled as a band-pass filter, in order for the attack to succeed, the attacker-desired signal $m(t)$ must be recovered by the ADC when $v(t)$ passes through it. For the sake of simplicity, suppose that the attacker's goal is for the sensor's readings to be a constant DC value, so they choose $m(t)=M$. When this signal passes through the ADC, the ADC will sample the signal at a sample rate \(F_s\), resulting in a sampled signal with samples at \(t_0=0, t_1=\frac{1}{F_s}, t_2=\frac{2}{F_s}...,  t_i=\frac{i}{F_s}\). Suppose the attacker chooses the carrier frequency \(f_c\) to be an integer multiple of the sampling frequency, \(F_s\), such that \(f_c=n\cdot F_s\), where $n$ is some positive integer. Then the output of ADC can be expressed as:
\begin{equation} \label{eq:eq5PluggingIn}
V[i]=KMsin(2\pi ni+\phi), \;\;\;\;\; i\in \mathbb{Z}^+ 
\end{equation}
where \(K\) is a parameter integrating the effects of attenuation and the band-pass filter.

Therefore, when the carrier frequency is chosen to be a multiple of sampling frequency, the samples fall on every $n$'th peak of the modulated injected signal, causing the ADC to act as an envelope detector, effectively demodulating the signal. As long as the ADC sampling frequency, \(F_s\), is constant, any $m(t)$ can be effectively demodulated by the ADC.

This effect is known as \textit{signal aliasing}, and is caused by under-sampling the analog signal. The Nyquist sampling theorem states that in order to effectively recover the shape of an analog signal by sampling, the sample rate must be at least twice the largest frequency component of the input signal; if this requirement is unsatisfied, the digitized signal will be aliased. In signal injection attacks where the ADC is used as a demodulator, the attacker intentionally causes the injected signal to be aliased to directly control the sensor readings.

Therefore, there are two general requirements for an attacker to effectively use the ADC to demodulate their injected signal:

\textit{(1) The sample rate of the ADC, \(F_s\), is time-invariant}

\textit{(2) The carrier frequency \(f_c\) satisfies \(f_c \gg F_s\) and \(f_L \leq f_c \leq f_H\)}

The second requirement essentially states that the attacker must be able to find some out-of-band carrier frequency that is a multiple of the ADC sample rate, which still resides within the frequency response of the unintended receiver.

\textbf{Amplifiers as Demodulators:}
\begin{figure}
\includegraphics[width=8.4cm]{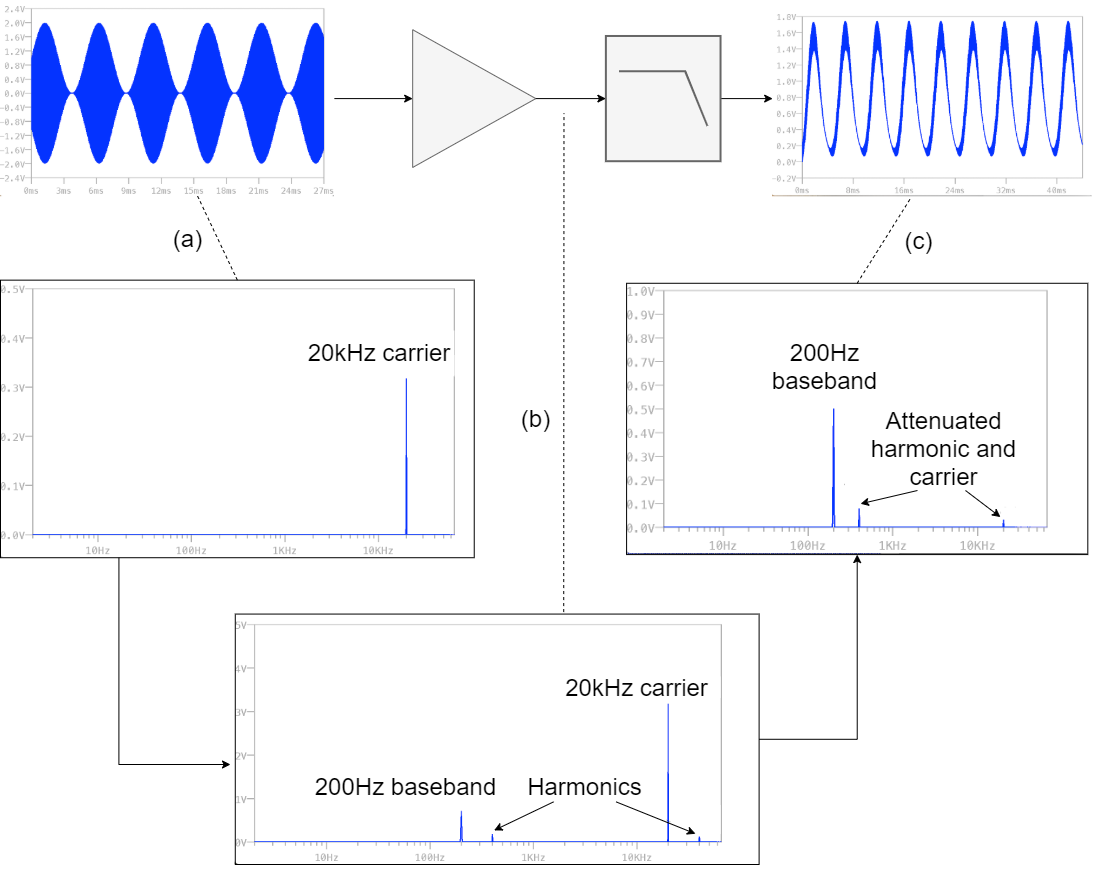}
\caption{Simulation results of demodulating out-of-band signals. (a) The attacker's injected signal $a(t)$ and its spectrum; (b) spectrum post-amplification; (c) recovered $m(t)$ post LPF and its spectrum.}
\label{fig:fig7AmplifierDemodulation}
\end{figure}
The non-linear characteristic of amplifiers described by Eq. (\ref{eq:eq2Amplifier}) can allow the amplifier to demodulate the attacker's signal. Suppose the injected signal takes a similar form to the one shown in Eq. (\ref{eq:eq1ModulatedSignal}), while the attacker's intended sensor output signal is \(m(t)=sin(2\pi f_mt)\). 

The Fourier transform of this attack signal contains a single spectral component at the carrier frequency, \(f_c\). Plugging Eq. (\ref{eq:eq1ModulatedSignal}) into Eq. (\ref{eq:eq2Amplifier}) describes the amplifier's response to this specific input signal. Taking the Fourier transform of the resulting output signal results in a spectral component at the in-band frequency, \(f_m\), a spectral component at the carrier frequency, \(f_c\), spectral components at integer multiples of these two frequencies (harmonics), and spectral components at intermodulation products of these two frequencies. Thus, after the signal passes through the non-linear amplifier, its spectrum contains the frequency components of the attacker's intended sensor output signal. Once the output of the amplifier is low-pass filtered, frequency components above \(f_m\) will be highly attenuated, resulting in a spectrum effectively only containing a component at \(f_m\). Thus, the non-linearity of amplifiers can demodulate the attacker's injected signal.

This is shown in Fig. \ref{fig:fig7AmplifierDemodulation}, for which an LTSpice schematic is used to simulate this model. In the model, a sine-wave voltage source is used, with $f_m=200Hz$ and $f_c=20kHz$. The non-linear amplifier is implemented using an arbitrary behavioral voltage source, which transforms the input voltage using Eq. (\ref{eq:eq2Amplifier}), with $A=10$ and $B=1$. This signal is then fed into an RC low-pass filter with a cutoff frequency of 250Hz. The Fast-Fourier transform (FFT) is applied to the signals that before the amplification stage; after the amplification stage and before the low-pass filter; and after the low-pass filter. The resulting frequency spectra shows the demodulating properties of non-linear amplifiers when the input is judiciously modulated as illustrated above. Note that a first-order RC low-pass filter is used, so the harmonics and carrier frequency components still remain post-filtering, but are highly attenuated. In real systems, a more complex, higher-order low-pass filter is likely used, resulting in more significant attenuation of the harmonics and carrier frequency components.

Another important consideration is the aforementioned non-linearity factor of the amplifier, \(\alpha\). As this factor approaches larger values, the non-linear characteristics of the amplifier are increasingly dominant, and the in-band signal $m(t)$ will be more effectively recovered.

Based on the above results, there are two general requirements for an attacker to effectively use an amplifier to demodulate their injected signal:

\textit{(1) The non-linearity factor, \(\alpha\), must be sufficiently large to demodulate the attacker's injected signal}.

\textit{(2) The cutoff frequency of the low-pass filter, \(f_{cut}\), must be low enough to sufficiently attenuate frequency components at \(f_c\), and high enough to not attenuate the important frequency components of m(t)}.

If the above requirements are met, $m(t)$ can be successfully recovered by an amplifier in the signal conditioning path.

\subsection{Signal Injection Attacks: Known Attacks}\label{subsec:knownsiginj}
This section will provide a survey of the existing literature of signal injection attacks on the sensors of cyber-physical systems. Here we summarize the existing signal injection work in Table \ref{tab:signalInjection}, which are discussed in more detail in the following. Section \ref{subsubsec:knowninband} will give an overview of existing in-band signal injection attacks, while Section \ref{subsubsec:knownoobsiginj} will provide an overview of existing out-of-band signal injection attacks. 

\input{SignalInjectionTable}

\subsubsection{In-Band Attacks}\label{subsubsec:knowninband}

By injecting in-band signals, an attacker can perform either \textit{sensor spoofing} or \textit{denial of service} attacks. The attacker can inject \textit{optical, electromagnetic,} or \textit{acoustic} signals to attack the camera, radar, MEMS motion sensor, ultrasonic sensor, LiDAR, GPS, infrared sensor, microphone, or magnetic sensor of cyber-physical systems.

\paragraph{\textbf{Electromagnetic Attacks}}
To meet the \textit{plausible input} requirement, the attacker must have knowledge of the communication protocol used, or the control logic that translates sensor readings into meaningful inferences about the system's surrounding environment. Thus, electromagnetic spoofing attacks tend to be either \textit{gray-box} or \textit{white-box} attacks.

Furthermore, utilizing injected electromagnetic signals to induce denial of service is generally referred to as \textit{signal jamming} attacks. To perform a jamming attack, the adversary must generate an attack signal $a(t)$ that is significantly larger than the voltage $v(t)$ generated by the sensed physical signal $s(t)$. The only necessary prior knowledge for the attacker is the operating frequency range of the sensor; thus, jamming attacks are generally \textit{black-box} or \textit{gray-box} attacks.

\textbf{Radar as Target:} 
A radar transmits radio waves and analyzes the reflected waves to infer details about the distance, velocity, and orientation of surrounding objects. One of the demonstrated attacks is conducted on the millimeter-wave (MMW) radar of an automobile~\cite{yan2016can}, where the primary function of the radar is for collision avoidance and pedestrian and cyclist detection~\cite{thing}. Due to the lack of protection, an attacker can analyze the emitted radar signal to spoof legitimate reflections. As a result, they are able to cause the radar to report false distances to objects in its trajectory.

\textbf{Magnetic Sensor as Target:} In~\cite{10.1007/978-3-642-40349-1_4}, Shoukry et al. use electromagnetic actuators placed in the wheel of a car to inject signals into magnetic field sensors for anti-lock braking systems (ABS). They perform two types of attacks, one that is simply disruptive to the system and unrefined, and another that spoofs magnetic signals such that a false wheel rotation speed is detected, leading to life-threatening situations.

\begin{figure}
\includegraphics[width=8.4cm]{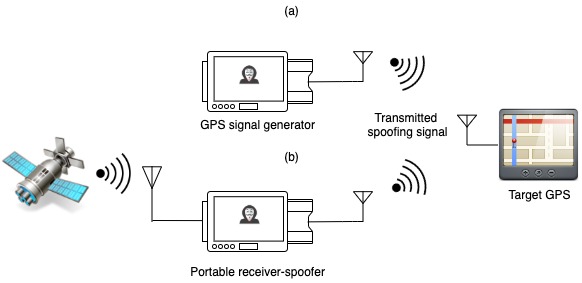}
\caption{Two typical ways to spoof GPS signals: (a) use a signal simulator to generate legitimate spoofing signals~\cite{warner2002simple}; (b) build a portable GPS receiver-spoofer to receive legitimate signals and manipulate the parameters~\cite{kerns2014unmanned}~\cite{humphreys2008assessing}.}
\label{fig:figGPSInjection}
\end{figure}

\textbf{GPS as Target:} The GPS receivers of GPS-based navigation systems have been shown to be vulnerable to spoofing attacks. The protocol for the communication of GPS signals is highly complex, and relies on specific timing and position data related to the orbital motion of multiple transmitting satellites. Given the complexity of GPS communication, an attacker can take two approaches to effectively spoof legitimate GPS signals as shown in Fig. \ref{fig:figGPSInjection}. The first approach shown in Fig. \ref{fig:figGPSInjection}(a), as demonstrated in \cite{warner2002simple}, uses a commercially available GPS signal simulator to generate legitimate GPS signals. The simulator's output is connected to a transmitting antenna, which transmits the spoofed signal to the victim's GPS receiver. In \cite{warner2002simple}, this technique is used to relay a false location to a truck's external tracking system, demonstrating that an attacker could stealthily steal a vehicle without alerting the operators of the tracking system. The main drawback of using a GPS signal simulator is its cost; the simulator can only be rented for around \$1000/week. Zeng et al. \cite{217476} build a cheaper spoofer with a cost at \$223 that could cause the driver to reroute to a different destination without realizing it. Using a portable GPS spoofer and the algorithm they developed, they show that it is possible to give fake turn-by-turn directions to drivers navigating using GPS.

The second approach shown in Fig. \ref{fig:figGPSInjection}(b) is to build a portable receiver-spoofer, as proposed in \cite{humphreys2008assessing}. This device consists of a receiver to receive legitimate GPS signals, a software layer to modify parameters of the received signal, and a transmitter to transmit the spoofed signal. In \cite{kerns2014unmanned} and \cite{2018}, this technique is used to spoof the position estimation systems of UAV drones and the navigation system of a yacht. Specificly, Kerns et al.~\cite{kerns2014unmanned} demonstrate that GPS receivers of UAV drones can be spoofed, causing them to measure an attacker-specified location estimate. This can allow the attacker to exhibit \textit{motion control} over the UAV and alter its course. In \cite{2018}, GPS spoofing is used to alter the trajectory of a large yacht, and thus gaining full control over its navigation system.

\textbf{EM DoS Attacks:} GPS receivers are shown to be vulnerable to jamming attacks~\cite{1338639}~\cite{806243}~\cite{coffed2014threat}. In cyber-physical systems, GPS works by decoding signals broadcast by multiple satellites orbiting the Earth. Due to their large distance from the Earth's surface, GPS signals suffer a transmission loss of 152dbW to 163dbW \cite{schmidt2016survey}. Thus, an attacker's local jamming signal can easily overpower the strength of legitimate signals transmitted by the satellites. For this reason, GPS jamming attacks represent a serious threat; simple jammers can be bought for less than \$10, and can significantly interfere with receivers up to a radius of 300 meters from the jamming device \cite{marks}. Generally, these low-cost GPS jammers are used by individuals to \textit{mislead system operators} by hiding their whereabouts from tracking systems \cite{marks}. The intention of the attacker can be relatively harmless, such as avoiding tolls or hiding their whereabouts from their employers' tracking systems. However, it can also have more severe consequences on critical infrastructure, such as aircraft~\cite{2011}~\cite{schmidt2016survey}, train \cite{volpe2001vulnerability}, and ship \cite{volpe2001vulnerability}~\cite{risk} navigation systems.

Radar sensors of autonomous vehicles are targeted for jamming attacks as well. Yan et al.~\cite{yan2016can} conduct signal jamming attacks against the MMW radar sensors of autonomous vehicles. While being jammed, the radar receiver is completely blinded and reports no objects in its surrounding environment. 

\cbox{Lessons learned: The majority of existing in-band EM attacks target on GPS. Since GPS signals can be easy to jam, DoS often doesn't require expensive equipment. The signals transmitted in the air also do not provide any authentication service, thus they can be easily spoofed. An important future direction of research is the defense against GPS signal spoofing.}

\paragraph{\textbf{Optical Attacks}} Existing work shows that optical sensors including cameras, LiDAR, and infrared sensors are vulnerable to spoofing attacks.

\textbf{Infrared Sensor as Target:} Park et al. \cite{park2016ain} demonstrate that the infrared drop sensor in medical infusion pumps can be exploited by performing spoofing attacks. They observe from analyzing the firmware of the device that drops are detected by falling edges in the readings of the infrared receiver. Thus, by saturating the receiver sensor with the infrared laser, then periodically turning off the laser, falling edges in the intensity of the received IR radiation are detected, allowing the researchers to spoof drops. This allows the attacker to under-infuse the medicine by spoofing fake drops. 

\textbf{LiDAR as Target:} LiDAR systems in autonomous vehicles work by transmitting laser pulses and analyzing the reflected pulses to infer the position and location to surrounding objects. They are vulnerable to spoofing attacks \cite{petit2015remote}~\cite{shin2017illusion}~\cite{tu2020physically}~\cite{cao2019adversarial} as well. Petit et al. \cite{petit2015remote} first explore the LiDAR spoofing attack by replaying the transmitted pulses from different locations with a pulse generator, with which the attacker can generate fake echoes and cause the receiver to detect real objects as closer or further away than they actually are. They can also use multiple attacking pulse generators to create multiple fake objects at different distances that can be classified by the control software as different objects, such as cars and walls. Therefore, these attacks targeting LiDAR can be critical because they can have a significant impact on the autonomous vehicle's mission planning. Shin et al. \cite{shin2017illusion} utilize a photodiode, a delay component, and an infrared laser to capture the victim's laser pulses and craft fake point clouds with delay, showing that the attacker is able to spoof a maximum of 10 dots in a horizontal line that appear even closer than the spoofer. Following this work, Cao et al. \cite{cao2019adversarial} adopt a faster emitting rate and a lens to focus the beam, and successfully improve the performance by achieving around 100 spoofing points in broader horizontal and vertical angles at a distance larger than 10 meters. Furthermore, they propose a new adversarial machine learning methodology called \textit{Adv-LiDAR}, which constructs the input perturbation while taking the post-processing process into consideration, and thus practically spoof an obstacle within a range of 2-8 meters in front of the Baidu Apollo. More recently, Tu et al. \cite{tu2020physically} attack LiDAR by placing a well-designed 3D adversarial object on the roof of the autonomous vehicle, and they are able to erase the whole vehicle from LiDAR perception with a success rate of 80\%.

\textbf{Optical Flow Algorithm as Target:} Spoofing attacks on cameras mostly target the image processing algorithms that are used to process the sensor data, such as optical flow algorithms, facial recognition, and object recognition. In \cite{davidson2016controlling}, Davidson et al. demonstrate that an adversary can gain control over UAV drones by exploiting the optical flow algorithm used in the controller to prevent the drone from drifting. The optical flow algorithm takes input from the camera, and compares successive frames to detect if the UAV is drifting. First, the algorithm detects features in the ground plane below that are amenable for tracking using the Shi-Tomasi corner detection algorithm \cite{shi1993good}. Then, the Lucas-Kanade optical flow algorithm \cite{lucas1981iterative} calculates the deviation in the horizontal position of the drone, \((\Delta x,\Delta y)\), using the previously calculated features. The drone's controller compensates for this deviation by adjusting its horizontal position by \((-\Delta x,-\Delta y)\). Thus, by injecting optical signals that are interpreted by the corner detection algorithm as features, the attacker can cause the optical flow algorithm to detect false changes in ground plane features, which are interpreted as drifts in the drone's horizontal position. The researchers use laser grids projected onto the ground plane in varying ground plane textures to cause the detection of fake features. They then move this projected image to cause the optical flow algorithm to falsely detect drifts in the drone's horizontal position, causing the controller to compensate by moving the drone in the opposite direction. Therefore, by moving the projected ground plane image, the attacker can obtain \textit{motion control} over the drone.

\textbf{Facial Recognition as Target:} Some optical spoofing attacks on cameras target machine learning algorithms used for facial recognition. This is closely related to \textit{adversarial machine learning}~\cite{huang2011adversarial}, which focuses on the vulnerabilities of machine learning algorithms. Sensor spoofing attacks on cameras that target ML algorithms fall into the category of after-training attacks using adversarial examples, which are inputs crafted by adding small perturbations that cause an already trained ML model to misclassify \cite{dreossi2018semantic}. In \cite{DBLP:journals/corr/abs-1803-04683}, Zhou et al. demonstrate an attack in which infrared LEDs are stealthily embedded into a hat, and are used to strategically project dots onto the attacker's face. By adjusting the sizes, intensities, and positions of the dots, the attacker's facial features are altered so that the facial recognition algorithm misclassifies the attacker as another individual. This allows the attacker to bypass facial recognition based authentication systems by impersonating a target individual. Sharif et al. \cite{sharif2016accessorize} also show a similar attack using specially designed eyeglasses. They propose an optimization problem, which finds the minimal physically realizable perturbation to an individual's face such that the facial recognition model misclassifies this individual as another, specifically chosen individual. This perturbation takes the form of an RGB pattern, which can be printed on a pair of eyeglasses and worn by the attacker. By wearing these eyeglasses, an attacker can impersonate another individual with a high accuracy to trick the facial recognition algorithm. A simpler, ``replay attack" based approach is proposed in \cite{7029643}, where a printed out image of a target individual is used to bypass a facial recognition authentication system. 

\textbf{Object Recognition as Target:} Some optical spoofing attacks target object recognition algorithms. This can pose a serious threat to cyber-physical systems such as autonomous vehicles and unmanned aircraft, which apply machine learning algorithms to optical sensor data for purposes such as object recognition and airborne collision avoidance \cite{julian2018deep}. Eykholt et al. \cite{eykholt2018robust} add small perturbations to a stop sign, which take the form of black-and-white tape attached to the sign. By arranging the tape in a specific pattern, they cause the object recognition algorithm to misclassify the stop sign as a 45mph speed limit sign. In \cite{dreossi2017compositional}, Dreossi et al. model the automatic emergency braking system of an autonomous vehicle. The system uses machine learning algorithms to classify objects in its trajectory, and initiates an emergency brake when an object is detected. They model this as a closed-loop system, and design an algorithm that searches the input space of the classifier to find sets of inputs that are misclassified. In \cite{10.1145/3319535.3354259}, Zhao et al. devise a new methodology for generating adversarial examples to trick machine learning based object detectors. They perform two types of attacks: a hiding attack where the object is not recognized, and an appearing attack where the object is classified incorrectly. For the hiding attack, they use feature-interference reinforcement and enhanced realistic constraints generation to improve attack performance, and in the appearing attack they use a nested approach that takes long and short distances into account separately and combines them.

\begin{figure*}
\centering
\includegraphics[scale=0.5]{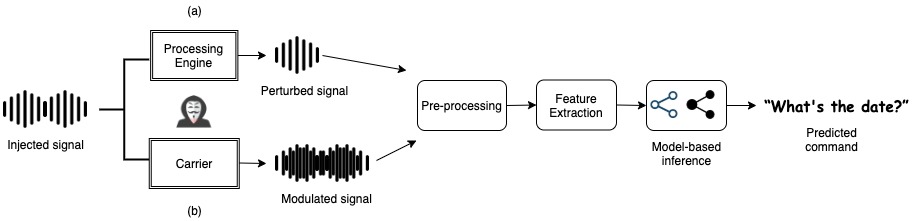}
\caption{Two typical ways to inject in-band signals into VCAs: (a) use a well-designed processing machine to generate a perturbed signal, such as cutting the signal into pieces~\cite{vaidya2015cocaine} or manipulating signal properties~\cite{abdullah2019practical}; and (b) modulate the injected command with a carrier wave such as music~\cite{yuan2018commandersong}. The crafted signal will then go through signal path consisting of: pre-processing that conducts speech/non-speech segmentation, signal processing that extracts Mel-Frequency Cepstral
Coefficients (MFCC)~\cite{DBLP:journals/corr/abs-1003-4083} features, and finally the extracted features are matched against pre-trained machine learning models for inference.}
\label{fig:figInbandInaudible}
\end{figure*}

\textbf{Optical DoS Attacks:} Existing optical DoS attacks have focused on cameras, LiDAR, and infrared sensors. In \cite{petit2015remote} and \cite{yan2016can}, the researchers perform DoS attacks on cameras of autonomous vehicles by using bright LEDs to blind the camera. By doing so, they can hide actual objects and fool auto-controls. Petit et al. \cite{petit2015remote} also demonstrate a DoS attack on the LiDAR sensor of an autonomous vehicle. By jamming the receiver with laser pulses, it reports no surrounding objects. Camera and LiDAR based optical attacks on autonomous vehicles represent a significant threat, since these sensors are responsible for detecting objects and road signs, and thus are critical for mission planning. Besides, infrared sensors are also vulnerable to DoS optical attacks. In \cite{park2016ain}, a medical infusion pump falls victim to an optical DoS attack using an infrared laser. The infusion pump functions by counting drops, and an IR transmitter and receiver are used to detect the drops by measuring beam intensity. The attackers use an infrared laser to drive the receiver into saturation, and therefore it is no longer responsive to changes in infrared light, leading to false reading of drops. This causes the infusion pump to over-infuse the medicine, due to the perceived lack of drops. 

\cbox{Lessons learned: One main direction of in-band optical spoofing attacks target facial and object recognition systems. The attackers do not intentionally transmit optical signals to the target CPS, but instead modify the objects to be probed. The same opportunity might exist for other signals, where the modification of physical objects could affect perceptions of machines only.}

\paragraph{\textbf{Acoustic Attacks}} Existing in-band acoustic attacks mainly focus on ultrasonic sensors in autonomous vehicles, and microphones embedded in voice-controlled assistants.

\textbf{Ultrasonic Proximity Sensor as Target:} Yan et al. \cite{yan2016can} demonstrate a spoofing attack on the ultrasonic sensors. Ultrasonic sensors are used in autonomous vehicles as short-range proximity sensors, and are generally used in parking assistance and self-parking systems to detect nearby objects. They are active sensors that require an ultrasonic transducer to transmit ultrasonic signals and a receiver to receive the reflected waves. In this attack, they are able to spoof reflected pulses in a scenario where there are no actual objects in the detection range, causing the ultrasonic receiver to detect and display pseudo objects. The researchers also conduct an ``acoustic cancellation" attack, in which the spoofed reflected signals are crafted in such a way that they destructively interfere with the legitimate reflected pulses from nearby objects, causing the system to fail to detect these objects.

\textbf{Voice Assistant as Target:} Fig. \ref{fig:figInbandInaudible} depicts one major direction within the field of in-band acoustic attacks. The attackers aim to inject perturbed voice commands that are unintelligible to humans, but still understood by the speech recognition algorithms used in cyber-physical systems. Most of these attacks target voice-controlled assistants (VCAs), such as Amazon Echo and Google Home~\cite{roy2018inaudible}~\cite{yuan2018commandersong}~\cite{vaidya2015cocaine}~\cite{197215}.
  
As shown in Fig. \ref{fig:figInbandInaudible}, two types of techniques are generally adopted to conduct in-band inaudible attacks against VCAs. The first one shown in Fig. \ref{fig:figInbandInaudible}(a) is to maliciously craft perturbed signals via a processing machine. Vaidya et al.~\cite{vaidya2015cocaine} demonstrate this technique, where the actual audio is ``mangled", such that it is unintelligible to humans, but still recognizable by the speech recognition algorithm. In order to ``mangle" the voice, they first extract the Mel-Frequency Cepstral Coefficients (MFCC)~\cite{DBLP:journals/corr/abs-1003-4083}, which are the features commonly used to characterize human speech. They perform transformations on the extracted MFCCs, such that the speech recognition still correlates them with the original speech. They then inverse the MFCC extraction process, obtaining an audio signal that is ``mangled" due to the transformed MFCC coefficients. This work takes a \textit{black-box} approach, assuming the adversary has no prior knowledge of the details of the speech recognition algorithm.
  
In \cite{abdullah2019practical}, Abdullah et al. attempt a \textit{black-box} attack, by focusing more on applying signal processing techniques to the original signal to maintain its MFCC features, but still have the resulting audio be highly unintelligible. To do so, they propose four types of perturbation: (1) modify the time domain signal, while maintaining the frequency domain spectrum; (2) add random phase information to the phase spectrum that still maintains the same magnitude spectrum; (3) add high-frequency components to the signal that can overpower the signal in the time domain, while the original signal's frequency component still remains in the spectrum; and (4) speed up the audio. By combining these operations, they develop a \textit{black-box} attack against speech recognition systems where the mangled audio is completely unintelligible to humans.

Carlini et al.~\cite{197215} perform both \textit{black-box} and \textit{white-box} attacks, again by maintaining MFCC features. They demonstrate the feasibility of performing \textit{black-box} attacks, and additionally show that with more knowledge of the system (\textit{white-box}), functional audio commands that are not understandable to humans can be generated. To counter the proposed attacks, the researchers also explore various defenses. A passive notification defense would alert the user when a voice command is given, but is likely to be ignored by the user. An active confirmation defense would impact usability of the system. However, machine learning algorithms could be trained to recognize the difference between computer-generated and human audio samples.

Most recently, Chen et al.~\cite{chenmetamorph} propose a similar attack, again by perturbing acoustic signals, where they successfully inject inaudible commands over-the-air into a speech recognition system. In their attack, a domain adaptation algorithm is utilized to refine the perturbation so as to improve the attack distance and reliability.

While in \cite{yuan2018commandersong}, a different approach shown in Fig. \ref{fig:figInbandInaudible}(b) is taken, where instead of generating a mangled speech signal unintelligible to humans, Yuan et al. use music as a carrier for the speech. By hiding voice commands in songs, the researchers demonstrate that it is possible to inject commands into speech recognition systems, maintaining the integrity of the original music such that surrounding individuals are not alerted of the attack. The proposed attack is a \textit{white-box} attack, in which knowledge of the underlying voice recognition system allows for mangled speech that is much less intelligible to humans than in \cite{vaidya2015cocaine}. Since the attacker knows the details of the speech recognition algorithm, the problem can be framed as an optimization problem, which is solved using the gradient descent algorithm~\cite{graddescent}. The mangled speech generated using this technique is fully effective against VCAs, while remaining completely unintelligible to humans. 

\textbf{Acoustic DoS Attacks:} In \cite{yan2016can}, Yan et al. demonstrate an acoustic DoS attack on the ultrasonic sensors of an autonomous vehicle. They show that by transmitting ultrasonic signals at the operating frequency of the receiver, the object detection system is blinded to nearby objects and thus halting its operation. 
\newline

\cbox{Lessons learned: Many of the existing in-band acoustic attacks target voice-controlled systems, and have attempted to hide malicious commands by perturbing them while maintaining the MFCC features recognized by algorithms. One possible direction for research in this domain could be attacks where other features such as the Linear Predictive Coefficient
(LPC)~\cite{itakura1975line}, Perceptual Linear Predictive (PLP)~\cite{hermansky1990perceptual}, or the combination of these features are maintained to craft hidden commands intelligible to CPS.}

\subsubsection{Out-of-Band Attacks}\label{subsubsec:knownoobsiginj}

Existing literature has demonstrated out-of-band signal injection attacks using \textit{acoustic}, \textit{electromagnetic}, and \textit{optical} signals to attack microphones, MEMS accelerometers, MEMS gyroscopes, temperature sensors, and magnetic sensors of CPS. In the following, we will discuss different types of out-of-band signal injection attacks based on the exploited physical property.

\paragraph{\textbf{Electromagnetic Attacks}} 
As discussed in Section \ref{subsubsec:sysmodelForOOBSigInj}, attackers can inject out-of-band electromagnetic signals by exploiting components in the \textit{cyber-physical layer} circuitry that act as unintentional low gain antennas. PCB traces in the signal conditioning path are susceptible to intentional electromagnetic interference (IEMI), but the specific range of frequencies for which a PCB trace behaves as a receiving antenna is highly system-dependent, and is determined by factors such as the length of the trace and potential coupling with other components in the circuit. Since these parameters vary greatly across different systems, the resonant frequency can be anywhere from the kHz to GHz range, so it is impractical for the attacker to simply sweep through all possible frequencies. Thus, out-of-band signal injection attacks using IEMI are always \textit{grey-box} attacks, since the attacker must disassemble the device in order to experimentally determine the frequency response of the unintended receiving trace.  

\textbf{Medical Device as Target:} In \cite{kune2013ghost}, Kune et al. demonstrate a low-frequency IEMI attack on implantable medical devices (IMD). This attack targets the leads of electrocardiograms (ECGs) and cardiac implantable electrical devices (CIEDs). ECGs are used to monitor cardiac activity by taking voltage readings on the surface of the skin, while CIEDs are used to monitor and regulate cardiac activity by sending small electrical signals to the heart. By injecting IEMI into the leads of these devices, they are able to spoof cardiac activity in the case of ECGs (\textit{misleading system operators}), and prevent and deliver defibrillator shocks in the case of CIEDs (\textit{physical harm to humans}). It is noteworthy that the leads are not intended to be receivers for remotely transmitted EM signals, thus the intentional receiving band for electromagnetic signals can be considered as zero.  

\textbf{Temperature Sensor as Target:} Tu et al.~\cite{10.1145/3319535.3354195} show that they can use EMI signals to inject spoofed readings into various temperature sensors. In this attack, they are able to manipulate temperature sensor readings by taking advantage of the amplifier without tampering with the system or triggering any alarms related to abnormal temperatures. The attack is executed remotely, from up to 16.2 meters away. These spoofed temperature sensor readings could have serious consequences for temperature-sensitive systems, as they would be unable to properly regulate the temperature without accurate readings.

\textbf{Modulating With Resonant Frequency:} As discussed in Section \ref{subsubsec:sysmodelForOOBSigInj}, if the attacker modulates the desired sensor output, $m(t)$, over the resonant frequency, it can potentially be realized at the output of the sensor given there exists a demodulating element in the circuit. Using this method, Kune et al. \cite{kune2013ghost} demonstrate a sensor spoofing attack on a vulnerable Bluetooth headset. By amplitude modulating voice signals over the resonant EM carrier, they are able to inject inaudible voice commands into the headset's microphone. In this attack, a hypothetical victim uses the headset paired with a cell phone to make a call to an automated dial-in system, which uses the tones generated by the keypad button presses to infer the button dialed. By injecting IEMI, they are able to generate arbitrary tones, which are interpreted by the dial-in system as their respective buttons being pressed. They also demonstrate a session hijacking attack on a video chat session, in which the Bluetooth headset is paired to the victim's laptop. In this attack, they are able to hijack the session by injecting voice signals that are loud enough to overpower the victim's voice.

A similar attack is demonstrated in \cite{kasmi2015iemi}, in which Kasmi et al. find that the microphone built into a pair of headphones is vulnerable to sensor spoofing via IEMI. After experimentally determining the resonant frequency of the trace connecting the microphone to the microcontroller, they modulate voice signals over this frequency, and transmit the resulting signal using an antenna. The signal is injected into the trace, demodulated by an audio amplifier, and is realized at the output of the sensor as a regular voice signal. It is demonstrated that this can be used to inject voice commands into the voice-controlled assistants on smartphones, such as Siri and Google Voice. This gives the attacker control over various application layer software, which can be used to perform tasks such as visiting websites, sending text messages, making phone calls, etc. A main difference between the methodologies of \cite{kune2013ghost} and \cite{kasmi2015iemi} is that in \cite{kasmi2015iemi}, the unintentional receiver is external to the system (front-door coupling), while in \cite{kune2013ghost} the unintentional receiver is a PCB trace internal to the system (back-door coupling).

\textbf{Conducted IEMI:} A recent electromagnetic signal injection attack \cite{latestemi} also targets voice-controlled systems, but uses conducted IEMI. In a conducted attack, the attacker is on the same power network as the target device, meaning that the injected signal is transmitted via a wired connection. In practice, this wired connection is through a USB charger or a power charger, and the attacker's point of entry is somewhere in the power network behind the power socket. The researchers use a specially designed injection probe to inject amplitude-modulated voice commands into the power network, through which the injected signal eventually arrives at the target device, where it is demodulated by the non-linear amplifier of the microphone circuit. This conducted attack offers a few advantages over the transmitted attacks demonstrated in \cite{kasmi2015iemi} and \cite{kune2013ghost}. First, since the attacker's signal is injected via a shared wired connection, the attenuation factor \(A_d\) is much lower, meaning the power of the transmitted signal can be much lower than in transmitted attacks. Second, in the conducted attack, the bandwidth of the unintended receiver's frequency response is much wider, giving the attacker a much wider range of frequencies for which the attack signal will not be severely attenuated before reaching the target sensor.

\textbf{EM DoS Attacks:} EM waves can also be used for DoS attacks. In \cite{kasmi2015iemi}, Kasmi et al. inject audio signals into the microphones in Bluetooth headsets and webcams via IEMI, by exploiting the fact that the PCB traces connecting the microphone to the microcontroller are resonant to EMI at frequencies in the MHz and GHz range. Induced voltages in this frequency range are down-converted by the audio amplifier in the microphone circuit, and are manifested as low-frequency noise at the output of the sensor. It is demonstrated that if the induced oscillating voltage has a sufficiently large amplitude, the resulting down-converted noise will be loud enough to overpower legitimate audio inputs to the microphone. Thus, the system taking input from the microphone cannot use it reliably during the attack. A similar attack is proposed in \cite{kune2013ghost}, where Kune et al. inject high power Weezer waveformed EMI signals to overwhelm original acoustic signals without causing distortion in the demodulated audio. As a result, they successfully block a Skype session without being detected.  

\cbox{Lessons learned: Existing work has demonstrated the feasibility of injecting out-of-band EM signals into various applications. Many of them modulate signals onto a carrier wave of unintended range, yet there is less attention given to the impact of different signal modulation techniques on the success of the attack. Therefore, it would be an important area to investigate and understand the potential impact of different techniques.}

\paragraph{\textbf{Acoustic Attacks}}Existing out-of-band acoustic attacks either target microphones that are sensitive to ultrasound, or MEMS accelerometers, gyroscopes, and HDDs that are sensitive to acoustic signals due to resonance in their mechanical structures.  
  
\begin{figure}
\includegraphics[width=8.8cm]{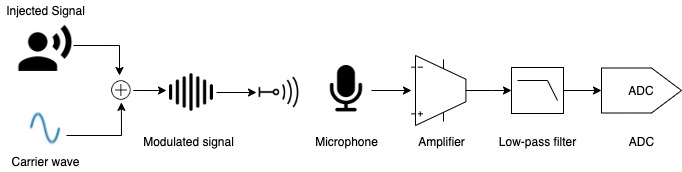}
\caption{General model for out-of-band signal injection attacks against microphones~\cite{zhang2017dolphinattack}~\cite{SurfingAttack}~\cite{roy2018inaudible}~\cite{yan2019feasibility}. Injected signals (voice commands) are modulated with a high-frequency carrier wave (e.g., ultrasound wave) and sent to the microphone. From the victim side, the microphone as an unintended receiver receives the modulated signal, the amplifier serves as a demodulator due to its non-linearity, and the low-pass filter removes the ultrasound frequency. Therefore, the injected signal is able to survive the signal conditioning path and induce effects. }
\label{fig:figMicrophoneInject}
\end{figure}

\textbf{Microphone as Target:} Sensor spoofing attacks using acoustic signals have been demonstrated on microphones, as shown in Fig. \ref{fig:figMicrophoneInject}. In the \textit{dolphin attack} proposed in \cite{zhang2017dolphinattack}, the voice signals that are amplitude-modulated over ultrasonic carriers can be demodulated by the amplifiers connected to smartphone microphones. Therefore, the researchers show that using this method, inaudible voice commands can be injected into voice-controlled assistants, such as Siri and Google Voice, using ultrasonic transducers. Following this work, Roy et al. \cite{roy2018inaudible} demonstrate that it is possible to increase the attack range of ultrasonic signal injection on microphones. Due to non-linearities in the speakers, the injected signal can only be amplified to a certain extent before these non-linearities start causing sound leakage at the transmitter side. To remedy this issue, the researchers show that by splitting the intended voice command's frequency spectrum into $N$ separate ``bins", modulating the contents of each bin over its own ultrasonic carrier, and playing each one of these modulated signals through a separate ultrasonic speaker, the attacker can significantly increase the range of the signal injection attack without any sound leakage at the transmitter. In \cite{yan2019feasibility}, the same researchers from \cite{zhang2017dolphinattack} propose an improvement to their attack, increasing the attack range to 1.75m by using an ultrasonic transducer array consisting of 40 speakers. Using this technique, they are able to increase the attack distance significantly, making long range attacks on VCA feasible. 

The \textit{surfing attack} proposed by Yan et al. \cite{SurfingAttack} takes a different perspective. Instead of transmitting hidden commands in the air, they perform attacks on smartphone voice recognition systems using ultrasonic waves transmitted through solid materials and surfaces. This attack is inaudible to humans, and they further explore attacks that interact with the target systems for multiple rounds, performing attacks like making fraudulent phone calls and stealing SMS passcode information without alerting the user. The attack is successful on some surfaces with a distance of 30 ft between the attacker's emitter and the target smartphone.

\textbf{Motion Sensor as Target:} Following the work of \cite{son2015rocking}, which demonstrates DoS attacks on the MEMS gyroscope sensor on a drone, Trippel et al. \cite{trippel2017walnut} find a similar acoustic vulnerability in MEMS accelerometers, which are also resonant to acoustic noise at certain frequencies, often less than 10kHz. By amplitude-modulating their desired sensor output over the resonant carrier wave, and using the ADC as a demodulator, the researchers show that they can inject arbitrary signals into the output of the sensor; though, due to drifts in the sample rate of the ADC, the attack can only be carried out for a few seconds. This attack is conducted on a smartphone app that uses the accelerometer to determine the motion of an RC car, for which the attacker is able to gain \textit{motion control} of the car. This attack is also used to register fake steps into a FitBit; due to FitBit's rewards program, an attacker could use this attack for financial gain. In \cite{217565}, Tu et al. build on the previous work, and propose attacks on MEMS gyroscopes and accelerometers. It shows that the problem of a drifting sample rate can be partially remedied by changing the frequency of the injected signal slightly in response to sample rate drifts, which results in an effective phase change in the digitized sensor output. This allows the attacker to carry out the attack for a longer duration. They implement this attack on numerous gyroscope-equipped systems, such as human transporters, self-balancing robots, camera stabilizers, anti-tremor devices, 3D mouses, and VR headsets. 

\textbf{Acoustic DoS Attacks:} Acoustic signals are also used for DoS attacks. The first paper to demonstrate that MEMS structures can be exploited with acoustic interference is \cite{son2015rocking}, which targets the MEMS gyroscope sensors on drones. The researchers show that MEMS gyroscopes are resonant to acoustic signals at frequencies generally in the ultrasound range (more than 20kHz). By using ultrasonic transducers to inject signals at these frequencies, an oscillating voltage appears at the output of the sensing mechanism. These signals are demodulated by the amplifier in the signal conditioning path, resulting in high amplitude noise in the digitized signal. In drones that use the feedback from the gyroscope system in a closed-loop fashion, this noise can cause the drone to crash. This attack is demonstrated on a drone constructed by the researchers, which uses a PID (proportional-integral-derivative) controller to achieve motion control. When subject to acoustic noise at the gyroscope's resonant frequency, the drone crashes. In \cite{shahrad2018acoustic}, Shahrad et al. show that the read/write heads of HDDs are also resonant at certain acoustic frequencies, and an attacker can inject acoustic signals into a HDD to cause faults. These faults can lead to interruption of the process of reads and writes. The attack is conducted on a CCTV DVR security camera system, for which the attack results in permanent data loss due to a memory buffer overflow. The researchers also test the attack on a PC with different operating systems, for which the attack causes a system reboot in the case of Windows or a fully unresponsive OS in the case of Linux-based distros. In \cite{roy2017backdoor}, Roy et al. propose a DoS attack that exploits amplifier non-linearity to inject out-of-band ultrasonic signals into microphones. 
By modulating in-band sound noise over ultrasonic frequencies, they generate an in-band ``shadow" of the original modulated signal. They show that this noise can serve as a jamming mechanism against unwanted eavesdroppers.  

Tu et al. \cite{217565} also show practical attacks to implicitly control or cause DoS across many different systems using MEMS gyroscopes or accelerometers. They use a methodology similar to previous research using acoustic signals. To improve attacks and make them more practical, they use two output control methods: digital amplitude adjusting and phase pacing. As a result, the control systems are either deceived by the MEMS sensor readings or rendered non-functional in a DoS attack.

\cbox{Lessons learned: A major research direction within out-of-band acoustic injection attacks is to use high-frequency acoustic signals to induce vibration of microelectromechanical components in sensors. Most of the attacks are demonstrated on motion sensors, but this attack can be extended to other systems where there are sensitive components. Furthermore, researching into potential defenses is also important.}

\paragraph{\textbf{Optical Attacks}}The field of out-of-band optical signal injection attacks is less explored until a recent study named \textit{LightCommands} proposed in \cite{Sugawara2020LightCommands}, where Sugawara et al. remotely inject inaudible and invisible commands into voice-controlled systems using light. The proposed attack is based on the \textit{photoacoustic effect}, with the specific application that MEMS microphones may respond to light signals as if it is sound. To exploit this, the attacker first modulates laser light with an intentional audio signal, and then transmits it by aiming the laser beam at microphone ports of the target devices. The injected optical signals will be converted back to the original audio signal inside the target system, which will result in the voice-controlled system executing arbitrary commands issued by the attackers. The main limitation of this work lies in the lack of stealthiness and difficulty of precise aiming: the laser beam is visible to humans or detectable to devices, and aiming at the small aperture of the target system with high precision is difficult while it is stationary, let alone for moving targets. 
\cbox{Lessons learned: Out-of-band optical injection attacks are relatively less explored. However, light, which is one of the most common energies and information transmission carriers, still presents a large attack surface for both injecting signals into light sensors of different frequency ranges or other sensors that react to light.}
 
\paragraph{\textbf{Attacking Sensor Fusion}} Many systems in practice leverage sensor fusion from multiple instruments to construct their perception of the physical world, such as estimating motion or orientation from both GPS and IMU. As a result, while it is intellectually interesting to explore new avenues to exploit individual sensors, it is often insufficient to inject signals into a single sensor, since the overall state estimation is generally resilient to large variations in a single sensor. The standard in-state estimation using sensor fusion is the Kalman Filter~\cite{welch1995introduction}, which estimates a target state of a system in the presence of errors. For example, in open-source drone software, such as Ardupilot \cite{ardupilot_2019} and Multiwii 
\cite{multiwii_2016}, the Extended Kalman Filter is used for sensor fusion based state estimation. Recently, Nashimoto et al.~\cite{nashimoto2018sensor} attempt to overcome sensor fusion in signal injection attacks. In this paper, they target the Kalman Filter based sensor fusion algorithm for measuring a system's inclination, which is the degree to which an object is inclined in a global reference frame. This is an important algorithm, often used in motion tracking and altitude control for robots and drones. It fuses data from the accelerometer, gyroscope, and magnetometer to estimate the system's overall inclination. The paper demonstrates that by performing spoofing attacks with both in-band (magnetic) and out-of-band (acoustic) vectors on all three sensors in a specific sequence, the attacker can defeat the sensor fusion algorithm, and control the system's estimation of inclination. In \cite{mo2010false}, a similar work is carried out, in which the effect of sensor spoofing attacks on closed-loop control systems is explored. Mo et al. demonstrate that for linear time-invariant Gaussian control systems equipped with Kalman Filters for state estimation, corrupting a subset of sensor readings is able to destabilize the system and bypass the failure detection mechanism.
\cbox{Lessons learned: Sensor fusion serves as an efficient way to improve CPS performance and raise the bar for simple signal injection attacks. However, the existing work shows that a well-designed injection input can still corrupt the fusion algorithm and induce malicious effects. Potential future research for defense could include developing more robust fusion algorithms, fusing redundant sensors, optimizing sensors to be fused, etc.}

\subsection{Limitations and Opportunities}

This section will systematically analyze the existing research related to signal injection attacks, outlining the limitations, identifying gaps in research, and suggesting potential directions for future work.  

\textbf{Spoofing encrypted or obfuscated signals: }Spoofing attacks become more difficult in the case of encrypted or hidden signals. Take GPS spoofing as an example, general GPS spoofing attack techniques are challenged by military-used \textit{GPS P(Y)}~\cite{scott2001anti} and \textit{hidden markers}~\cite{kuhn2004asymmetric}, which are proposed to hide GPS signals in the noise level with secret keys unknown to the attackers. To reveal the signals, the keys used for encryption are required, after which the received signals can be validated. As a result, it becomes challenging for the attackers to recover original navigation signals, let alone launch spoofing attacks. 

\cbox{Lessons Learned: Therefore, the combination of signals and cryptography is another interesting direction for effective defenses against signal spoofing attacks.}

\textbf{Overpowering high power signals: }This limitation widely exists in jamming attacks. To achieve a DoS attack, the attacker generally uses more powerful signals to overwhelm the legitimate signals. However, it will raise the bar for the attacker if the legitimate signal is already very powerful. Take positioning system jamming attacks as an example, Becker et al. \cite{becker2009efficient} demonstrate that jamming a terrestrial positioning system like eLORAN is much harder than jamming a satellite based positioning system. Compared to low-powered GPS signals, eLORAN uses high-power, low-frequency signals. Therefore, the attacker is dependent on physically tall or long antennas to efficiently transmit low-frequency signals \cite{lo2009assessing} so as to overcome the strong eLORAN signal. 

\cbox{Lessons Learned: Though it may be challenging to directly overpower all signals, the attacker can still interfere with the normal functionality of the target system if it can send well-crafted signals to trigger DoS at upper layers.}

\textbf{Lack of stealthiness: }Although most signal injection attacks are contactless to the target and are conducted remotely, it is possible for some attacks to be either noticed by humans or detected by machines. Attacks that leverage visible light~\cite{yan2016can}~\cite{petit2015remote}~\cite{Sugawara2020LightCommands} or wearing colorful masks~\cite{sharif2016accessorize} can lead to the suspicion of other users. For instance, Sharif et al. \cite{sharif2016accessorize} utilize well-designed glasses to help the attacker evade face recognition, but the glasses might stand out. A potential solution to this limitation is shown in another optical attack~\cite{eykholt2018robust} where visible perturbations are attached to road signs to mislead object recognition systems, and the perturbations are intentionally designed to mimic graffiti so as to ``hide in the human psyche." This idea can be potentially applied to other optical attacks for improvements as well, where the attacker may disguise the visible optical signal to be something common in daily life. Acoustic attacks against voice-controlled systems face a similar limitation. Existing attacks try to hide malicious commands by either perturbing audio signals~\cite{vaidya2015cocaine}~\cite{abdullah2019practical} or modulating commands with music~\cite{yuan2018commandersong} so as to avoid raising suspicion. However, the effect of the injection, such as a response from the voice assistant, may still reveal the attack. To tackle this problem, a command can be injected to minimize the observable result of the injection, such as turning down the volume at the beginning, so that the following responses will not be noticed by the victim users~\cite{SurfingAttack}.

\cbox{Lessons Learned: Stealthiness is a key factor in real world attacks, and should be taken into consideration when developing attacks. } 

\textbf{Short attack range: }Short distance requirements between an attacker's equipment, which is generally signal generating and transmitting devices, and the target CPS can also be a limitation. Take ultrasonic injection attacks for example: due to atmospheric attenuation, the ultrasonic DoS attacks proposed in~\cite{yan2016can} are normally conducted within 1 meter with the assistance of high jamming noise amplitude, and the attack range for spoofing is several meters. 
The \textit{dolphin attack} proposed by Zhang et al. \cite{zhang2017dolphinattack} achieves an attack range of 1.75m for ultrasonic signal injection, but is still restricted by the power of amplifier. In recognition of the short range limitation, \textit{Metamorph} proposed by Chen et al.~\cite{chenmetamorph} achieves an attack range of 6m, and \textit{LipRead} proposed by Roy et al.~\cite{roy2018inaudible} extends the attack range to 7.62m by aggregating ultrasound signals from an array of speakers. Yan et al. take a different perspective, where they leverage solid surfaces to propagate the malicious ultrasound signal up to 30 ft~\cite{SurfingAttack}.

Electromagnetic injection attacks face the same limitation. EMI signal injection attacks targeting implantable medical devices proposed in \cite{kune2013ghost} achieve an attack distance of 1 to 2 meters when placing devices in the open air, but the effective distance decreases to under 5cm when immersing devices in a saline bath approximating human body. Similarly, in a GPS spoofing attack targeting trucks proposed in \cite{warner2002simple}, the attack range is limited to 30 feet due to the attack signal strength. 

For optical injection attacks, one attack targeting LiDAR proposed by Petit et al.~\cite{petit2015remote} faces a similar challenge in which the attacker signal will fail to reach the victim photodetector of the target LiDAR at a distance. Park et al. \cite{park2016ain} succeed in an optical-based spoofing attack targeting infrared sensors at a distance of 12 meters using a 30mW IR laser, but face the challenge of precise aiming at the target.

\cbox{Lessons Learned: Finding new methods to enable the extension of the attack distance without significantly raising the attack power can be an opportunity to overcome existing limitations. }

\textbf{Moving targets: } This limitation exists in injection attacks because the transmitted signals need to aim at the target, thus constantly tracking a moving target becomes more difficult. 

In \cite{yan2016can}, the best performance for an ultrasonic spoofing attack is achieved at a perpendicular angle, because the longitudinal sound wave will project most of its energy in the forward direction. Though this may not be a problem for a still target CPS and attacker, the attack becomes significantly more challenging when the target is constantly moving.

Also, the spoofing attacks targeting LiDAR sensors of autonomous vehicles face a similar limitation. Both \cite{petit2015remote} and \cite{shin2017illusion} mention that aiming is one of their main obstacles in deploying practical attacks, since the attacker has to track the target LiDAR with their device.

Similar limitations are present in the spoofing attack on MEMS gyroscopes proposed in \cite{217565}, where it becomes difficult for the attacker to follow and aim the moving target while manually tuning acoustic signals. Though movement prediction may help mitigate these limitations, it is still restricted to certain scenarios; an automatic tracking and aiming system seems to be an ideal solution theoretically, but it will significantly raise the attack bar and may also raise the limitation of costly equipment. 

\cbox{Lessons Learned: Existing work has demonstrated that aiming at moving targets is a general challenge for signal injection across different physical waveforms. Potential future directions of research may include developing better tracking algorithms or attacks that do not require continuous interaction with the victim.}

\textbf{Costly equipment: }This limitation refers to the fact that costly equipment is required in order to achieve some of the proposed attacks. For instance, Warner et al.~\cite{warner2002simple} rent a GPS satellite simulator for \$1000 per week to generate and transmit the spoofed signal to the GPS receiver, with an additional antenna for broadcasting and two GPS signal amplifiers costing \$400 in total. Similarly, in order to generate signals at the operating frequency of the MMW radar that is attacked in \cite{yan2016can}, the attacker must make use of a function generator that can generate signals up to 81GHz, which has significant costs. It is also an important issue for the attacks that use injected EMI; generally, the resonant frequency of the PCB traces used as unintentional receivers is in the GHz range, and signal generators capable of producing those frequencies reliably are in the range of thousands of dollars. 

\cbox{Lessons Learned: The limitation of costly equipment exists in some of the developed attacks, thus developing defense solutions that will drive the cost higher as well as low-cost defenses are important directions for future research.}

\textbf{Physical hazard: }This limitation applies to certain attack vectors such as infrared lasers that may cause physical harm to human bodies. In \cite{park2016ain}, Park et al. discuss that in order to achieve larger attack distances, the required infrared laser with higher power is dangerous in that it can cause serious eye damage. Zhou et al.~\cite{DBLP:journals/corr/abs-1803-04683} use an infrared laser to project dots onto the attacker's face, which they state that may lead to potential health concerns. Sugawara et al. \cite{Sugawara2020LightCommands} also use a laser beam as an attack vector to inject commands, which may result in eye damage as well.

\cbox{Lessons Learned: It is often important to consider the potential health hazards for both the attacker and user when considering attack and defense mechanisms.}

\textbf{Generalizing for environmental variability: } The ambient environment can be an important factor affecting the effectiveness of some attacks. For example, both \cite{yuan2018commandersong} and \cite{abdullah2019practical} propose acoustic attacks targeting voice-controlled systems, and they can both be impacted by environmental noise. Both papers indicate that ambient noise will interfere with over-the-air attack audio and degrade the malicious perturbations. 

The optical attacks that mislead recognition systems using adversarial machine learning methods face similar challenges. For example, an optical attack targeting facial recognition is developed in~\cite{sharif2016accessorize}, in which the researchers propose to reduce lighting variations by conducting the attack indoors. 

Recognizing the lack of generalization, there are some efforts to take the variances into consideration. For instance, Eykholt et al. in \cite{eykholt2018robust} consider the distance and angle of a camera in an autonomous vehicle with respect to a road sign. Similarly, Shahrad et al. in~\cite{shahrad2018acoustic} also investigate the distance and angle factors influencing the proposed out-of-band acoustic DoS attack targeting HDDs, and they discover that the maximum distance and effective angles vary dramatically between different HDDs. 

\cbox{Lessons Learned: In order to make the attack or defense more realistic, it is important to take the possibility of environmental variations into consideration when designing attack and defense mechanisms.}

%% file: section4.tex
\section{Cyber to Physical: Information Leakage Attacks}\label{sec:infoleakageatks}
In an information leakage attack, the sensors of a cyber-physical system are exploited by an attacker to covertly measure signals leaked from an individual or a system through a \textit{vibrational, optical, acoustic,} or \textit{magnetic} side channel that can be processed to recover private information. Compared to signal injection attacks, these attacks leverage digital signal data to eavesdrop on physical activities, therefore they can be regarded as \textit{cyber-to-physical} attacks. Also, instead of actively manipulating physical world signals as is done in signal injection, these attacks tend to passively receive and analyze the signals emanated from target systems. In this sense, they can be regarded as \textit{passive} attacks.

\begin{figure}
\centering
\includegraphics[width=8.4cm]{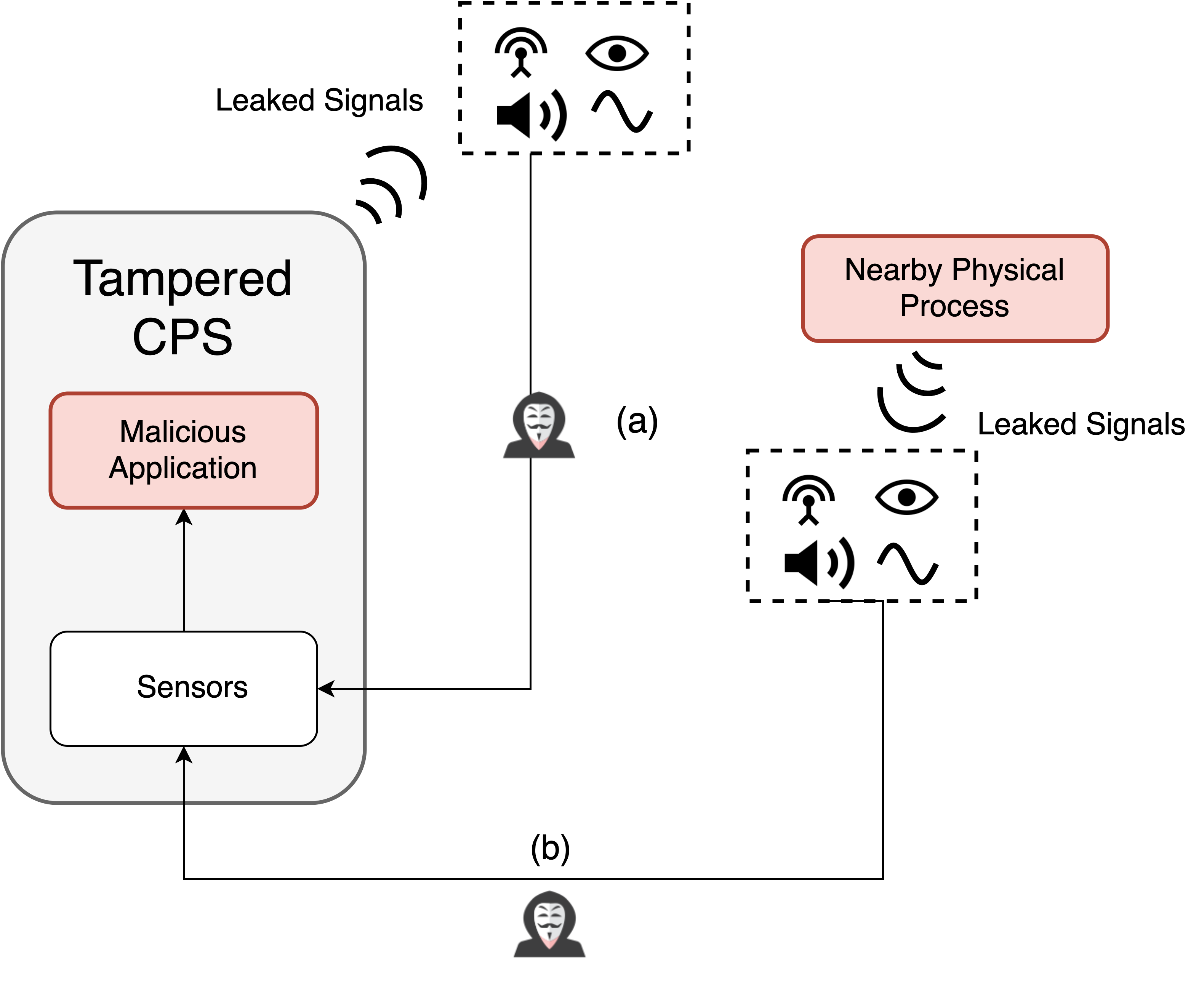}
\caption{Two types of information leakage attacks categorized by the consistency between the locations where the information leaked from and where the attacker collects data. (a) \textit{Co-located attacks}, where the signals leaked from the target CPS are collected by sensors embedded within the same CPS. (b) \textit{Remote attacks}, where the sensors are remotely measuring signal leakage from a nearby physical process.}\label{fig:fig8InfoLeakTypes}
\end{figure}

In general, information leakage attacks can be divided into two distinct categories: \textit{co-located attacks} and \textit{remote attacks}.

\begin{itemize}
    \item \textbf{Co-located attack: }This refers to attacks in which the source of the leaked signal is \textit{co-located} with the sensor used to measure it within a single computing system. In this scenario, the target CPS is maliciously utilized as the attacker's tool as well.

    \item \textbf{Remote attack: }This refers to attacks in which the source of the leaked signal is \textit{external} to the attacker's CPS, which measures said signals with its sensors.
\end{itemize}

These two types of attacks are shown in Fig. \ref{fig:fig8InfoLeakTypes}. Both attacks are normally carried out through trojan applications installed on the device that collects data from the sensors, which is processed and transmitted to the attacker. The main difference lies in whether the attack and the victim reside in the same computing system or not. In a \textit{co-located attack} shown in Fig. \ref{fig:fig8InfoLeakTypes}(a), the sensors within the CPS are maliciously utilized to collect information emitted from CPS itself, which means that the victim CPS itself serves as the attack tool at the same time. For a \textit{remote attack} shown in Fig. \ref{fig:fig8InfoLeakTypes}(b), the CPS together with its embedded sensors are external to the target, and are used by the attacker to collect nearby leaked information.

Modern smart devices such as smartphones are a common target for information leakage attacks since they are equipped with a wide range of high-precision sensors \cite{nield_2018}, and are owned by over 67\% of the global population \cite{howmanyphones}, making them a highly attractive target for attackers. Furthermore, third-party applications often have direct access to the high-precision sensors of the device without requiring any specific permissions. Therefore, a large portion of existing attacks in the literature are designed with iOS or Android apps.

\subsection{Threat Model}\label{subsec:InfoLeakThreatModel}
This section will provide a generalized discussion of the attack model for information leakage attacks on CPS, defining the attacker requirements, attack vectors, a generalized attack model, and attack goals. 

\subsubsection{Attack Requirements}
We propose three general attacker requirements for information leakage attacks in the following:

\begin{itemize}
    \item \textbf{Covert extraction: } This requires the adversary to have access to the stream of sensor data. This can be accomplished by various methods such as developing apps being secretly installed on the device through a remotely accessible backdoor, or embedding malicious code in a harmless-looking web page.
    \item \textbf{Data exfiltration: } This requires that the collected sensor data can be sent to a remote server, either directly through the attacker's software or through another application.
    \item \textbf{Information recovery: } This requires that the attacker can recover the desired information from the collected sensor data by either obtaining or training a classification model.
\end{itemize}

\subsubsection{Attack Model}\label{subsec:atkModel}
\begin{figure}
\includegraphics[width=8.4cm]{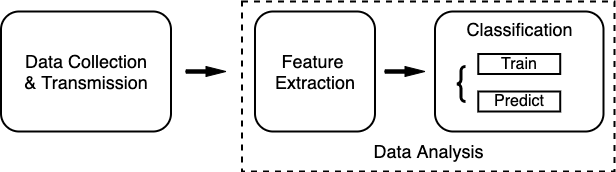}
\caption{Workflow of an information leakage attack. The leaked signals are collected by nearby sensors and transmitted to the attacker for further analysis. The analysis generally contains feature extraction and classification stages.}
\label{fig:fig9InfLeakage}
\end{figure}

Fig. \ref{fig:fig9InfLeakage} shows the overall workflow of an information leakage attack, which generally consists of the following two stages:

\begin{itemize}
    \item \textbf{Data collection and transmission: }In this stage, the attacker's malware collects raw data from the sensor. In the context of smartphones, an important consideration is the level of user permissions required by the platform for accessing the targeted sensor's data. In Android and iOS, permissions of sensitive sensors such as microphones, GPS sensors, and cameras are restricted, while some other sensors such as ambient light sensors, gyroscopes, accelerometers, and magnetometers can be accessed by an application without explicit user permissions. The collected sensor data will then be transmitted to the attacker for further analysis.
    \item \textbf{Data analysis: }The collected and transmitted data will be processed to retrieve the attacker's desired private information. In the vast majority of existing literature, machine learning techniques are used to extract the desired information from the collected sensor data. Thus, this stage can be further divided into \textit{feature extraction} and \textit{classification}. Feature extraction refers to identifying and extracting features from the raw data; the quality of the methods used to extract features plays a large role in the accuracy of the eventual classification. The data is then fed to the classifier, which is first trained with a chosen subset of the data. Once trained, the classifier can be used by the attacker to recover the sensitive information from the collected data.  
    
\end{itemize}

\subsubsection{Attack Vector}

The attack vector for an information leakage attack is categorized by the type of signal measured by the sensor. The sensor can measure signals leaked by \textit{vibrational}, \textit{electromagnetic, acoustic,} and \textit{optical} side channels. Besides, based on the \textit{locality} of the attacker, the attack can either be \textit{co-located} with the source of leaked information, or \textit{remote}. 

\subsubsection{Attack Goals} 

The attack goal refers to the information that is being leaked. Most of the existing literature has demonstrated inference on \textit{keystroke, task, location}, and \textit{speech}. 

\begin{itemize}
    \item \textbf{Keystroke inference:} Users' input, either via a mechanical or touch-based keypad, generates \textit{optical, electromagnetic, acoustic}, and \textit{vibrational} side-channel leakage that can be detected by the sensors of CPS.
    
    \item \textbf{Task inference:} Ongoing tasks or applications on CPS, or physical activity by the user, can generate \textit{vibrational} and \textit{electromagnetic} side-channel leakage that can be detected by the sensors of CPS.
    
    \item \textbf{Location inference:} \textit{Acoustic} side-channel leakage can be detected by the sensors of CPS, potentially revealing details about a user's location.
    
    \item \textbf{Speech extraction:} Human speech that generates \textit{vibrational, optical,} and \textit{electromagnetic} signals can be picked up by the sensors of CPS.
\end{itemize}

\subsection{Techniques for Information Leakage}
As discussed in Section \ref{subsec:atkModel}, successful implementation of an information leakage attack consists of two steps, \textit{data collection and transmission} and \textit{data analysis}. These are the core processes of an information leakage attack, and also are the main factors determining an attacker's requirements and challenges. The first challenge that the attacker faces is to stealthily obtain data from targeted CPS. For \textit{co-located} attacks, sensors embedded in targeted CPS are used for signal measurements and are the sources of raw data for attackers as well. As malware becomes more widespread and affects more users~\cite{2019_8million}, it would be easy for the attacker to obtain selected data within CPS through installed malicious software~\cite{damopoulos2013keyloggers}. Especially for some data that is generally not considered highly sensitive, such as accelerometer data on smartphones~\cite{Owusu:2012:API:2162081.2162095}~\cite{liu2015good}~\cite{han2012accomplice}, less protection or isolation is granted, thus making it easier for the attacker to access sensor data without raising suspicion. However, for \textit{remote} attacks, the attacker collects data in a different way: signals originating from user inputs are captured and measured by nearby CPS belonging to the attacker. In this case, the main problem for the attacker is to deploy the proper CPS at a target location~\cite{marquardt2011sp}~\cite{genkin2014rsa}~\cite{vuagnoux2009compromising}. 

To further infer private information, the main processes can be summarized as three steps: \textit{pre-processing, feature extraction,} and \textit{classification}. \textit{Pre-processing} is required as the extracted data often contains a significant amount of noise, and it mainly includes selecting, normalizing, and pre-calculating to generate a modified data matrix to be used in the following steps. \textit{Feature extraction} refers to the attacker extracting core features derived from the pre-processed data matrix that are intended to be informative and non-redundant. This step is intended to raise attack accuracy by discarding less relevant data and simultaneously facilitating the following learning step with dimensional reduction. \textit{Classification} is the core part of the data analysis, from which the attacker eventually gets results with the help of machine learning based methods. Techniques from other domains such as dictionary mapping~\cite{zhuang2009keyboard}, Hidden Markov Models (HMM)~\cite{halevi2012closer}~\cite{asonov2004keyboard}, and image deconvolution~\cite{4531151}~\cite{loughry2002information} have also been used in some works for private information inference. 

\subsection{Information Leakage: Known Attacks}
Existing literature on information leakage using signals will be discussed in this section. We summarize the existing information leakage attacks in Table \ref{tab:infoLeak} and more details will be presented in the following sections. To be more specific, sections \ref{subsubsec:KeyInfAtk} to \ref{subsubsec:speechExtraction} will cover \textit{keystroke inference, task inference, location inference,} and \textit{speech extraction} respectively. This section is divided according to the attack goal because methodologies used in various information leakage attacks are more dependent on the information being leaked.

\input{InformationTable}

\subsubsection{Keystroke Inference Attacks}\label{subsubsec:KeyInfAtk}
\begin{figure}
\centering
\includegraphics[width=8.4cm]{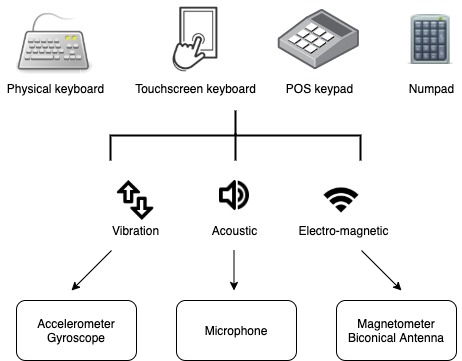}
\caption{Overview of existing work on keystroke inference attacks. Physical keyboards, touchscreen keyboards, POS keypads, and numpads have been explored as target devices, from which vibrational, acoustic, and electromagnetic information is leaked and picked up by corresponding sensors.}\label{fig:figKeystrokeInfer}
\end{figure}

Users' input to a cyber-physical system, either via a mechanical or touch-based keyboard, can generate \textit{acoustic}, \textit{electromagnetic}, and \textit{vibrational} side-channel leakage. Existing research explores using sensors to measure this side-channel leakage to recover the user's keystrokes. As is shown in Fig. \ref{fig:figKeystrokeInfer}, keystrokes on various devices including physical keyboards, touchscreen keyboards, POS keypads, and numpads have been explored as leaked information and are recovered from different sensor data.

\paragraph{\textbf{Keystroke Inference through Vibrational Leakage}}

In the case of mechanical keyboards, the vibrational side channel can be measured \textit{remotely} by the accelerometer and/or gyroscope of CPS that share a common surface with the keyboard. Each keystroke generates a unique, acoustically conducted, vibrational signature. Thus, by analyzing raw data from these sensors, an attacker can infer the keystrokes being typed. In the case of touch-based keyboards, the user's input changes the orientation of the device as well as generating vibrations. Thus, similarly to mechanical keyboards, this side channel can be measured by a \textit{co-located} gyroscope and/or accelerometer to infer keystrokes on a touch-based keypad.

\textbf{Co-Located Keystroke Inference:} Cai et al.~\cite{cai2011touchlogger} attempt at inferring keystrokes based on co-located motion sensors including accelerometer and gyroscope. They use a covert application installed on a victim's smartphone to measure raw device orientation data based on readings from the accelerometer. This data is used to train a Gaussian classifier, which can predict keystrokes with an overall accuracy of 71.5\% on a 10-digit numeric touch keypad. Following this work, Owusu et al. \cite{Owusu:2012:API:2162081.2162095} describe a similar attack, but develop two separate test modes, \textit{area} and \textit{character}, and use a random forest classifier. More importantly, the granularity of the attack is much higher, meaning the device's screen is divided into up to 60 squares in \textit{Area} mode and 29 squares in \textit{character} mode, rather than 10 used in \cite{cai2011touchlogger}. Using their techniques, they are able to break 59 out of 99 passwords logged during text entry on smartphones. Both of these attacks are performed on an Android device. 

In \cite{miluzzo2012tapprints}, the gyroscope and accelerometer are used in combination to predict keystrokes across both Android and iOS devices, including smartphones and tablets. In this paper, Miluzzo et al. achieve a prediction accuracy up to 90\%. In the same vein, Xu et al. \cite{xu2012taplogger} demonstrate a similar attack, making use of acceleration and gesture changes caused by tap events but using a different feature extraction scheme. Furthermore, they focus on the detection of tap events themselves with the obtained noisy motion sensor data, which can greatly improve the accuracy of keystroke inference in noisy environments, for example, even when the individual is walking. 

Al-Haiqi et al. \cite{al2013keystrokes} explore the topic of sensor fusion for keystroke inference, using features extracted from different combinations of the motion sensors (linear accelerometer, gyroscope, rotation vector sensor) and the combined accelerometer and magnetometer synthetic sensor data that are fed to the classifier. They discover that contrary to what is assumed in previous papers, using only the gyroscope's data is beneficial to the accuracy of the keystroke prediction, compared to fusing sensor readings. They are able to correctly predict up to 95\% of keystrokes by using only the gyroscope. 

\textbf{Remote Keystroke Inference:} It should be noted that the attacks discussed above are using data from \textit{co-located} inertial sensors to infer keystrokes. In other researches~\cite{marquardt2011sp}~\cite{wang2015mole}~\cite{liu2015good}~\cite{beltramelli2015deep}~\cite{maiti2018side}~\cite{ha2017side}, \textit{remote} attacks are demonstrated in which the vibrational leakage generated by user input is measured by the inertial sensors of a nearby system. This attack can generally be described by one of two attack scenarios: first, the inertial sensors of a wearable device, such as a smartwatch, are used to infer the keystrokes either on a mechanical or touch-based keypad~\cite{liu2015good}~\cite{wang2015mole}~\cite{maiti2018side}~\cite{beltramelli2015deep}; second, the inertial sensors of a CPS that share a surface with a mechanical keyboard, most commonly of a smartphone, are used to predict keystrokes on the keyboard~\cite{marquardt2011sp}~\cite{bartlett2017acctionnet}~\cite{Zhu:2014:CAU:2660267.2660296}.

The first remote keystroke inference attack exploiting vibrational side channels targets on mechanical keyboards with inertial sensors of smartphones \cite{marquardt2011sp}. One of the major limitations of remote information leakage attacks is that the attacker must place their sensor-equipped CPS within a reasonable distance of the target device, such that the leaked side channel can be measured with a sufficient degree of precision. This often requires special equipment, such as parabolic microphones or high-quality antennas, so that the attack can be launched from a far distance to remain concealed. Marquardt et al. \cite{marquardt2011sp} address this limitation by suggesting an attack in which the victim's smartphone shares a surface with the target keyboard, and infers keystrokes based on the leaked vibrational side channel. It is a common scenario for an individual to put their smartphone on the same table as their keyboard, so this attack can be both effective and covert. In this attack, the researchers manage to recover up to 80\% of the typed content using this method. 

Following this paper, more remote keystroke inference attacks adopt similar methodologies of exploiting a sensor-equipped CPS that the victim themselves brings into the vicinity of the keyboard. Most notably, smartwatches are equipped with inertial sensors, and can be used to carry out keystroke inference attacks in this fashion. In \cite{wang2015mole}, Wang et al. show that it is indeed possible for a smartwatch to leak vibrational information about a user's keystrokes. By combining Bayesian inference techniques with an awareness of the structure of the English language, the researchers are able to predict the words being typed 30\% of the time ideally and 10\% to 20\% practically depending on the subject. The main limitation of this technique is that the smartwatch is only worn on a single hand, introducing a significant upper limit on the accuracy of predictions. A similar attack is carried out in \cite{liu2015good} and \cite{beltramelli2015deep}, where the techniques are extended to POS terminals, such as credit card machines and ATM keypads. In \cite{maiti2018side}, these techniques are also applied to the numeric touch-based keypads of smartphones; in this case, a user wearing a smartwatch can leak a vibrational side channel, which can be analyzed to infer their pin. Further experiments are conducted for comparison with different smartwatches, motion sensors, and fusing mobile data. This attack is also extended to the application of mobile keypads with QWERTY layout.

In \cite{ha2017side}, a unique attack is proposed in which the password to a smart door-lock can be inferred by measuring a vibrational side channel, where the sensor-equipped device is placed by the attacker between the door and its frame. In this work, Ha et al. design the device such that it is discreet but still has the ability to exfiltrate the collected data to the attacker's remote server. By measuring the leaked vibrational side channel, the researchers demonstrate that it is possible to infer the password to the door lock.

\paragraph{\textbf{Keystroke Inference through Acoustic Leakage}}

Keystroke inference through acoustic side-channel leakage has also been researched. It is shown that on mechanical keyboards, there is enough of a correlation between the spectral signature of a keystroke and the actual key pressed for an attacker to predict keystrokes by recording their acoustic emanations. Asonov et al. \cite{asonov2004keyboard} are the first to demonstrate this attack; by simply recording the sound of typing, extracting time and frequency domain features via FFT, and training a neural network, they show that it is possible to predict keystrokes. Following this work, Zhuang et al.~\cite{zhuang2009keyboard} use MFCC features, unsupervised learning techniques (clustering), and a HMM-based language model to predict keystrokes. They are able to show a significant improvement in the accuracy of the predictions, predicting 96\% of typed characters correctly. Furthermore, Halevi et al. \cite{halevi2012closer} assume that the attacker has no access to the language model, and introduce the uncertainty of different typing styles. The techniques used to account for different typing styles bring the attack much closer to being feasible in a real-world scenario.

Similarly, a context-free attack is demonstrated in \cite{Zhu:2014:CAU:2660267.2660296} by using \textit{Time Difference of Arrival} (TDoA), which essentially takes the time between consecutive keystrokes into consideration. It also takes a ``context-free" approach, meaning that language models are not used. This way, the attack is also effective against random words that do not follow the English language structure, such as passwords. Using this technique, the researchers are able to recover more than 72.2\% of keystrokes.

\paragraph{\textbf{Keystroke Inference through EM Leakage}}

In \cite{vuagnoux2009compromising}, Vuagnoux et al. show that it is also viable to eavesdrop on keystrokes on physical keyboards (wireless or wired) by leveraging electromagnetic emissions. Their attack successfully recovers keystrokes 95\% of the time at distances up to 20 meters between the keyboard and antennae used to eavesdrop. The researchers use raw signals from the antennae for their analysis, and further classify four different identifying types of emanations from physical keyboards. They are able to successfully perform their attack in four different scenarios, ranging from within a shielded room to outside the building the keyboard is in.

\cbox{Lessons Learned: The majority of existing keystroke inference attacks focus on vibrational signals, but the possibility of optical leakage has not been explored. It remains an open research problem how keystrokes may create recoverable optical emanations in certain scenarios.}

\subsubsection{Task Inference Attacks}
In a task inference attack, the adversary measures some leaked side channels to gather information about the state of the physical process from which it is emanated. The recovered information can either be related to some tasks or applications running on a device being attacked or the activity of a nearby user. Many of the papers discussed in this section are not security research, but still have implications in the context of security. For example, a large body of research explores the use of motion sensors for user activity recognition on smartphones for use by context-aware applications. Though this is not directly related to security, it still describes a specific method used to recover potentially sensitive information from a leaked vibrational side channel. 

\paragraph{\textbf{Task Inference through Acoustic Leakage}}
Existing work has revealed that acoustic leakage contains abundant information which can be utilized to recover cyber data (e.g., cryptographic keys~\cite{genkin2014rsa}), optical information (e.g., LCD screen content~\cite{genkin2018synesthesia}), and even physical motions (e.g., printer nozzle motion~\cite{song2016my}).

Genkin et al. \cite{genkin2014rsa} demonstrate an acoustic cryptanalysis attack, in which a microphone is used to record acoustic signals that are leaked from the vibration of electronic components in a computer. When performing cryptographic operations, the CPU's power draw changes as a function of the specific operations being performed, which is reflected in the vibration of the electronic components. Thus, the private keys used for encryption can potentially be recovered by observing the acoustic side-channel leakage. 

More recently, the same researchers from \cite{genkin2014rsa} show a new attack~\cite{genkin2018synesthesia}, where a microphone is used to record acoustic emanations from an LCD computer screen that can be used to recover the contents of the screen. The momentary power draw of the monitor's digital circuits varies as a function of the screen content being processed in raster order, which affects the electrical load on the power supply components, causing the components to vibrate and emit acoustic noise. Thus, the acoustic signals can be processed to recover the original contents of the screen. Advanced signal processing techniques are utilized to extract various time and frequency domain features from the recorded acoustic leakage, and these features are used to train a convolutional neural network (CNN) classifier, which can eventually disclose information such as text on the screen and websites being visited.

Backes et al. \cite{backes2010acoustic} show that the acoustic emanations from a dot-matrix printer can be measured to infer the text being printed. This attack starts with a training phase, where known words are printed through the printer and features extracted from the resulting acoustic leakage are stored in a dictionary. The word-based approach is preferred to the letter-based approach, since the emitted sound is significantly blurred across adjacent letters. In addition to using the dictionary, they also use HMM-based speech recognition techniques to improve the accuracy of the attack. In a similar attack, Song et al.~\cite{song2016my} demonstrate that acoustic side-channel leakage from a 3D printer can be measured by using a microphone and magnetic sensor together to infer the motion of the nozzle, which can leak the designed object.

\paragraph{\textbf{Task Inference through Vibrational Leakage}}

Data of motion sensors in smartphones can be used to infer users' acivities. This is referred to as \textit{activity recognition}, and is generally useful when developing context-aware applications. Thus, some existing literature detailing activity recognition using motion sensors is not security research. Nonetheless, this could pose a side-channel leakage threat in which an attacker can potentially track a victim's activities covertly through a malicious application that measures this leakage. 

In \cite{kwapisz2011activity} and \cite{shoaib2014fusion}, the researchers propose various machine learning techniques to use accelerometer and gyroscope data to classify human activity. Current research is able to classify activity as walking, running, climbing stairs, descending stairs, jogging, sitting, and standing. Though leaking this information does not pose an immediate threat to the user, this technique could once again be used as a single step of a larger, more substantial information leakage attack. In \cite{bartlett2017acctionnet}, Bartlett et al. propose to build a large-scale dataset called \textit{AcctionNet}, which is essentially a dictionary mapping inertial sensor readings to human activities. If an attacker chooses to exploit vibrational leakage for task inference, this dataset could serve as a great resource in carrying out the attack.

\paragraph{\textbf{Task Inference through EM Leakage}}

The first work in this domain is proposed by Van et al. \cite{van1985electromagnetic}, where they show that electromagnetic radiation generated by CRT TVs when displaying video can be used to recover what was being displayed remotely. Using a TV receiver, antennae, a signal amplifier, and a synchronization method, they are able to recover video information from up to a kilometer away.

In \cite{gandolfi2001electromagnetic}, Gandolfi et al. are able to use microscopic copper coils as probes to read electromagnetic power radiation, and in doing so successfully attack DES, COMP128, and RSA on different CMOS chips. In this attack, they are able to use the small amount of voltage induced in the coils to successfully recover key information from these algorithms. 

Another attack demonstrating task inference through magnetic leakage exploits hard disk drives by measuring magnetic leakage with the magnetic sensor of a nearby smartphone \cite{biedermann2015hard}. And thus this is a \textit{remote} attack. Hard drives contain different magnets which rapidly move the read/write head to a target position on the disk to perform a read or a write. The magnetic field due to the moving head can be picked up by a nearby smartphone; by analyzing the collected data, Biedermann et al. show that an attacker can detect the operating system that is used, distinguish between known applications being started, distinguish virtual machine activity on a server, match ongoing network traffic to a server, and detect file caching based on disk activity. 

\paragraph{\textbf{Task Inference through Optical Leakage}}
Optical leakage exploited to infer tasks can be categorized into two types. One is through visible light such as a blinking LED~\cite{loughry2002information}~\cite{10.1145/3133956.3134039} or screen~\cite{kuhn2002optical}, while another is reflections on smooth surfaces~\cite{4531151}~\cite{backes2009tempest}.

\textbf{Visible Light Inference:} Loughry et al. \cite{loughry2002information} show that LED indicators of various devices can act as sources of optical side-channel leakage. By using a photodetector, they find that by measuring leakage from the LEDs on the backside of a wireless router, information about the transmitted and received packets can be extracted. In \cite{kuhn2002optical}, Kuhn et al. use a photodetector to measure optical side-channel leakage from a CRT display at a distance. They demonstrate that since the intensity of the light emitted by a raster-scan screen as a function of time corresponds to the video signal convolved with the impulse response of the phosphors, the intensity readings measured by the photodetector can be deconvolved to recover the text on the screen.

In \cite{10.1145/3133956.3134039}, Tajik et al. use optical probing of FPGA boards to infer bitstreams of data. By using precise optical equipment, they improve on previous attacks by requiring no prior preparation or modification of the FPGA boards. In this attack, the plain text of securely encrypted bitstreams being processed on the device is recovered within ten days of acquiring the board.

\textbf{Reflection Inference:} More recently, Backes et al. \cite{4531151} make use of reflections from computer monitors and screens detected by cameras or telescopes, revealing the contents of the screen. They show that teapots, eyeglasses, the human eye itself, wine glasses, and glass bottles can reflect optical emanations from screens, which can be picked up by an optical sensor. In \cite{backes2009tempest}, the same researchers from \cite{4531151} improve this attack by using image deconvolution techniques to remove the blur caused by observing optical leakage from moving objects. Using this technique, they are able to greatly increase the attack distance of attacks exploiting compromising reflections from the human eye. 

\cbox{Lessons Learned: The majority of existing task inference attacks exploiting acoustic, electromagnetic, and optical leakage are \textit{remote attacks}, where the attackers observe and collect the emanations using their own sensing systems at a distance. It would be interesting to explore the possibility of \textit{co-located attacks} in this domain.}

\subsubsection{Location Inference Attacks} This kind of attacks aim to infer the location of the victim via \textit{acoustic} or \textit{vibrational} side-channel leakage.

\paragraph{\textbf{Location Inference through Acoustic Leakage}}
One of the methods used for location inference through acoustic leakage is the analysis of electro network frequency (ENF) signals in recorded audio. ENF refers to background signals that reflect the supply frequency of electric power in distribution networks of a power grid, commonly with a mean value of 60Hz in the United States or 50Hz in most other parts of the world. Thus, the vibrations of ENF contain time-varying and location-sensitive information of the original audio or video recordings. This is exploited by Jeon et al. \cite{jeon2018m} to infer a system's location based on data collected from a co-located or nearby microphone. The first step of the attack is to scrape the web for audio recordings that contain ENF signals and build a map correlating ENF signals with physical locations. Thus, when the attack is actually carried out, the measured ENF signature can be compared against the map to infer the victim's location. 

\paragraph{\textbf{Location Inference through Vibrational Leakage}}
Existing work that falls into this category generally exploits motion sensors such as gyroscopes and accelerometers, which are sensitive to vibrations, to infer the location or routine leaked from users' movements.

Han et al. \cite{han2012accomplice} first demonstrate the possibility of using accelerometers and gyroscopes to infer location. The attackers take the approach of using motion sensor data to infer a trajectory through space. Since idiosyncrasies in roadways create globally unique constraints, the calculated trajectory can then be mapped onto a global map, in which the closest existing trajectory is found. Using this technique, the researchers show that it is possible to infer location by measuring vibrational side-channel leakage. In a similar attack proposed in~\cite{narain2016inferring}, Narain et al. use a sensor fusion based approach for location inference, and use the accelerometer, gyroscope, and magnetometer to infer routes of the victim user. Then, in a similar manner to the previous paper, they compare this route to existing routes, returning the top matches for potential locations. 

It is worth mentioning that some works not focusing on security also imply the risk of a possible attack. In \cite{lee2002activity}, inertial sensors are used to infer an individual's position indoors. The approach taken can be summarized as: based on a known starting position, software can be implemented that calculates the number of steps taken based on motion sensor data; once the user stops, their path and location can be inferred given a limited knowledge of the architecture of the building that they are inside. In \cite{gao2014elastic}, which also does not focus on proposing new information leakage attack, a method called ``elastic pathing" is proposed to use velocity data to reconstruct a user's routes, and eventually their location. This paper focuses on the potential privacy risk posed by insurance companies monitoring a user's speedometer, showing the possibility for an insurance company to track their customers using this data. In general, the techniques used for route reconstruction and route mapping are similar to those used in the 
proposed attacks. 

\cbox{Lessons Learned: Most of the existing location inference attacks focus on predicting location or routine by analyzing motion sensor data that imply the user's movement. However, the approach taken by \cite{jeon2018m} to leverage geo-specific signals to recover victim locations can be generally applied beyond ENF. New attacks may pay attention to signals such as light or EM waves that vary in different regions, with which the attack performance can be improved.}

\begin{figure*}
\centering
\includegraphics[width=11.4cm]{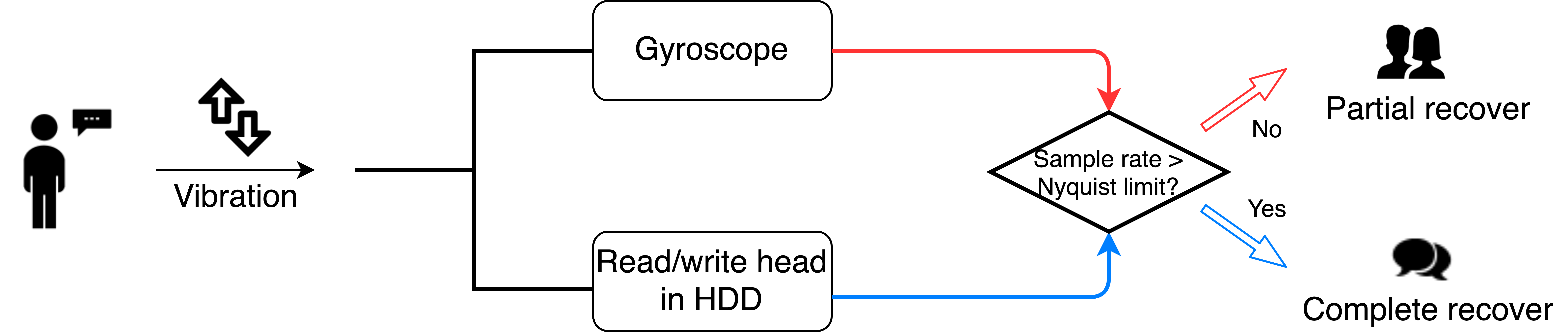}
\caption{Overview of existing work on vibrational speech extraction attacks. The main idea is to collect vibrational data from sensitive sensors and recover speech content~\cite{michalevsky2014gyrophone}~\cite{kwonghard}. However, if the sample rate of the adopted sensor is below the Nyquist limit for human speech (e.g., gyroscope), it will only recover partial information such as the gender of the speaker~\cite{michalevsky2014gyrophone}.}\label{fig:figSpeechExtract}
\end{figure*}

\subsubsection{Speech Extraction}\label{subsubsec:speechExtraction}

In this section, we choose to focus on unintended speech content disclosure via physical side channels. Therefore, speech extraction attacks accomplished by compromising a system to access the microphone are not covered. 

\paragraph{\textbf{Speech Extraction through Vibrational Leakage}}
Fig. \ref{fig:figSpeechExtract} presents an overview of existing attacks for human speech vibrational inference. The basic attack principle is that the acoustic speech will cause sensitive components (such as gyroscopes~\cite{michalevsky2014gyrophone} and HDD heads~\cite{kwonghard}) to vibrate, and this vibration can be recovered from sensor data. However, depends on whether the sampling rate exceeds the Nyquist limit, the attacker is able to recover either complete speech~\cite{kwonghard} or just information of the speaker~\cite{michalevsky2014gyrophone}.

Michalevsky et al. \cite{michalevsky2014gyrophone} show that smartphone gyroscope sensors can act as unintentional microphones. In this work, it is shown that gyroscopes in smartphones are sensitive to acoustic vibrations, and can thus potentially be used to recover speech. The main limitation of this approach is that the sample rate of the gyroscope is often below the Nyquist limit for human speech; thus, the attack does not recover actual intelligible speech, but rather recovers details about the speaker such as their gender.

To address this limitation, Kwong et al.~\cite{kwonghard} utilize a hard disk drive as an unintentional receiver for acoustic signals. They show that since the read/write head of a hard disk drive is sensitive to acoustic frequencies, it can also be used as a microphone. Since the sample rate is higher, intelligible speech is recoverable from the HDD that is even clear enough for recorded music to be recognized by song-recognition apps such as Shazam. 

\paragraph{\textbf{Speech Extraction through Optical Leakage}}
Davis et al.~\cite{Davis2014VisualMic} also show the possibility of eavesdropping on human speech via optical leakage. Since many objects can be considered as receivers of acoustic signals, vibrations in these objects due to sound waves can be observed by a high-precision camera. For example, a high-speed camera is used to take a video of a bag of potato chips. By observing the slight movement and changes in reflection patterns on the bag, the researchers are able to recover audible sounds such as speech and music. 

More recently, Nassi et al.~\cite{nassilamphone} state that acoustic signals will cause fluctuations of the air pressure on the surface of the hanging bulb, therefore leading to the vibration of the bulb. In this attack, an electro-optical sensor is placed remotely to analyze a hanging light bulb's frequency response to sound. With their developed algorithm, they are able to recover speech and singing in real time with an intelligibility over 53\%.

\paragraph{\textbf{Speech Extraction through EM Leakage}} One work that extracts human speech through EM leakage is proposed by Wang et al.~\cite{wang2016we}. In the proposed \textit{WiHear} attack, the attacker is able to capture the micro-movements of human mouths by collecting and analyzing WiFi radio reflections, with which they successfully infer up to 91\% of speech content. Furthermore, with leveraged multi-input multi-output (MIMO) technology, they are able to infer multiple people's speech simultaneously with an accuracy of 74\%. In this attack, a multi-cluster/class feature selection scheme~\cite{gheyas2010feature} is utilized as a feature extractor, and dynamic time warping (DTW)~\cite{wang2013dude} is adopted for classification.

\cbox{Lessons Learned: Existing attacks on speech extraction have exploited the vibrational and optical leakage caused by human speech. Given the threat of these attacks, it is important to develop novel defense solutions in this area. }

\subsection{Limitations and Opportunities}\label{subsec:infoLimitation}
In this section, we will examine some limitations and opportunities of information leakage attacks.

\textbf{Dependency on stealthy malware: }To successfully conduct an attack, the first step is to obtain raw data collected from the target CPS. To increase the stealthiness of the attack, the malware installed in the targeted system is designed to behave as explicitly benign and only acquires insensitive data as well as network access for offloading the user data. However, there are limitations on the level of stealthiness the malware can accomplish. These limitations are mentioned in one paper~\cite{maiti2018side}, where the malicious application installed on smartwatch runs in the background collecting data at a high frequency, leading to high power consumption. This can be significant as the battery capacity of smartwatch is far smaller than that of smartphone, thereby raising more suspicion.

\cbox{Lessons Learned: This is often a limitation for many information leakage attacks, as they are generally dependent on malware to collect sensor data. To this end, one possible defense against information leakage attacks could be monitoring the power consumption or resource usage for unique characteristics of the malware.}

\textbf{Lack of generalization: }Similar to the same limitation of signal injection attacks, the attacker has to address variations in both the user and targeted device to infer accurate information. 

The proposed models in some existing research face the challenge that they are not generalized for general targets with large variations between them. The adversary has to train the model separately for each possible configuration and pick the most suitable one which generates the most sensible output. This is further validated in \cite{beltramelli2015deep}, where the keystroke inference accuracy drops to 19\% when data sets used for training and evaluation are from different keypads, comparing to 73\% and 59\% achieved in the same work when using the same keypads. In recognition of this limitation, some researchers test their models on several experimental devices \cite{miluzzo2012tapprints} or different users \cite{wang2015mole} to validate the generalizability.

Halevi et al. \cite{halevi2012closer} take the first investigation into the variation of typing styles affecting acoustic-based keystroke inference attacks and draw the conclusion that the strength of the proposed acoustic eavesdropping attack is limited due to the uncertainties brought about by different typing styles. Similarly, Maiti et al. \cite{maiti2018side} also investigate popular typing styles used by the public and discuss how various typing scenarios can affect the effectiveness of vibrational keystroke attacks, though they do not investigate the case where the user holds the smartphone and types using both hands. In \cite{marquardt2011sp}, the orientation of the attacker's deployed iPhone is shown to bring uncertainty to the results of keystroke inference via vibrational leakage. 

\cbox{Lessons Learned: To the scope of this limitation, one direction for attackers may be developing more robust methodologies that can be applied to a wider range of targets; or on the contrary, recognizing the relationship between variations and user habits, collected sensor data can be used to infer user identity as well.}

\textbf{Short distance for remote attacks: }This limitation affects only \textit{remote} attacks. Compared to \textit{co-located} attacks where the attackers collect insensitive data from local sensor readings, the remote attackers need to deploy their own sensing systems within a certain distance of the target CPS. This, however, violates the physical security of the area near the targets and brings about a lack of stealthiness. More importantly, the distance between deployed CPS and the target plays an essential role that considerably affects the attack: even a small increase in distance will dramatically reduce its effectiveness and accuracy. Therefore, the attacker has to make a trade-off between stealthiness and attack effectiveness. Generally, the attacker has to place their CPS at a relatively short distance from the target so as to reach a satisfactory attack accuracy. 

Genkin et al. \cite{genkin2014rsa} conduct an acoustic-based task inference attack by placing a mobile phone next to the target laptop or a sensitive microphone at a distance of 4 meters away. Backes et al. \cite{backes2010acoustic} show the results that the recognition rate will drop to 4\% with a distance of only 2 meters. Asonov et al. \cite{asonov2004keyboard} achieve an acoustic-based keystroke inference attack at a maximum distance of 15 meters without physical invasion of the target CPS. However, as is discussed in \cite{zhuang2009keyboard}, similar attacks can be defended against by ensuring the physical security of machine as well as its surroundings. Especially considering scenarios such as ATM pin pad entry, where the users are sealed into a small space with effective sound insulation, thus it becomes difficult to collect acoustic data from the outside at a distance. Marquardt et al. \cite{marquardt2011sp} conduct a remote vibration-based keystroke inference attack where they try to evade this problem by installing malware on the victim's iPhone placed nearby, which in turn plays the role of the attacker's sensing system by eavesdropping on a nearby keyboard. 

\cbox{Lessons Learned: The root cause of the limitations caused by large distance lie in the low signal-to-noise ratio (SNR) of the collected data. One method to compensate is to leverage machine learning to aid in estimating the missing data.}

\textbf{Environmental noise: }Apart from selecting equipment sensitive enough for capturing signals, the attacker also has to take ambient noise into consideration, which affects inference accuracy dramatically. Although some subtle or regular noises in the environment can be filtered out by the attacker's malware, there still exist scenarios where noises cannot be distinguished from target signals. Marquardt et al. \cite{marquardt2011sp} point out that significant vibrations caused by the user bouncing their knees or tapping the desk will result in a loss of useful vibrational data for keystroke inference. Though related research such as \cite{liu2015good} mitigates the negative influence of unexpected movement of the user by integrating acoustic signals and vibrations, effective mitigation remains an open problem. Also, a task inference attack methodology is proposed in \cite{genkin2014rsa}, where Genkin et al. state that the typical noise environment is concentrated at low frequencies while the acoustic leakage is above this range, thus the attacker can easily filter out these background noises. 

\cbox{Lessons Learned: From the attacker's perspective, potential interference should be taken into consideration in the design. The very same noise, on the other hand, can be an effective defense for the users.}

\textbf{Signal quality: }For some attacks, the quality of signals that are emitted by the target or received by sensing systems is limited due to physical imperfections. 

Some are limited by the sensing systems on the receiver side. For example, in \cite{wang2015mole}, the effectiveness of the proposed keystroke inference attack is limited (30\% ideally and less than 20\% practically) because the wearable device (i.e., smartwatch) serving as sensing system is deployed on only one wrist of the typist. The main problem encountered in \cite{michalevsky2014gyrophone} is that the sample rate of the gyroscope embedded in user's phone lies below the Nyquist requirements for human speech, resulting in the failure to recover speech content. Instead, the attacker obtains information about the speaker, which requires less information recovered from raw data. 

Also, some attacks are limited by the target systems. One simple method to mitigate the acoustic-based keystroke inference attacks as discussed in \cite{asonov2004keyboard} is for the user to adopt a silent keyboard made of rubber, or a touch-screen keyboard. Because the key principle behind acoustic leakage is to differentiate sounds emanated by different keystrokes, the attack could easily fail if acoustic signal emissions are alleviated.

\cbox{Lessons Learned: This limitation can shed light on effective defenses against these attacks, where an effective mitigation strategy is to reduce the emanated signal strength, such as by placing the keyboard on a rubber layer or placing a film over the computer monitor.}

%% file: InformationTable.tex
\begin{table*}[t]
\small
\caption{Summary of Existing Information Leakage Attacks}
\centering
\setlength{\tabcolsep}{1mm}{
\begin{tabular}{|c|c|c|c|c|c|c|c|c|c|c|}
\hline
Exploited Sensor 
& \rotatebox{90}{co-located} & \rotatebox{90}{Remote}
& \rotatebox{90}{Vector} 
& Data Collection
& Leaked Information & Accuracy & Method & TE 
& Ref   \\
\hline

Acc
& \checkmark &
&   \faExchange    
& Android phone
& TS Keyboard Presses(29)
& 59\% & Random Forest & R
& \cite{Owusu:2012:API:2162081.2162095} \\
\hline

Acc
& & \checkmark
&   \faExchange    
& iPhone
&  PK on Keyboard(26)
& 80\% & L/R, N/F classifier & R
& \cite{marquardt2011sp} \\
\hline

Acc
& & \checkmark
&   \faExchange   
& Smartwatch
& PK on Keyboard, Numpad(26, 10)
& 65\% & K-Nearest Neighbor & I
& \cite{liu2015good} \\
\hline

Acc
& \checkmark&
&  \faExchange    
& Android Phone
& User's Physical Activities
& 90\% & LR, MlP, SM & I
& \cite{kwapisz2011activity} \\
\hline

Acc
& \checkmark &
&   \faExchange 
& iPhone
& User Location/Route
& - & Trajectory Mapping & R
& \cite{han2012accomplice} \\
\hline

Acc, Gyro
& \checkmark & 
&   \faExchange    
& Android phone
& TS Numpad Presses(10)
& 71.5\% & Supervised Learning & I
& \cite{cai2011touchlogger} \\
\hline

Acc, Gyro
& \checkmark &
&   \faExchange    
& Android \& iPhone
& TS Tap Location 
& 90\% & Ensemble Learning & I
& \cite{miluzzo2012tapprints} \\
\hline

Acc, Gyro
& \checkmark &
&   \faExchange    
& Android phone
& TS Numpad Presses(10)
& $>$80\% & K-means, LibSVM & R
& \cite{xu2012taplogger} \\
\hline

Acc, Gyro
& & \checkmark
&   \faExchange    
& Smartwatch
& PK on Keyboard(26) 
& 30\% & Bayesian & I
& \cite{wang2015mole} \\
\hline

Acc, Gyro
&  & \checkmark
&   \faExchange    
& Smartwatch
& TS Numpad, Keyboard(10, 26)
& ~30-90\% & Ensemble Learning & I
& \cite{maiti2018side} \\
\hline

Acc, Gyro
& & \checkmark
&   \faExchange    
& Smartwatch
& PK/TS Numpad Presses(10)
& 59-73\% & LSTM  & I
& \cite{beltramelli2015deep} \\
\hline

Acc, Gyro
& \checkmark &
&   \faExchange    
& Smartphone
& User's Physical Activities
& 54-90\% & CNN  & I
& \cite{bartlett2017acctionnet} \\
\hline

Acc, Gyro
& & \checkmark
&   \faExchange   
& EBIMU24G Sensor
& PK Door Code(10)
& 30-70\% & Logistic Regression & R 
& \cite{ha2017side} \\
\hline

Acc, Gyro, Mag 
& \checkmark &
&   \faExchange    \faWifi
& Android phone
& TS Numpad Presses(10)
& 95.63\% & Ensemble Learning & I
& \cite{al2013keystrokes} \\
\hline

Acc, Gyro, Mag
& \checkmark &
&   \faExchange    \faWifi
& Android Phone
& User's Physical Activities
& ~30-98\% & Multiple/WEKA & I
& \cite{shoaib2014fusion} \\
\hline

Acc, Gyro, Mag
& \checkmark &
&    \faExchange    \faWifi
& Android Phone
& User Location/Route
& 30-50\% & Graph Matching & R 
& \cite{narain2016inferring} \\
\hline

Acc, Gyro, Mag
& \checkmark &
&   \faExchange    \faWifi
& Custom Sensor Array
& User Activity/Location
& 86\% & K-Nearest Neighbor & R
& \cite{lee2002activity} \\
\hline

Antenna
& & \checkmark
&       \faWifi
& TV Receiver
& Screen Content
& - & Re-synchronization & R
& \cite{van1985electromagnetic} \\  
\hline

Antenna
& & \checkmark
&       \faWifi
&  Biconical Antenna
& PK on Keyboard(39)
& 95\% & STFT & R
& \cite{vuagnoux2009compromising} \\ 
\hline

Cam 
& & \checkmark
&     \faSunO 
& Cam, Telescope
& LCD Screen Content
& - & Human Readability & I
& \cite{4531151} \\
\hline

Cam
& & \checkmark
&    \faSunO  
& Cam, Telescope
& LCD Screen Content
& -  & Image Deconvolution & I
& \cite{backes2009tempest} \\
\hline

Cam
& & \checkmark
&     \faSunO  
& High Speed Cam
& Audio Eavesdropping
& - & Local Motion Signals & I
& \cite{Davis2014VisualMic} \\
\hline

EM Probe
& & \checkmark
&       \faWifi
& Copper Probes
& Encryption Keys
& -  & Key Inference & I
& \cite{gandolfi2001electromagnetic} \\
\hline

Gyro
&\checkmark &
&  \faExchange    
& Android Phone
& Audio, Speech
& 6-84\% & SVM, GMM, DTM & I
& \cite{michalevsky2014gyrophone} \\
\hline

HDD as Mic
& \checkmark &
& \faMicrophone      
& Hard Drive
& Audio Eavesdropping
& - & Audio Recognition & R
& \cite{kwonghard} \\
\hline

Light bulb
& &\checkmark
&  \faSunO    
& Electro-optical sensor
& Audio, Speech
& 53-70.4\% & Developed Algorithm & I
& \cite{nassilamphone} \\
\hline

Mag
& & \checkmark
&      \faWifi
& Android Phone
& Computer Activities/Behaviors
& 65-95\% & Enrollment Vectors & I
& \cite{biedermann2015hard} \\
\hline

Mag, Mic
& & \checkmark
& \faMicrophone     \faWifi
& Android Phone
& 3D Printer IP
& ~89-94\% & MTE & R 
& \cite{song2016my} \\
\hline

Mic 
& & \checkmark
& \faMicrophone   
& Microphone
& PK on Keyboard (26)
& 40-60\% & Time-Frequency & I
& \cite{halevi2012closer} \\
\hline

Mic
& & \checkmark
& \faMicrophone  
& Microphone
& DM Printed Text 
& 72-95\% & HMM & I 
& \cite{backes2010acoustic} \\
\hline

Mic
& & \checkmark
& \faMicrophone  
& Microphone
& PK on Keyboard, Numpad(30, 9)
& 79\%  & Neural Network & I 
& \cite{asonov2004keyboard} \\
\hline

Mic
& & \checkmark
& \faMicrophone     
& Microphone
& PK on Keyboards(26)
& 96\% & HMM & R
& \cite{zhuang2009keyboard} \\
\hline

Mic
& &\checkmark
& \faMicrophone      
& Android Phone, Mic
& RSA Key(4096 Bit)
& -  & Chosen Ciphertext & R
& \cite{genkin2014rsa} \\
\hline

Mic
& & \checkmark
& \faMicrophone      
& Microphone
& PK Keyboard
& 73-90\% & Dictionary & I 
& \cite{berger2006dictionary} \\
\hline

Mic
& & \checkmark
&  \faExchange    
& Smartphone
& PK Keyboard
& 72.2\% & TDoA  & R 
& \cite{Zhu:2014:CAU:2660267.2660296} \\
\hline

Mic
& \checkmark & \checkmark
& \faMicrophone    
& Android Phone, Mic
& LCD Screen Content
& 40-100\% & CNN & I
& \cite{genkin2018synesthesia} \\
\hline

Mic
&\checkmark &
& \faMicrophone      
& Video/Audio File
& User Location
& 76\% & K-Nearest Neighbor  & I
& \cite{jeon2018m} \\
\hline

Microscope
& & \checkmark
&     \faSunO 
& Optical Probing
& Encrypted Bitstream
& - & Plaintext Extraction & I
& \cite{10.1145/3133956.3134039} \\
\hline

NIC
& & \checkmark
& \faWifi
& Desktop Computer
& Audio, Speech
& 91\%
& MCFS, DTW & R
& \cite{wang2016we}\\
\hline

Photodiode
& & \checkmark
&     \faSunO  
& LED/PIN Photodiode
& Processed Data on Devices
& - & Classification & R
& \cite{loughry2002information} \\
\hline

Photosensor
& & \checkmark
&     \faSunO  
& Photosensor Module
& CRT Screen Content
& - & Image Deconvolution & I
& \cite{kuhn2002optical} \\
\hline

\end{tabular}}
    \begin{tablenotes}
      \small
      \item TE = Testing Environment; I = Ideal; R = Replicated Real-World Conditions; TS = Touchscreen; PK = Physical Keystrokes; IP = Intellectual Property; Acc = Accelerometer; Gyro = Gyroscope; Mag = Magnetometer; Cam = Camera; Mic = Microphone; NIC = Network Interface Controller; (X) = X number of unique keys; LR = Logistic Regression; SM = Straw Man; MlP = Multilayer Perception; MTE = Mean Tendency Error; \faSunO = Optical; \faExchange = Vibrational; \faMicrophone = Acoustic; \faWifi = Electromagnetic;
    \end{tablenotes}
\label{tab:infoLeak}
\end{table*}

%% file: section5.tex
\section{Defense against Cyber-Physical Attacks}\label{sec:defense}
As cyber-physical attacks show an increasing threat across CPS, it is as essential to investigate corresponding defenses as to understand the principles of cyber-physical attacks. Existing literature has demonstrated various ways to detect and prevent attacks in the \textit{physical}, \textit{cyber}, \textit{cyber-physical}, and \textit{application} domains. In this section, we will present existing defense techniques corresponding to these categories.

\begin{figure*}
\centering
\includegraphics[width=12.4cm]{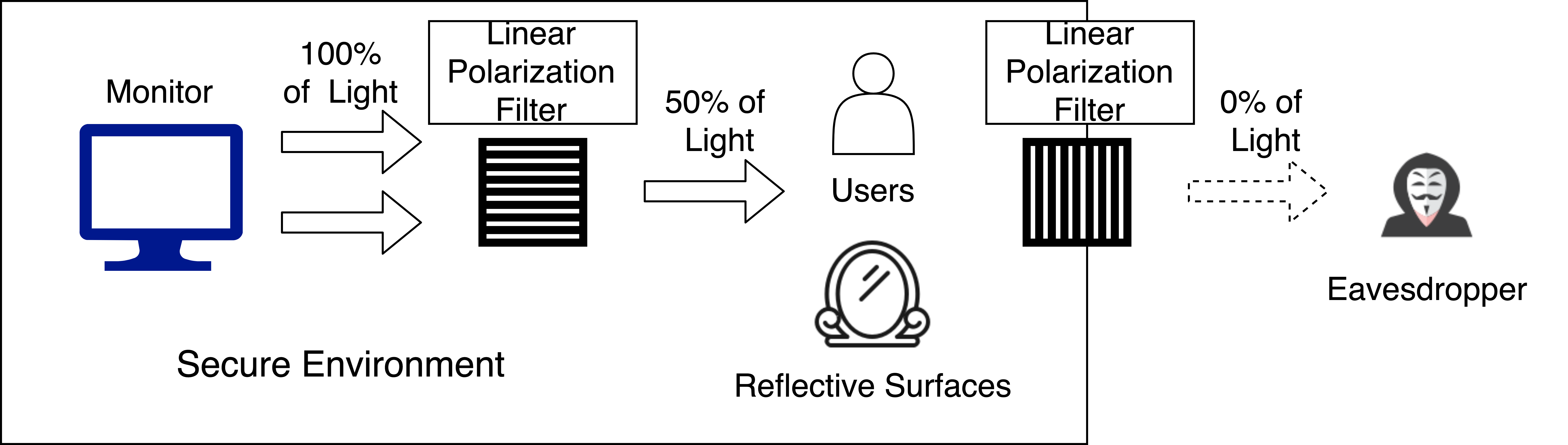}
\caption{Filtering defense against optical information leakage using polarization mirrors. This phenomenon shows that two linear polarization filters aligned at 90 degrees will block all light, while a single filter will let 50\% of original unpolarized light pass through. With proper set-up of polarization mirrors (one at the window and one at the screen), it will keep attackers from eavesdropping without blocking the user's normal use~\cite{backes2009tempest}.}\label{fig:figPolarization}
\end{figure*}

\subsection{Physical-Domain Defense}

In the physical domain, the biggest challenge to defend against cyber-physical attacks lies in the incapability of manipulating physical phenomena while probing them. This brings two challenges to conventional defenses: (1) it becomes difficult to validate the authenticity of passively-sensed signals via protected protocols or certificates like is done in networks or software, and (2) emitted signals from CPS cannot be encrypted or stopped from leaking sensitive information.

To handle these two challenges, two types of defenses are proposed, which we refer to as \textit{environment verification} and \textit{physical barriers}. For the first challenge, \textit{environment verification} proposes to actively probe the surrounding environment and validate signals with responses, while \textit{physical barriers} aim to prevent CPS from receiving specific types of signals (e.g., EM signals) via physical isolation. On the other hand, to prevent leaking information through signals, \textit{environment verification} suggests that users clear the surrounding area before interacting with CPS, and \textit{physical barriers} propose to isolate the CPS from open environments. More details about these two types of defenses are discussed in the following.

\subsubsection{\textbf{Environment verification}}
As signal injection attacks cause sensor readings to deviate from environment measurements, one way to detect attacks is to actively probe the environment and compare the probe reading with the expected response. With this method, any major difference between received signals and the expected response will be considered as an indication of an attack. Kune et al.~\cite{kune2013ghost} adopt this principle and design a defense where some investigative actuation is cast on cardiac tissues, and cardiac signal readings are compared to their expected values. Similarly, Muniraj et al.~\cite{muniraj2019detection} propose an ``active detection" method where judiciously designed excitation signals are superimposed on control commands, thus leading to a set of acceptable behavior for the system. Then the sensor will probe the environment, and the aforementioned effect is expected to be reflected in the sensor measurements if the system functions normally.

Also, for information leakage attacks, which generally suffer from the limitation of \textit{short distance} as discussed in Section \ref{subsec:infoLimitation}, verifying the security of surrounding areas may be a good choice before using devices. Zhuang et al.~\cite{zhuang2009keyboard} propose to ensure the physical security of the machine and room by clearing bugging devices and using quieter keyboards to prevent keystroke inference targeting physical keyboards. Similarly, to counter the task inference attack proposed in \cite{backes2009tempest} that exploits reflections in human eyes, Backes et al. suggest securing the surroundings and avoid any suitable hiding places for a spying telescope.

\subsubsection{\textbf{Physical barrier}}
A physical barrier exterior to the target CPS may also provide protection by offering better isolation from the environment. To be more specific, the physical barrier can be used both to reject exotic injected signals and to contain sensitive information from leaking out. 

An example is to use Faraday cage to protect against EM interference, this technique is also referred to as \textit{shielding}. Besides, existing work has demonstrated some other shielding techniques such as acoustic damping~\cite{son2015rocking}~\cite{217565}~\cite{trippel2017walnut}~\cite{bolton2018blue}, optical isolation~\cite{park2016ain}, physical separation~\cite{shahrad2018acoustic}~\cite{khazaaleh2019vulnerability}, and EM shielding~\cite{giechaskiel2019framework}~\cite{kune2013ghost}~\cite{kasmi2015iemi}~\cite{selvaraj2018electromagnetic}. 

Also, physical barriers have been shown to be effective at preventing surrounding maliciously-used sensors from collecting sensitive data, thus mitigating information leakage threats~\cite{marquardt2011sp}~\cite{genkin2014rsa}~\cite{backes2010acoustic}~\cite{kuhn2002optical}. For example, Marquardt et al.~\cite{marquardt2011sp} suggest placing mobile phones away from keyboards (physical separation); Backes et al.~\cite{backes2010acoustic} propose to mitigate acoustic leakage from printers by either covering them with acoustic shielding foam (acoustic damping) or keeping 2 meters of distance from sensors (physical separation); and Genkin et al.~\cite{genkin2014rsa} propose to use acoustic absorbers to attenuate leaked signals (acoustic damping). This principle has been adopted to prevent optical leakage as well. Fig. \ref{fig:figPolarization} shows a well-known phenomenon that two linear polarization filters aligned at 90 degrees will block all light, while a single filter will let 50\% of original unpolarized light pass through. With the proper set-up of polarization mirrors, it will keep an attacker from eavesdropping without blocking the user's normal use~\cite{backes2009tempest}.

On the other hand, however, the development of shielding techniques is limited by three main drawbacks, which weaken the effects of defenses and may even affect the normal functionality of target CPS.

\begin{itemize}
    \item \textit{Imperfect isolation.} In most scenarios, the isolation provided by shielding is not perfect due to physical arrangement and sensor functionality. For instance, Kune et al.~\cite{kune2013ghost} build a conducting shield exterior for webcams but have to leave large holes for a number of components including the camera lens, two large buttons on the side, the microphone, and the mechanical stand. Similarly, Tesche et al.~\cite{tesche1996emc} point out that the holes left for wires to pass through cause a major degradation to the shielding.
    \item \textit{Signal blocking.} People have found that while the shield blocks external interference, it may negatively impact the normal functionality of CPS as well. For example, it is noted in \cite{selvaraj2018electromagnetic} that magnetic shielding would prevent light from reaching the sensor, and dampening foam utilized to protect an HDD from side acoustic attacks will significantly reduce the HDD's susceptibility to write blocking~\cite{bolton2018blue}.
    \item \textit{Scenario restriction.} As this defense adds an exterior barrier or physical distance to the CPS, it may not be applicable for some scenarios. Marquardt et al.~\cite{marquardt2011sp} state that although distancing can achieve good performance in preventing side-channel inference attacks with surrounding phones, it may be too restrictive for most corporate and home environments where mobile phones are commonly used, and thus isolating phones from keyboards becomes impractical.
\end{itemize}

\cbox{Lessons learned: These two types of defenses are good initial steps to leverage the physical world to defend against cyber-physical attacks. However, challenges still remain in two main aspects: (1) how to verify passively sensed signals where active probing is infeasible (e.g., accelerometer and gyroscope); and (2) how to prevent CPS from leaking signals without blocking their normal functionalities.}

\subsection{Cyber-Domain Defense}\label{subsec:CyberDefense}

Conventional cyber defenses such as firewalls~\cite{nessett1999multilayer} and trusted execution environments (TEE)~\cite{vetillard2015system} are commonly adopted to secure cyber data. However, they cannot be directly applied to effectively defend against cyber-physical attacks because of two challenges: (1) lack of the ability to distinguish maliciously injected signals from benign input, as a result, the malicious inputs cannot be processed and eliminated; (2) inability to prevent signal leakage, especially those previously considered as non-sensitive such as accelerometer and gyroscope data, which has been demonstrated to have great potential for physical activity inference.

To solve each challenge correspondingly, new methodologies have been proposed, which we summarize as three general directions: (1) use \textit{signal analysis} techniques to extract the unique differences between injected signals and benign input, therefore distinguishing the injected signals; (2) change the way that digital signal data are processed and sent for actuation, such as \textit{data filtering} and \textit{sensor fusion}, in an effort to prevent malicious signals from inducing effects on CPS. In this way, although the injected signals are received by CPS as \textit{plausible input}, the other attack requirement of \textit{meaningful response} will not be met (Section \ref{subsec:SigInjThreatModel}); and (3) mask the signal data to prevent leaking physical information, generally via \textit{adding noise for obfuscation}. These mechanisms will be discussed in more detail in the following.

\subsubsection{\textbf{Signal analysis}}
Researchers have investigated the differences between attack signals and legitimate signals, thus coming up with another defense by analyzing the received signal data to only pick the legitimate parts. \textit{Machine learning} and \textit{frequency-domain analysis} are the two commonly adopted methods for distinguishing benign and malicious signals.
\begin{itemize}
    \item \textit{Machine learning.} In this field, machine learning techniques commonly act as tools to distinguish malicious and legitimate signals. For instance, \cite{zhang2017dolphinattack} and ~\cite{yan2019feasibility} extract 15 features from audio data and utilize a supported vector machine (SVM) as a classifier, and achieve a result of 100\% true positive and true negative in detecting a \textit{dolphin attack}. Also, Tharayil et al.~\cite{tharayil2019sensor} propose two \textit{sensor defense in-software} (SDI) methods, one of which is designed to distinguish the benign and malicious sensor output with one-sided or two-sided classifiers pre-trained on benign and spoofed traces.
    \item \textit{Frequency analysis.} Besides machine learning methods, researchers have discovered unique characteristics of frequency distributions of attack signals that help with distinction. To extract and analyze frequency properties such as harmonics and frequency components, commonly utilized techniques include spectrograms~\cite{roy2018inaudible} and frequency-domain analysis~\cite{yan2019feasibility}~\cite{blue2018hello}. For instance, Roy et al.~\cite{roy2018inaudible} manage to detect the injected signal modulated with ultrasound signals because of the well-understood patterns of fundamental frequencies shown in the spectrogram. Similarly, Blue et al.~\cite{blue2018hello} differentiate the audio created by humans and electronic devices by identifying the presence of low-frequency components inherent to the speakers but outside of the human voice range.
\end{itemize}

\subsubsection{\textbf{Data filtering}}
Filtering is designed to mitigate malicious signal values in the cyber domain while minimizing changes to legitimate signals. Due to the simplicity and low cost of implementation, the adoption of filtering has been seen across literature defending against EM~\cite{giechaskiel2019framework}~\cite{kune2013ghost}~\cite{selvaraj2018electromagnetic}, optical~\cite{petit2015remote}, and acoustic~\cite{217565}~\cite{trippel2017walnut} attacks. Overall, this technique can be categorized based on the filter type utilized for defense as follows:
\begin{itemize}
    \item \textit{Cutoff filtering.} This refers to the defense where low-pass filters (LPF), high-pass filters (HPF), and band-pass filters (BPF) are adopted to filter out unwanted content in the received signal data. For example, Kune et al.~\cite{kune2013ghost} successfully attenuate the injected EM signal by 40dB with the help of LPF, while still allowing the audio signals to pass through Bluetooth headsets. Trippel et al.~\cite{trippel2017walnut} suggest using LPF with a cut-off frequency of less than half of the ADC sampling rate to prevent signal aliasing, thus effectively preventing signal injection attacks. 
    
    Moreover, this technique will filter out leaked sensitive data when it passes through. Michalevsky et al.~\cite{michalevsky2014gyrophone} state that the usage of LPF will help to filter out the unintentionally collected speech data in motion sensors, and therefore making it difficult to extract speech from filtered signals. Another similar example is that the notion of filtering can be applied to optical signals. In~\cite{backes2009tempest}, optical notch-filters that serve as a band-stop filter are adopted to filter out the narrow spectrum of LEDs.
    \item \textit{Adaptive filtering.} Different from cutoff filtering where the threshold for filtering is fixed, adaptive filtering first extracts certain characteristics within attack signals, and then filters out signals using these features. Following the proposed cutoff filtering method, Kune et al.~\cite{kune2013ghost} also propose to utilize an additional wire to probe the ambient EM level, and then cancel out the injected signals with collected EM values. 
\end{itemize}

\subsubsection{\textbf{Sensor fusion}}
One of the most popular defenses is to fuse sensor data collected from multiple distinct sensors~\cite{brooks1998multi}, and this technique has been applied in various CPS such as robots~\cite{554212}~\cite{681240}, UAVs~\cite{6696917}~\cite{nemra2010robust}, medical body sensor networks (BSN)~\cite{gravina2017multi}, and most commonly, autonomous vehicles~\cite{bento2005sensor}~\cite{li2013sensor}. In this category, people either add \textit{redundant sensors} of the same type, or \textit{fuse heterogeneous sensors} of different types. As such, even if the attacker successfully spoofs one target sensor, the inconsistency of other sensor readings will reveal the existence of an attack. Therefore, sensor fusion requires the attacker to attack multiple sensors almost simultaneously and thus raises the bar for an attack.

\begin{itemize}
    \item \textit{Redundant sensors.} One method is to add redundant sensors with the same type to detect and monitor unexpected signals. Bolton et. al.~\cite{bolton2018blue} propose to add an additional microphone or fuse several vibration sensors to detect an out-of-band ultrasound injection attack. Similarly, Shin et al.~\cite{shin2017illusion} propose to add multiple LiDAR sensors to form an overlapping field of view, and thus mitigating the effect of saturating and spoofing to some extent. Though this kind of defense is not perfect and may not prevent well-designed attacks, it will significantly increase the efforts required for a successful attack.
    \item \textit{Fusing heterogeneous sensors.} Another method is to fuse data collected from sensors of different types, such as combining the data of a gyroscope and a magnetometer. One typical example is given in \cite{tharayil2019sensor}, where Tharayil et al. determine the mathematical relationships between the measured magnet and vibration data collected from the magnetometer and gyroscope, with which the fused data will identify injection attacks in the cyber domain.
\end{itemize}

\subsubsection{\textbf{Adding noise for obfuscation}} Intentionally combining noise with signals has been shown to be one of the simplest and most effective methods to prevent information leakage attacks~\cite{ha2017side}~\cite{narain2016inferring}~\cite{liu2015good}~\cite{genkin2014rsa}~\cite{kuhn2002optical}. For instance, Ha et al. state that adding random noise to the vibration signals makes the buttons indistinguishable for the keystroke inference attack proposed in~\cite{ha2017side}. Their experiments show that by adding uniform random noise varying from half to double the magnitude of the vibration signal, the performance of the attack sharply deteriorates. Similarly, Liu et al.~\cite{liu2015good} propose to have the keyboard or PC/laptop speakers generating white noise during the user’s typing, which can disturb the malicious app which requires an accurate acoustic data stream for inference. In the same vein, Genkin et al.~\cite{genkin2014rsa} propose to defend against RSA acoustic inference by masking the leaked signals with a carefully designed acoustic noise generator.

\cbox{Lessons learned: Different from traditional cyber defenses, these defenses implemented in the cyber domain are more effective in detecting and mitigating malicious injections, as well as masking leaked information. With the rapid development of sensors and IoT devices, sensor data will become increasingly intertwined to better sense the physical world. However, as it raises the bar for injecting malicious commands via signals, it also leaves more attack surfaces for eavesdropping. Therefore, how to preserve privacy in the context of sensor fusion leaves an open question.}

\subsection{Cyber-Physical-Domain Defense}
As sensors serving as interfaces between cyber and physical domains are widely exploited for attacks, people are seeking to prevent attacks by combining physical stimuli with cyber processing techniques to reduce the system's susceptibility to cyber-physical attacks. It should be noted that, however, defending against information leakage by improving the sensing layer can be tricky, due to the fact that sensing surrounding physical phenomena is what sensors are designed for.

For defending against signal injection attacks, the challenges faced by defenders are similar to what is mentioned in previous sections, i.e. how to distinguish maliciously injected signals and how to eliminate them to prevent \textit{meaningful response}. To solve these challenges, numerous defenses have been proposed that can be categorized into two general categories: (1) adding non-spoofable patterns to the signal itself and the conditioning path, such as \textit{sampling patterns} and \textit{stimulus randomization}, aiming to distinguish malicious signals via users' enriched knowledge (i.e., signal patterns) over the attackers~\cite{trippel2017walnut}~\cite{shoukry2015pycra}; and (2) improving the design of the \textit{cyber-physical layer} to mitigate injected signals, as described in \textit{improve sensor design} and \textit{symmetric receiver}, again by preventing the attackers from satisfying \textit{meaningful response}~\cite{shahrad2018acoustic}~\cite{kune2013ghost}.

\subsubsection{\textbf{Sampling patterns}} As discussed in Section \ref{subsubsec:DCModOOBSigs}, in existing literature on signal injection attacks, ADCs are commonly exploited as demodulators due to the intentionally caused \textit{signal aliasing} effects. However, simple changes to the sample rate (e.g., increase or decrease the sample interval) may not be effective, as the attacker can still craft corresponding signals because the demodulated signal can be predicted according to the fixed sample rate. Therefore, researchers have added patterns to the sampling process, most commonly randomized sampling patterns, to prevent attacks with unpredictable demodulated signals. It is first proposed by Trippel et al.~\cite{trippel2017walnut} where they protect the accelerometers by using randomized ADC sampling, and thus the predictability of ADC's sampling regime is eliminated. Following this work, Tu et al.~\cite{217565} propose to adopt \textit{dynamic sampling} which has similar functionality to randomized sampling but has the advantage of maintaining the accuracy of inertial measurements. Besides, \textit{out-of-phase sampling} proposed by Trippel et al.~\cite{trippel2017walnut} also adopt a special sampling pattern related to the frequency such that the frequency components near the resonant frequency are removed. 

For information leakage attacks, \cite{marquardt2011sp},~\cite{miluzzo2012tapprints},~\cite{Owusu:2012:API:2162081.2162095}, and \cite{han2012accomplice} propose that simply reducing the sensor sampling rate can help mitigate side-channel inference attacks by giving out fewer data. One extreme case is, as proposed by \cite{miluzzo2012tapprints}, to completely pause the functionality of motion sensors when the user is typing on the phone. However, it should be noted that this approach may downgrade the normal functionality of legitimate use of sensors.

\subsubsection{\textbf{Stimulus randomization}}
There is also a direction of defense in which the users add random patterns to the emitted stimulus in the physical domain and expect to detect reflected patterns in the received signals within the cyber domain~\cite{shin2017illusion}~\cite{shoukry2015pycra}~\cite{zhang2020detection}~\cite{xu2018analyzing}. With this method, an attack will be detected if the received signal does not match the pattern due to the attacker being unable to predict them, and thus being unable to craft plausible signals. 
\begin{itemize}
    \item \textit{Add operating pattern.} One way to add patterns is to manually set operating patterns in sensors and systems. Physical challenge-response authentication (PyCRA) is a typical scheme that falls into this category where the emitter is turned off at random instants, and any received signals during this pause indicate a signal injection attack~\cite{shoukry2015pycra}. Also, Shoukry et al.~\cite{shoukry2015pycra} state that the attacker will find it difficult to take real-time actions to blend in with the pattern (e.g., seize the transmitted spoof signals during the pause period of sensors), as they are limited by physical and computational delay. In the context of medical infusion pumps, Park et al. \cite{park2016ain} follow PyCRA and detect spoofing attacks by turning off the drop sensor at random instants. More recently, Zhang et al.~\cite{zhang2020detection} propose a similar pattern to detect electromagnetic interference attacks.
    \item \textit{Manipulate signal parameters.} On the other hand, crafted patterns can be achieved by manipulating signal parameters. For example, Shin et al. in~\cite{shin2017illusion} propose to transmit pulses with randomized waveforms and reject pulses different from the transmitted ones to prevent lidar spoofing attacks and mitigate inter-lidar interference. Similarly, Xu et al.~\cite{xu2018analyzing} use this scheme to enhance security of the ultrasound sensors in autonomous vehicles by randomly manipulating several physical parameters of the waveform.
\end{itemize}

\subsubsection{\textbf{Improve sensor designs}}
Besides adding more redundant sensors to the target CPS, people have proposed ways to improve the sensor design to prevent attacks. For example, Trippel et al.~\cite{trippel2017walnut} suggest designing a secure amplifier by either enlarging its tolerance of large amplitude inputs leveraged by acoustic interference, or adding an LPF or band-stop filter to conduct pre-filtering before the signal enters the amplifier. Son et al.~\cite{son2015rocking} propose to adopt an additional feedback capacitor connected to the sensing electrode, with which the resonant frequency and the magnitude of the resonance effect can be tuned to prevent attacks. 

Moreover, some vulnerable sensors can be replaced with ones that are less susceptible to attacks but show similar functionalities. For instance, both \cite{bolton2018blue} and \cite{shahrad2018acoustic} propose to replace HDDs with Solid-State Drives (SSDs) to mitigate acoustic attacks exploiting the resonance of moving components that are susceptible to vibration.

\subsubsection{\textbf{Symmetric receiver}}
As unintended receivers play an important role in out-of-band signal injection attacks to meet the \textit{plausible input} requirement (Section \ref{subsec:techOOBSigInj}), researchers have found a way to mitigate injected signals by adopting symmetric differential circuits. The key idea is to make the target CPS receive injected signals from two ends, thus canceling the common mode interference present in differential inputs. One typical example is shown in the design of a pacemaker, where the original design of one single conductor in lead connected to the cardiac tissue is phased out by a bipolar lead~\cite{hayes2000cardiac}. Following this work, Kune et. al \cite{kune2013ghost} conduct experiments under similar conditions and observe an attenuation of the injected signal by 30dB using a bipolar lead.

\cbox{Lessons learned: These proposed defenses in the cyber-physical layer primarily focus on improving hardware design in the signal conditioning path. While they are effective in differentiating and mitigating the effects of the injected signals, it remains an open question how this principle can be leveraged to defend against information leakage attacks.}

\subsection{Application-Domain Defense}

As mentioned in Section \ref{subsec:SensorFeed}, sensor readings can directly influence application behaviors. While signal injection attacks generally happen within hardware to craft malicious sensor output, defenses in the \textit{application layer} is tricky as sensor readings have already been disrupted. However, techniques introduced in this domain can still be used to defend against information leakage attacks.

\subsubsection{\textbf{Access permission}} One key cause of the big threat brought about by information leakage attacks is that access to the ``less" sensitive sensor data is not well protected. Therefore, to address this problem, a proper method of authentication and access permission control are proposed to prevent these attacks~\cite{Owusu:2012:API:2162081.2162095}~\cite{liu2015good}~\cite{ha2017side}~\cite{maiti2018side}~\cite{narain2016inferring}. 

Also, Liu et al.~\cite{liu2015good} propose two ways to prevent attacks by either introducing new permission controlling into Android OS and leaving the permission choice to the users, or by adding dynamically-granted permissions mediating access to sensors. Similarly, Ha et al.~\cite{ha2017side} propose to add a second authentication system such as an ID card system, thereby adding another layer of security to prevent keypress inference attacks targeting door locks; and Xu et al.~\cite{xu2012taplogger} propose to modify the system such that motion sensors are considered sensitive and require permissions to access.

\subsubsection{\textbf{Reduce application sampling}} This refers to the defense where the user can reduce the rate that applications sample data collected from sensors~\cite{maiti2018side}. Different from the aforementioned permission control method that forbids unwanted data access, this defense allows sensors to collect data but reduces the application's access to them and therefore brings the challenge of inferring useful information with pieces of fragmented data. 
It should be noted that this defense will be effective for applications that occasionally use sensor data at a low frequency, such as text editors, but may not be appropriate for applications such as games which demand high-frequency usage of sensor data for accurate and prompt feedback.

\subsubsection{\textbf{Parallel task masking}} As signal leakage is hard to completely eliminate, another direction to defend against information leakage is to add parallel workloads on the target CPS. The key idea is to mix the sensitive signals with leakage of less important tasks or dummy tasks, thus the inference will be inaccurate or even fail. In \cite{genkin2014rsa}, the countermeasure against RSA key acoustic inference is to induce additional load on the CPU, thus the extra computation in parallel will mask the leakage of the decryption operation. Similarly, Song et al.~\cite{song2016my} propose to inject dummy tasks for printer nozzles, and thus the additional tasks with the same nozzle speed will be convoluted with the protected printing process.

\cbox{Lessons learned: Most existing application-domain defenses focus on the prevention of information leakage attacks by restricting applications' behaviors. These measures show effectiveness for \textit{co-located attacks}, but the attacker may still be able to launch the attack via crafted applications on their malicious devices. However, given the ease deployment, these defenses are favored for real-world implementations. }

\input{SensorTable}

\subsection{Real World Countermeasures}\label{subsec:RealWorldCounter}

Besides the aforementioned efforts made by academia, industrial companies have become increasingly aware of cyber-physical threats and multiple measures have been taken to secure devices. As discussed in Section \ref{sec:infoleakageatks}, motion sensors such as gyroscopes and accelerometers have been exploited to infer sensitive information such as human speech and keystrokes, and these data can be easily accessed via a browser's API. We investigated this issue on IOS 9/10/11/12 and found that default access to motion sensors in Safari has been removed since version 12.2~\cite{ios12_2}. Additionally, Apple has implemented \textit{adding noise for obfuscation} (Section \ref{subsec:CyberDefense}) by adding random noise to sensor outputs in an effort to defend against \textit{information leakage attacks}. Google Chrome also developed a new feature to inform the user of sensor data being extracted, enabling users to block a site's use of sensors and manage associated permissions~\cite{Google_MotionDefense}. Other popular browsers such as Firefox~\cite{Firefox_MotionDefense} and Vivaldi~\cite{Vivaldi_MotionDefense} have taken similar measures for defense as well.

However, current real-world defenses against cyber-physical attacks still face two challenges.

First, they rely heavily on \textit{application-domain defenses} to protect against cyber-physical attacks. However, the effectiveness of these defenses is limited. For signal injection attacks, they are unable to prevent injected signals from being received by CPS, since the malicious commands are not distinguished at the application level. For information leakage attacks, restricting applications' sensor access on mobile apps can mitigate \textit{co-located attacks}, but the adversary may still launch the attack with their own devices in physical proximity with the victim. Compared to traditional cyber attacks, developers and users have less knowledge of the low-level sensor structure and signal processing~\cite{sadeghi2015security}, which is abstracted away intentionally to simplify the development process. However, such a level of understanding is critical for developing effective defenses.

Second, cyber-physical attacks vary greatly in terms of working principles and techniques. For some attacks, the attacker and victim are on the same physical computing device. For instance, the current restrictions on users' access to motion sensors will mitigate the threat of information leakage via vibrational signals~\cite{michalevsky2014gyrophone}~\cite{kwonghard}, but it will not prevent the attacker from eavesdropping through optical~\cite{Davis2014VisualMic} or EM leakage~\cite{vuagnoux2009compromising}. Also, the attacker is still able to manipulate motion sensor readings by injecting acoustic signals~\cite{son2015rocking}. Compared to the rich diversity in attack vectors, existing defenses often focus on mitigating specific vectors. This is because defenses against each vector can be significantly different from each other, and defenders have to resort to prioritizing high-risk vectors first.

Therefore, a generalized defense that tackles these challenges will likely require efforts in both academia and industry, not only to come up with the solution but also to deploy it.

\section{Future Outlook of Cyber-Physical Security}\label{sec:futureOutlook}

Previous sections have demonstrated recent advances in the field of cyber-physical attacks and defenses. However, given the limitations of existing attacks and remaining challenges of defenses, there is still a large space left for research. Therefore, in this section we propose three questions to inspire new research: (1) what potential threats can cyber-physical attacks post to new technologies; (2) what attacks have yet to be explored; and (3) what are the potential directions for developing novel defenses. Although these three remain open questions and are yet to be answered, we believe there are large opportunities left that will require efforts from both academia and industry to be fully explored.

\subsection{Cyber-Physical Threats in New Era}

The integration of the cyber and physical world is progressing rapidly due to recent advances in related fields such as computer communications. While new technology such as 5G will bring unparalleled benefits to society, it also enables new attacks. In the following we will use 5G and ransomware as examples to discuss how these two emerging technologies could impact the attack landscape of CPS.

\subsubsection{5G communication}
5G being one of the major developments in communication technologies will have a large impact and deployment on cyber-physical systems~\cite{singh20205g}~\cite{ahmad2019deployment}, such as virtual reality~\cite{driscoll2017enabling}, autonomous vehicle~\cite{molina2017lte}, and smart grid~\cite{zerihun2020effect}. As such, it will bring about significant changes in the threat landscape~\cite{lai2020security}, as well as the new defenses~\cite{hussain2020deep}~\cite{ansari2020chaos}. From the communication perspective, millimeter-wave communication and massive MIMO technology will bring revolutionary changes to how content can be delivered. From the network layer, software-defined networks will catalyze service-oriented architecture to enable network utilization more efficiently than ever before. The massive bandwidth and low latency communication from service-oriented architecture will change not only the richness and amount of content that can be delivered to a network node, such as a self-driving car, but also change the granularity and speed of sensing. It becomes possible for a self-driving car to receive live updates of detailed traffic information miles away in a crowded city and take corresponding action~\cite{Tmobile20215G}.

From the security perspective, there are two key changes. First, 5G has become the new frontier of wireless attacks. Existing attacks on cellular networks can still be attempted by attackers, such as attempting to obtain authentication without the right credentials~\cite{ferrag2018security}. The massive MIMO and SDN may also allow network attacks to aim to degrade performance~\cite{choudhary2019survey}. Second, this new architecture may change the scale of sensor attacks, since using networks, CPS can not only gain more sensing input, it might also utilize distributed computing elements such as the edge and cloud to aid in the decision process~\cite{singh20205g}~\cite{ahmad2019deployment}. In order to mislead such CPS, a single sensor injection will most likely be unsuccessful. On the other hand, from the information leakage perspective, the amount of sensor data will allow for much better extraction of physical world activity. While it helps CPS in enabling a better decision, the amount of fine-grained data can pose a significant threat to individual privacy~\cite{borgaonkar2019new}~\cite{kumar2018user}. In the recent pandemic caused by COVID-19, many regions and countries have adopted aggressive contact tracing~\cite{li2020covid}~\cite{ahmed2020survey} and surveillance technology~\cite{calvo2020health} to fight the spread of the disease. A potential fruitful direction is to employ privacy-preserving computation techniques to extract knowledge from data without leaking information on user private data~\cite{privacyguard}~\cite{xiao2020survey}. In a 5G environment, it is possible to take such surveillance to a different level, and many have raised concern over this privacy aspect~\cite{zhang2020americans}. It is important to find the delicate balance between degradation in privacy and societal benefits when it comes to 5G-enabled sensing. We believe secure communication technology will pave the path in this front towards achieving the best of both.

\subsubsection{Ransomware}
Another potential frontier of cyber-physical security is an attack from cyber space to the physical world in the form of ransomware~\cite{richardson2017ransomware}. As discussed in Section \ref{sec:infoleakageatks}, existing work has revealed the threat where attackers can eavesdrop on physical phenomenon via leaked signals. Besides implicitly containing information about the physical world, cyber data can be more valuable if they are critical to maintaining the functionality of various devices and infrastructures. Recently, a patient lost her life due to a ransomware attack locking up the hospital system~\cite{NYTimes2020CyberAttack}. There have also been demonstrations of ransomware on automobiles~\cite{bajpai2020ransomware}~\cite{Naveen2017Connected} as well as hardware-harden ransomware~\cite{zhang2018memory}. With the rapid demand and deployment of sensors in various devices and facilities, sensor data will be playing an increasingly important role in the control loop of CPS. As such, ransomware attacks targeting sensor data are likely to occur, and could cause fatal outcomes to individuals and serious impacts on society as a whole. It is highly probable that in the future where our society increasingly relies on computing systems and sensing components, an attack from cyberspace can easily place a ransom on the user's well-being. It is important to start seeking mitigation methods not just in ensuring the availability of the data, but also the availability of the system.

\subsection{Evolution of the Attack Landscape}
Although the ultimate goal of systemizing cyber-physical attacks is to develop effective defenses, the exploration of new attacks is also a vital role as it enables a better understanding of the threats at play, thus inspiring novel defenses. Therefore, in this section we will discuss the unexplored areas from an offensive perspective.

As presented in Section \ref{sec:siginject} and \ref{sec:infoleakageatks}, existing cyber-physical attacks can be categorized based on attack vectors. Since attack vectors are closely related to the target sensors, summarizing existing attacks by sensors can provide a clear high-level view of research in this area. This is shown in Table \ref{tab:SummarySensor}, which shows that some sensors such as microphones, accelerometers, and gyroscopes have been exploited for both signal injection and information leakage attacks. However, some other sensors such as temperature sensors and ultrasonic sensors have less research on them. This leads to three directions of potential development, which will be discussed in the following.

Firstly, attackers may develop unexplored attacks targeting known sensors listed in Table \ref{tab:SummarySensor} (marked as $\times$). For instance, there is only one work on injecting in-band signals into infrared sensors, but is it possible to inject out-of-band signals without using the optical vector? Also, how can attackers leverage infrared sensors to eavesdrop on physical information? Similarly, the ambient light sensor has only been exploited to steal physical information, but how can attackers manipulate system behavior by injecting signals into ambient light sensors? These remain open questions.

Secondly, attackers may target new sensors. For example, the MEMS barometer widely exists in smartphones to provide altitude information and weather forecasting, and attackers may exploit this sensor to conduct signal injection or information leakage attacks. Given the rapid development and wide deployment of new sensors, we believe new attacks can be proposed targeting CPS and IoT devices equipped with them.

Lastly, when developing new attacks, the selection of the attack vector is of vital importance and determines the methodology to be adopted. Therefore, relevant physical correlations must be considered. As an example, the correlation between acoustic and vibrational signals has been fully researched, where sensitive audio can be inferred from less-protected vibrational signals~\cite{Davis2014VisualMic}~\cite{michalevsky2014gyrophone}~\cite{kwonghard}, while at the same time injected acoustic signals can be leveraged to spoof motion sensor output for malicious goals~\cite{son2015rocking}~\cite{217565}. Researchers may attempt to discover other correlations between physical emanations and exploit them as new attack surfaces. For example, optical light contains energy that has been maliciously crafted to induce voice commands on VCAs~\cite{Sugawara2020LightCommands}. How can attackers leverage this correlation to inject other out-of-band signals?

To sum up, new attacks can be inspired by unexplored attack types, new sensor targets, and physical correlations. These directions leave open questions to the research community.

\subsection{Directions of Defenses}

In Section \ref{sec:defense}, we present multiple limitations faced by conventional cyber defenses when dealing with cyber-physical attacks. For these reasons, researchers have proposed defenses in each layer of CPS architecture to protect systems. However, existing defenses can still be limited to certain application scenarios. In the following we will provide a new perspective to understand the existing defenses based on attack requirements and discuss potential future research directions.

As discussed in Section \ref{subsec:SigInjThreatModel} and \ref{subsec:InfoLeakThreatModel}, there are several necessary conditions for signal injection attacks and information leakage attacks. Effective defenses can be built by preventing the attacker from meeting one or more of these conditions.

For signal injection attacks, the attack requirements are \textit{plausible input} and \textit{meaningful response}. From this perspective, some existing defenses (e.g., \textit{physical barrier})~\cite{park2016ain}~\cite{son2015rocking} propose to block CPS from receiving external signals so that the malicious signals will not be taken by the system as plausible inputs, i.e. the first requirement will not be met. However, existing defenses in this direction are still primitive as they will harm the normal functionalities of CPS due to the blocked sensing components. To solve this challenge, many other defenses turn to the second requirement by detecting or eliminating the received injected signals. Commonly used techniques include \textit{signal analysis}, signal processing (e.g., \textit{data filtering}), and improved sensing-layer designs such as \textit{symmetric receivers} and \textit{sampling patterns}~\cite{trippel2017walnut}~\cite{kune2013ghost}. In this way, the malicious signals will not induce any meaningful effects even if they are taken as plausible inputs by the CPS.

For information leakage attacks, the corresponding attack requirements summarized in Section \ref{subsec:InfoLeakThreatModel} include \textit{covert extraction}, \textit{data exfiltration}, and \textit{information recovery}. As discussed in Section \ref{subsec:RealWorldCounter}, some proposed defenses (e.g., \textit{access permission} and \textit{reduce application sampling}) as well as industrial solutions put emphasis on the first requirement~\cite{Owusu:2012:API:2162081.2162095}~\cite{maiti2018side}. There are also efforts that focus on the third requirement, such as \textit{adding noise for obfuscation} and \textit{parallel task masking}~\cite{genkin2014rsa}~\cite{song2016my}. Therefore, the existing work has examined defenses in two categories: limit the applications' extraction of sensor data to prevent \textit{covert extraction}, and mask the sensor data to mitigate \textit{information recovery}.

In summary, although the implementation of defenses may vary, their working principles can be understood and categorized based on our proposed attack requirements. Therefore, an in-depth understanding of the attack requirements plays an important role in developing effective defenses. New defenses can be inspired by the remaining challenges and unexplored attack requirements. For signal injection attacks, existing defenses on \textit{plausible input} will block legitimate sensing together with malicious signals, which therefore degrades the functionality of CPS. As such, it could be a research direction that seeks to differentiate the malicious signals and block them before they reach the target CPS. For example, shielding microphones from ultrasound may serve as an effective defense against out-of-band acoustic attacks, which rely on ultrasound as the carrier wave. As a result, the functionality of these devices will not be affected as the audible sound can still be sensed. For information leakage attacks, defenses that are based on the second requirement, \textit{data exfiltration}, have not been explored. Following this direction, researchers may make use of network techniques to detect and intercept the network traffic that sends out unauthorized sensor data.

Additionally, it is noteworthy that when considering general defenses against various \textit{cyber-physical attacks}, there are two key factors: (1) the two attack categories, \textit{signal injection} and \textit{information leakage}, differ significantly from each other in terms of attack model, as they are conducted in reverse directions; (2) even within the same attack category, working principles and adopted techniques of different attacks vary according to the attack type (e.g., in-band and out-of-band signal injection) and vector (e.g., EM and optical). To this end, any single defense is not a once-and-for-all solution. Especially with an increasing number of threats that combine multiple attack vectors, it is important to consider the composition of defenses as well. This, therefore, raises some open questions to be answered. What is the most effective strategy for defense fusion? What is the optimal number of defenses? What is the best way to correlate and communicate among these defenses securely? These open questions are yet unexplored.

\section{Conclusion}\label{sec:conclusion}

In this paper, we systemize an emerging category of \textit{cyber-physical attacks} that exploit the interface between cyber and physical domains to perform \textit{signal injection} and \textit{information leakage} attacks. Given the challenges of conventional defenses when faced with \textit{cyber-physical attacks}, we discuss and map existing defenses in the literature to individual attack vectors. 

In light of this emerging threat, gaining a fundamental understanding of the existing cyber-physical attack surface is of vital importance, and paves the path towards generalized defense mechanisms for this cyber-physical interface. Particularly, during the effort of systemization, we find unexplored areas of attack, such as leveraging geo-specific signals to eavesdrop on physical information. We also provide an abstraction of necessary conditions for cyber-physical attacks, and discuss how defenses can be constructed by disabling one or more of these conditions for the adversary. Through this paper, we hope to offer our perspective on the emerging landscape of cyber-physical security, and call for joint efforts of academia and industry to fuel further research in the field to protect critical technologies of the future. 

%% file: SensorTable.tex
\begin{table*}[t]
\caption{Summary of Existing Literature by Sensors}
\centering
\begin{threeparttable}
\begin{tabular}{|c|c|c|c|c|}
\hline
Target Sensor & IB Injection & OoB Injection & Info Leakage & References\\ 
\hline

Microphone & \checkmark & \checkmark & \checkmark 
& \cite{yuan2018commandersong}, \cite{vaidya2015cocaine} \cite{abdullah2019practical}, \cite{197215} / \cite{zhang2017dolphinattack}, \cite{roy2017backdoor}, \cite{kasmi2015iemi}, \cite{Sugawara2020LightCommands}
\\ 
\hline

MEMS Accelerometer & $\times$ & \checkmark & \checkmark
& \cite{217565} \cite{nashimoto2018sensor}, \cite{trippel2017walnut} / \cite{al2013keystrokes}, \cite{cai2011touchlogger}, \cite{Owusu:2012:API:2162081.2162095}, \cite{miluzzo2012tapprints}
\\ 
\hline

MEMS Gyroscope & $\times$ & \checkmark & \checkmark
& \cite{217565} \cite{son2015rocking}, \cite{nashimoto2018sensor} / \cite{miluzzo2012tapprints} \cite{cai2011touchlogger}, \cite{xu2012taplogger}, \cite{al2013keystrokes}
\\ 
\hline

Camera & \checkmark & $\times$ & \checkmark
& \cite{davidson2016controlling}, \cite{petit2015remote}, \cite{sharif2016accessorize}, \cite{7029643}, \cite{eykholt2018robust}, \cite{10.1145/3319535.3354259} / \cite{4531151},  \cite{kuhn2002optical}
\\ 
\hline

Temperature Sensor & $\times$ & \checkmark & $\times$ 
& \cite{10.1145/3319535.3354195}\\ 
\hline

Photodetector/Ambient Light Sensor & $\times$ & $\times$ & \checkmark 
& \cite{kuhn2002optical}, \cite{loughry2002information}\\ 
\hline

Magnetometer & \checkmark & $\times$ & \checkmark
& \cite{10.1007/978-3-642-40349-1_4},  \cite{nashimoto2018sensor}  / \cite{biedermann2015hard}, \cite{song2016my}, \cite{narain2016inferring}\\ 
\hline

GPS Sensor & \checkmark & $\times$ & $\times$
& \cite{warner2002simple}, \cite{humphreys2008assessing}, \cite{kerns2014unmanned}, \cite{217476}\\ 
\hline

Infrared Sensor & \checkmark & $\times$ & $\times$
& \cite{park2016ain}\\ 
\hline

LiDAR & \checkmark & $\times$ & $\times$
& \cite{petit2015remote}, \cite{shin2017illusion}, \cite{cao2019adversarial}, \cite{tu2020physically}\\ 
\hline

Radar & \checkmark & $\times$ & $\times$
& \cite{yan2016can}\\ 
\hline

Ultrasonic Sensor & \checkmark & $\times$ & $\times$
& \cite{yan2016can}\\ 
\hline

\end{tabular}
\end{threeparttable}
    \begin{tablenotes}
      \item IB Injection = In-band signal injection, OoB Injection = Out-of-band signal injection, Info Leakage = Information leakage attack
      \item * References in this table serve as examples of papers belonging to the categories shown in left to right order.
    \end{tablenotes}

\label{tab:SummarySensor}
\end{table*}

%% file: bio.tex
\newpage
\begin{IEEEbiography}
[{\includegraphics[width=1in,height=1.25in,clip,keepaspectratio]{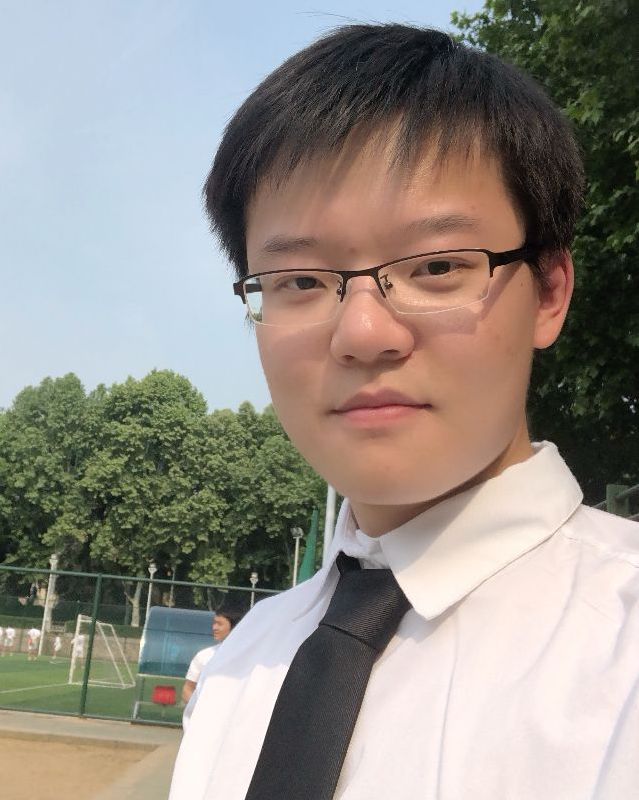}}]{Zhiyuan Yu} (S'20) is a first-year Ph.D. student in the Department of Computer 
Science and Engineering at Washington University in St. Louis. He has interests in IoT security, medical security, and social privacy. Before that, he earned his B.S. degree in Electrical Engineering from the Huazhong University of Science and Technology in 2019. 
\end{IEEEbiography}
\vskip -4in

\begin{IEEEbiography}
[{\includegraphics[width=1in,height=1.25in,clip,keepaspectratio]{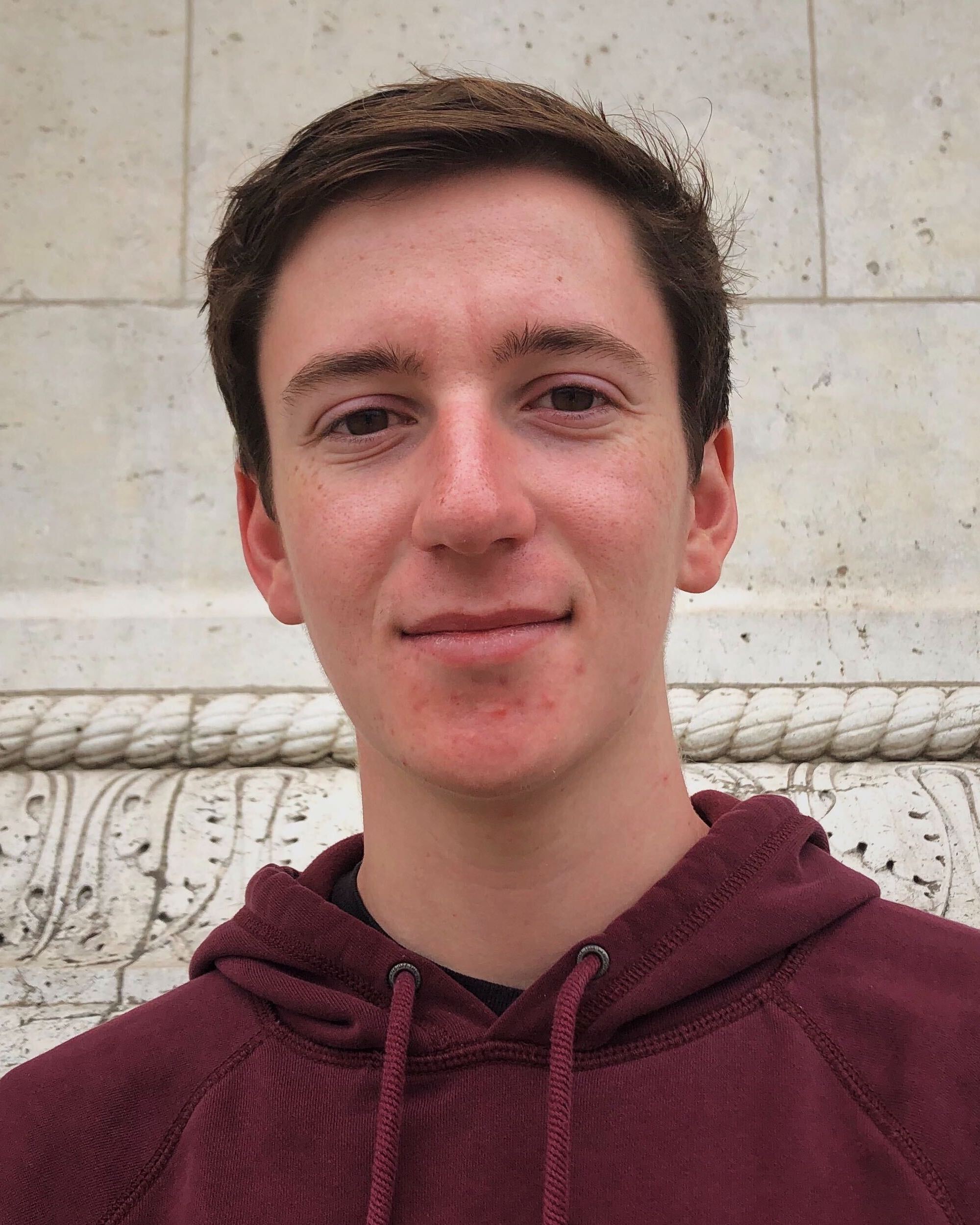}}]{Zack Kaplan} (S'20) is currently a B.S./M.S. student at Washington University in St. Louis studying computer science, psychology, and cybersecurity. He is a member of BEAR5HELL, a Capture the Flag club at Washington University, and has interests in social engineering and network security.
\end{IEEEbiography}
\vskip -4in

\begin{IEEEbiography}
[{\includegraphics[width=1in,height=1.25in,clip,keepaspectratio]{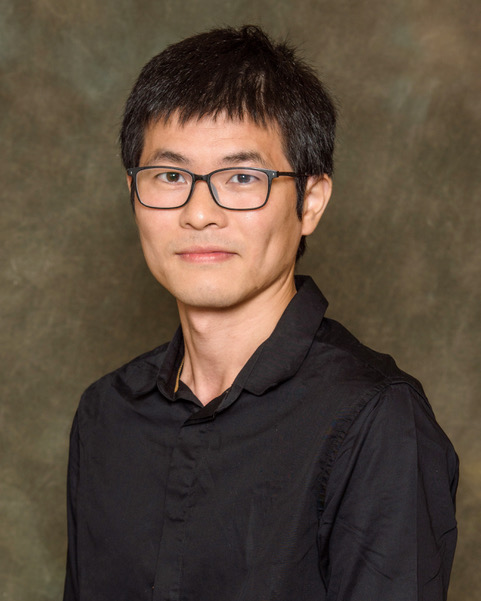}}]{Qiben Yan} (S’11–M’15) is an Assistant Professor in Department of Computer Science and Engineering of Michigan State University. He received his Ph.D. in Computer Science department from Virginia Tech, an M.S. and a B.S. degree in Electronic Engineering from Fudan University in Shanghai, China. He is a recipient of NSF CRII award in 2016. His current research interests include wireless communication, wireless network security and privacy, mobile and IoT security, and big data privacy.
\end{IEEEbiography}
\vskip -4in

\begin{IEEEbiography}
[{\includegraphics[width=1in,height=1.25in,clip,keepaspectratio]{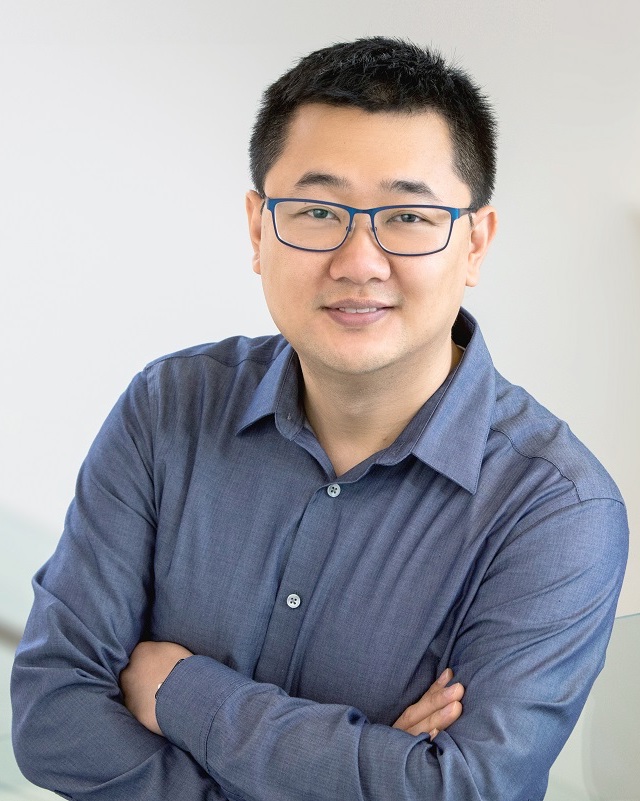}}]{Ning Zhang}
(M'11) received his Ph.D. degree from Virginia Tech in 2016. He is currently an Assistant Professor in the Department of Computer Science and Engineering at Washington University in St. Louis. Before that, he worked in the security industry for ten years. His research focus is system security. 
\end{IEEEbiography}